\documentclass[12pt,letterpaper]{article}

\usepackage[margin=15pt,font=small,labelfont={bf,sf}]{caption}
\usepackage{epsfig}
\usepackage{amsmath}
\usepackage{amssymb}
\usepackage{amsthm}
\usepackage{indentfirst}
\usepackage{xspace}
\usepackage{multirow}
\usepackage{hyperref}
\usepackage{xcolor}
\definecolor{darkred}{rgb}{0.5,0.15,0.15}
\hypersetup{colorlinks=true,urlcolor=darkred,linkcolor=darkred,citecolor=darkred}
\usepackage{verbatim}
\usepackage[letterpaper,margin=1in,headheight=15pt]{geometry}
\usepackage{mathpazo}
\usepackage{microtype}
\usepackage{authblk}
\usepackage{tikz-cd}
\usepackage{silence}
\usepackage[missing=]{gitinfo2}
\usepackage{enumitem}
\usepackage{blkarray}
\usepackage{siunitx}

\usepackage{tikz}
\usetikzlibrary{calc}   
\usetikzlibrary{topaths}
\usetikzlibrary{decorations}
\usetikzlibrary{decorations.pathmorphing}

\numberwithin{equation}{section}

\newcommand{\fg}{{\mathfrak g}}

\newcommand{\fX}{{\mathfrak X}}

\newcommand{\cB}{\ensuremath{\mathcal B}}

\newcommand{\cL}{\ensuremath{\mathcal L}}

\newcommand{\cF}{\ensuremath{\mathcal F}}
\newcommand{\cK}{\ensuremath{\mathcal K}}

\newcommand{\cM}{\ensuremath{\mathcal M}}

\newcommand{\cX}{\ensuremath{\mathcal X}}

\newcommand{\cW}{\ensuremath{\mathcal W}}

\newcommand{\cA}{\ensuremath{\mathcal A}}

\newcommand{\bZ}{\ensuremath{\mathbf Z}}

\newcommand{\bS}{\ensuremath{\mathbf S}}
\newcommand{\bT}{\ensuremath{\mathbf T}}

\newcommand{\R}{\ensuremath{\mathbb R}}
\newcommand{\C}{\ensuremath{\mathbb C}}
\newcommand{\bbP}{\ensuremath{\mathbb P}}
\newcommand{\Z}{\ensuremath{\mathbb Z}}

\newcommand{\bbH}{\ensuremath{\mathbb H}}

\newcommand{\CP}{\ensuremath{\mathbb{CP}}}

\newcommand{\bA}{\ensuremath{\mathbf A}}
\newcommand{\bB}{\ensuremath{\mathbf B}}
\newcommand{\bC}{\ensuremath{\mathbf C}}
\newcommand{\bD}{\ensuremath{\mathbf D}}

\newcommand{\half}{\ensuremath{\frac{1}{2}}}

\newcommand{\cN}{{\mathcal N}}

\newcommand{\hk}{hyperk\"ahler\xspace}

\newcommand{\formal}{{\mathrm{formal}}}

\newcommand{\cwarrow}{\text{\Large$\curvearrowright$}}

\newcommand{\I}{{\mathrm i}}
\newcommand{\E}{{\mathrm e}}
\newcommand{\de}{\mathrm{d}}
\newcommand{\ab}{\mathrm{ab}}

\newcommand{\intro}{\mathrm{intro}}

\newcommand{\abs}[1]{\lvert#1\rvert}

\newcommand{\IP}[1]{\langle#1\rangle}

\newcommand{\eps}{\epsilon}

\newcommand{\ti}[1]{\textit{#1}}

\newcommand{\tX}{\widetilde{X}}

\newcommand{\crem}{{\widehat\Phi}}

\newcommand{\RH}{{\mathrm{RH}}}
\newcommand{\smsol}{{\mathrm{sm}}}

\newcommand{\oper}{{\mathrm{oper}}}
\newcommand{\SL}{{\mathrm{SL}}}
\newcommand{\GL}{{\mathrm{GL}}}
\newcommand{\SU}{{\mathrm{SU}}}

\DeclareMathOperator{\im}{Im}
\DeclareMathOperator{\re}{Re}
\DeclareMathOperator{\Tr}{Tr}

\DeclareMathOperator{\Hol}{Hol}

\DeclareMathOperator{\Sym}{Sym}

\newcommand{\insfig}[2]{\begin{figure}[htbp] \centering \includegraphics[scale=0.3]{figures/#1.pdf} \caption{#2} \label{fig:#1} \end{figure}}

\newcommand{\insfigscaled}[3]{

\medskip
\noindent
\begin{minipage}{\linewidth}
\captionsetup{type=figure}
\makebox[\linewidth]{\includegraphics[keepaspectratio=true,scale=#2]{figures/#1.pdf}}

\captionof{figure}{#3}

\label{fig:#1}
\end{minipage}
\medskip

}

\begin{document}

\renewcommand{\sectionautorefname}{Section}
\renewcommand{\subsectionautorefname}{Subsection}

\title{Exact WKB and abelianization for the $T_3$ equation}
\date{{{\tiny \color{gray} \tt \gitAuthorIsoDate}}
{{\tiny \color{gray} \tt \gitAbbrevHash}}}
\author[1]{Lotte Hollands}
\author[2]{Andrew Neitzke}
\affil[1]{Department of Mathematics, Heriot-Watt University}
\affil[2]{Department of Mathematics, University of Texas at Austin}

\maketitle

{\abstract{We describe the exact WKB method from the point of view
of abelianization, both for Schr\"odinger
operators and for their higher-order analogues (opers). 
The main new example which we consider is the
``$T_3$ equation,'' an order $3$ equation on 
the thrice-punctured sphere, with regular singularities
at the punctures. In this case the exact WKB analysis
leads to consideration of a new sort of Darboux coordinate system on 
a moduli space of flat $\SL(3)$-connections. We give the simplest
example of such a coordinate system, and verify numerically that
in these coordinates the monodromy of the $T_3$ equation has 
the expected asymptotic properties. We also briefly revisit
the Schr\"odinger equation with cubic potential and the
Mathieu equation from the point of view of abelianization.}}

\setcounter{page}{1}

\tableofcontents

\section{Introduction}

\subsection{Exact WKB}

The \ti{exact WKB method} is a scheme for studying
the monodromy (or bound states, or more generally Stokes data) 
of linear scalar differential equations.
This method was initiated in
\cite{MR553093,MR644636,MR729194,MR819680} and subsequently developed
in a large body of literature.
Its origin is in the study of Schr\"odinger equations,
of the form
\begin{equation} \label{eq:schrodinger-intro}
	\left[ \hbar^{2} \partial_z^2 + P(z) \right] \psi(z) = 0,
\end{equation}
where $P(z)$ is holomorphic or meromorphic;
most of the literature is concerned with this case.
For some useful reviews see \cite{MR2182990,MR3706198,MR3280000}.
More recently the exact WKB investigation of higher-order analogues of Schr\"odinger
equations has been taken up, e.g. in 
\cite{MR3617726,MR3539387,MR3380962,MR2758912,MR2758896,MR2311539,MR2132714}.

\subsection{Abelianization}

The method of \cite{Gaiotto:2009hg,Gaiotto2012} leads to a new geometric reformulation of exact WKB, both for Schr\"odinger operators and
their higher-order analogues.
In this reformulation, the key step in exact WKB is a process
of ``abelianization''
which replaces a flat $\SL(K)$-connection $\nabla$ over a surface 
$C$ by a flat $\GL(1)$-connection $\nabla^\ab$ over a $K$-fold covering
$\Sigma \to C$.\footnote{Throughout this paper $\SL(K)$ means $\SL(K,\C)$,
and $\GL(1)$ means $\GL(1,\C) = \C^\times$.} Some aspects of this abelianization process and its relation to exact WKB 
have been further developed in
\cite{Hollands2013,Hollands2017,Coman2018,Nikolaev2019}.

In \autoref{sec:wkb-review} we review
the exact WKB method for Schr\"odinger operators, i.e. the $K=2$ case,
from the perspective of abelianization.
The aim is not to break any really new ground,
but just to explain the theory from the abelianization point of view, 
which is a bit different from the conventional language of exact WKB.

\subsection{Voros symbols for Schr\"odinger equations} \label{sec:intro-voros}

The exact WKB analysis of Schr\"odinger equations
revolves around certain complex-valued functions
$\cX_\gamma(\hbar)$ known as the \ti{Voros symbols}.\footnote{In the main text we will distinguish several different variants of the functions $\cX_\gamma$. The
functions $\cX_\gamma^{\intro}$ we use in the introduction are related to those
appearing in the main text by $\cX_\gamma^{\intro}(\hbar) = \cX_\gamma^{\vartheta = \arg \hbar}(\hbar)$.}
In the language of abelianization, the $\cX_\gamma(\hbar)$ are the holonomies of the $\GL(1)$-connection
$\nabla^\ab$ around $1$-cycles $\gamma$ on the double cover $\Sigma$.

The $\cX_\gamma(\hbar)$ can be expressed as
products of Wronskians of distinguished local
solutions $\psi_i(z,\hbar)$ of \eqref{eq:schrodinger-intro}. 
The solutions $\psi_i(z,\hbar)$ have a dual role:
\begin{itemize}
\item On the one hand, the $\psi_i(z,\hbar)$ are produced by Borel resummation of the 
perturbative WKB series. As a result, one has good control
over their behavior as $\hbar \to 0$, which gives good control
over the behavior of $\cX_\gamma(\hbar)$ as $\hbar \to 0$.
\item On the other hand, the $\psi_i(z,\hbar)$ can be characterized
intrinsically: either as asymptotically decaying solutions 
as $z$ approaches a singularity,
or as eigenvectors of the monodromy as $z$ goes around a loop.
This allows one to identify the $\cX_\gamma(\hbar)$ as
familiar coordinate functions on a moduli space of flat $\SL(2)$-connections.
In a generic enough situation these are 
the ``Fock-Goncharov coordinates'' introduced
in \cite{MR2233852}, as explained in \cite{Gaiotto:2009hg,MR3280000,Hollands2013,Allegretti2018}.
In less generic situations, as discussed in \cite{Gaiotto:2009hg,Hollands2013},
one can get the ``exponentiated 
complexified Fenchel-Nielsen coordinates'' studied in \cite{MR1266284,Kabaya,Nekrasov:2011bc}, 
or other slight variants.
\end{itemize}
The combination of these two points of view on the 
$\cX_\gamma(\hbar)$ is responsible for much of the power of
the exact WKB method.

In this paper we revisit this story in two examples, again with the aim
of showing how exact WKB works in the language of abelianization, and
paving the way for the higher-order case:
\begin{itemize}
\item First, in \autoref{sec:cubic}, we discuss the Schr\"odinger equation with 
cubic potential. This is an instance of \eqref{eq:schrodinger-intro} with 
$P(z) = z^3 - u$. We treat this example relatively briefly. 
We consider only the choice $u = 1$ and real $\hbar$, for which 
the $\cX_\gamma(\hbar)$ are Fock-Goncharov coordinates.

\item 
Second, in \autoref{sec:mathieu}, we discuss the Mathieu equation.
This is an instance of \eqref{eq:schrodinger-intro} with
$P(z) = \frac{1}{z^3} - \frac{2E-\frac14 \hbar^2}{z^2} + \frac{1}{z}$.
We focus on the cases of real $\hbar > 0$ and $E > 1$ or $E < -1$. For
$E < -1$ the $\cX_\gamma(\hbar)$ turn out to be Fenchel-Nielsen
coordinates, and we explain their application to the bound state
problem for the modified Mathieu equation; for $E > 1$ the $\cX_\gamma(\hbar)$
turn out to be a slight variant of the Fock-Goncharov coordinates,
and we explain their application to the quasiperiodic solutions
of the ordinary Mathieu equation.
In either case we do not do anything really new, except perhaps that we
give a new version of the exact quantization condition for the
Mathieu equation, \eqref{eq:mathieu-A-rewritten}, and use it to derive
the width of the gaps at small $\hbar$.
\end{itemize}

\subsection{Exact WKB for order 3 equations}

The next natural test bed is the case of $\SL(3)$-opers: this means 
order $3$ equations of the general form
\begin{equation} \label{eq:deg-3-oper-intro}
	\left[ \partial_z^3 + \hbar^{-2} P_2 \partial_z + (\hbar^{-3} P_3 + \frac12 \hbar^{-2} P'_2) \right] \psi(z) = 0.
\end{equation}

In \autoref{sec:wkb-higher} we describe an extension of the exact WKB method to
this case, again in the language of abelianization.
This extension comes from combining the methods of \cite{Gaiotto2012} with 
the scaling limit of \cite{Gaiotto2014}, now applied to families of $\SL(3)$-connections.

As in the order $2$ case, the theory is founded on the existence of
distinguished local solutions $\psi_i(z,\hbar)$ of \eqref{eq:deg-3-oper-intro},
with $\hbar \to 0$ asymptotics given by the WKB series.
In contrast to the order $2$ case, however, as far as we know, there are not yet theorems
guaranteeing the existence of these local solutions. Thus the higher-order exact WKB method
is not yet on solid footing.

Nevertheless, we press on, making the assumption that the
$\psi_i(z,\hbar)$ do exist.
Then, as before, one can use them to construct functions $\cX_\gamma(\hbar)$, which we call
\ti{spectral coordinates} because of their relation to abelianization; 
one might also have called them \ti{higher-order Voros symbols}.
Now the question arises: can we identify the $\cX_\gamma(\hbar)$ as some concrete
coordinate functions on moduli of flat $\SL(3)$-connections --- or, essentially 
equivalently, can we give an intrinsic characterization of 
the $\psi_i(z,\hbar)$ in terms of their monodromy properties?

For some examples of equations \eqref{eq:deg-3-oper-intro}, 
the expected picture is well understood, and similar to
the order $2$ case.
One such situation arises if $C$ is a punctured surface, $P_2$ is meromorphic and 
generic with generic residues, and 
$P_3$ is small compared to $P_2$. In this case
the $\psi_i(z,\hbar)$ in most of $C$ can be described 
by beginning with the filtrations induced by the asymptotic growth
rate at the punctures, and then using the linear algebra of ``snakes'' 
as introduced by Fock-Goncharov
\cite{MR2233852,Gaiotto:2012db}. The $\cX_\gamma(\hbar)$ then
turn out to be higher-rank Fock-Goncharov coordinates. 
Another such situation arises if $(P_2, P_3)$ is a \ti{generalized Strebel} pair
of \ti{length-twist type} 
as defined in \cite{Hollands2017}; then the $\psi_i(z,\hbar)$ can be characterized
as monodromy eigenvectors, and the $\cX_\gamma(\hbar)$ turn out to be 
``higher length-twist'' coordinates generalizing Fenchel-Nielsen.
In this paper we do not revisit these cases.

\subsection{The \texorpdfstring{$T_3$}{T3} equation}

Instead, in \autoref{sec:t3}, we turn our attention to the $T_3$ equation.
This is a specific instance of \eqref{eq:deg-3-oper-intro},
defined on the Riemann surface $C = \CP^1$ with three generic
regular singularities (``full punctures'' 
in the physics literature),
and depending on a parameter $u \in \C$: namely, we take
in \eqref{eq:deg-3-oper-intro}
\begin{equation} \label{eq:t3-equation-intro}
	P_2 = \frac{9 \hbar^2 z}{(z^3 - 1)^2}, \qquad P_3 = \frac{u}{(z^3-1)^2}.
\end{equation}
This equation is a particularly interesting test case.
One way to understand this is to remark that this family of opers
corresponds (in the sense of \autoref{sec:physics-intro} below)
to a specific $\cN=2$ superconformal quantum field theory, the Minahan-Nemeschansky
theory with flavor symmetry $E_6$ \cite{Minahan:1996fg,Gaiotto:2009we}, 
which is known to be difficult to study by conventional Lagrangian 
field theory methods.

As expected, in this case we meet new difficulties. One source of these difficulties
is that the Stokes graphs can be rather wild for general $(u,\hbar)$.
We thus restrict ourselves to only the simplest situation, which arises
when $u' = u / \hbar^3$ is real and positive; in this case the Stokes graph
is actually very simple. It is shown in \autoref{fig:circle-network} below.

Then we find that the $\psi_i(z,\hbar)$ are
solutions of an interesting linear algebra problem:
relative to the local basis $\{ \psi_1, \psi_2, \psi_3 \}$ near $z=0$,
the monodromies $\bA$, $\bB$, $\bC$ around the
three punctures must have zeroes in specific positions,
\begin{subequations} \label{eq:monodromy-constraints-intro}
\begin{alignat}{3}
	\bA &= \begin{pmatrix} * & 0 & * \\ * & * & * \\ * & * & * \end{pmatrix}, \quad & 
	\bB &= \begin{pmatrix} * & * & * \\ * & * & 0 \\ * & * & * \end{pmatrix}, &
	\bC &= \begin{pmatrix} * & * & * \\ * & * & * \\ 0 & * & * \end{pmatrix}, \\
	\bA^{-1} &= \begin{pmatrix} * & * & * \\ 0 & * & * \\ * & * & * \end{pmatrix}, &
	\bB^{-1} &= \begin{pmatrix} * & * & * \\ * & * & * \\ * & 0 & * \end{pmatrix}, \quad &
	\bC^{-1} &= \begin{pmatrix} * & * & 0 \\ * & * & * \\ * & * & * \end{pmatrix}.
\end{alignat}
\end{subequations}
The best approach we have found to this linear algebra problem involves a 
bit of algebraic geometry,
as we describe in \autoref{sec:enumerating-special-bases}:
we reduce the problem to finding fixed points of a certain degree $64$ birational
automorphism of $\CP^2$, and then identify these fixed points with singularities
in the fibers of a certain rational elliptic surface.

At any rate, once this problem has been solved, we can then compute
the spectral coordinates $\cX_\gamma(\hbar)$ for the $T_3$ equation.
The concrete formulas are given in \eqref{eq:spectral-coords-t3} below,
reproduced here: for a basis $\{\gamma_A, \gamma_B\}$ of $H_1(\Sigma,\Z)$,\footnote{Here and elsewhere in
this paper, unless explicitly noted, $\sqrt{\cdot}$ denotes the principal branch of the square root.}
\begin{subequations} \label{eq:spectral-coords-t3-intro}
\begin{align}
\cX_{\gamma_A} &= \frac{[\psi_2 , \psi_3 , \psi_1]}{[\bC^{-1} \psi_3 , \bA \psi_2 , \psi_1]}, \label{eq:XA-t3-intro} \\
\cX_{\gamma_B} &= \sqrt{- \frac{[\bC \psi_1 , \bB^{-1} \psi_2 , \psi_3][\bC \psi_1 , \psi_1 , \psi_3] [\psi_2 , \bA \psi_2 , \psi_1] [\bB \psi_3 , \bA^{-1} \psi_1 , \psi_2] [\bB \psi_3 , \psi_3 , \psi_2]}{[\psi_2 , \bB^{-1} \psi_2 , \psi_3][\bC^{-1} \psi_3 , \bA \psi_2 , \psi_1] [\bC^{-1} \psi_3 , \psi_3 , \psi_1] [\psi_1 , \psi_3 , \psi_2] [\psi_1 , \bA^{-1} \psi_1 , \psi_2]} } \label{eq:XB-t3-intro}
\end{align}
\end{subequations}
where $[\psi, \psi', \psi'']$ means the Wronskian of the three solutions
$\psi, \psi', \psi''$.
This gives a local Darboux coordinate system
on the moduli space of flat $\SL(3)$-connections with unipotent holonomy
on the thrice-punctured sphere. As far as we know, this coordinate
system has not been considered before. What our computations say is that these particular coordinates
arise naturally from the WKB analysis of the 
equation \eqref{eq:deg-3-oper-intro}, \eqref{eq:t3-equation-intro}.

Combining our conjectures and computations, one can extract a concrete
prediction: the quantities \eqref{eq:spectral-coords-t3-intro},
computed from the monodromy of the $T_3$ equation, should have a specific asymptotic series expansion, with leading behavior
$\cX_\gamma \sim \exp(Z_\gamma / \hbar)$, as $\hbar \to 0$ in an appropriate sector.
Here the constants $Z_\gamma \in \C$ are periods, $Z_\gamma = \oint_\gamma P_3^{1/3}$,
given explicitly in \eqref{eq:T3-periods} below.
We have implemented this computation numerically and find very good agreement
(see e.g. \autoref{fig:X-numerics} below).
We regard this as evidence that the higher-order exact WKB method indeed
works.

\subsection{Integral equations and analytic structures}

A relatively recent development in the exact WKB method is the discovery that
the functions $\cX_\gamma(\hbar)$ are, quite generally, solutions
of integral equations in the $\hbar$-plane.
A general form of these integral equations was formulated in \cite{Gaiotto2014}
(see \eqref{eq:rh-integral-equation} below),
generalizing some cases which had been known before.
In particular, the equations closely resemble the thermodynamic
Bethe ansatz, and some cases literally match with the
high-temperature (chiral) limit of the thermodynamic Bethe ansatz
for specific integrable models; these cases had been studied as part of
the ODE-IM correspondence, explained in e.g. \cite{Dorey1998,Dorey2007}.

One way to motivate these equations is to argue
that their solutions solve a certain Riemann-Hilbert problem: they
have the same analytic structure and $\hbar \to 0$ asymptotic properties 
as the desired functions $\cX_\gamma(\hbar)$ have.
One hopes that these properties are sufficient to characterize $\cX_\gamma(\hbar)$.\footnote{
For a more elementary example, if a function $x(\hbar)$ is known to be holomorphic for
$\hbar \in \C^\times$, $x(\hbar) \to c$
as $\hbar \to 0$, and $x(\hbar)$ is bounded as $\hbar \to \infty$, then we can conclude $x(\hbar) = c$.}

The general idea of determining the $\cX_\gamma(\hbar)$ from their analytic
properties has appeared before in the exact WKB literature, 
e.g. in \cite{MR729194} under the name ``analytic bootstrap.''
In another direction, the same Riemann-Hilbert problem has been studied recently
in relation to the topological string \cite{Bridgeland2016}.

In various sections of this paper we consider integral equations for our $\cX_\gamma(\hbar)$:
\begin{itemize}
	\item In \autoref{sec:integral-equations-cubic} we review the integral equations obeyed 
	by $\cX_\gamma(\hbar)$ for the Schr\"odinger equation with cubic potential. In this case
	the $\cX_\gamma(\hbar)$ are Fock-Goncharov coordinates. This case
	is by now reasonably well understood in the literature; it was discussed already in
	\cite{Dorey1998}, in \cite{Gaiotto2014}, and more recently in \cite{Ito2018}.

	\item In \autoref{sec:integral-equations-mathieu} we propose integral equations for
	$\cX_\gamma(\hbar)$ for the Mathieu equation, in the case where $\cX_\gamma(\hbar)$ are
	complexified exponentiated Fenchel-Nielsen coordinates. This case is somewhat more difficult;
	we find definite equations, which do seem to be satisfied by the $\cX_\gamma(\hbar)$ in 
	numerical experiments, but we
	are not able to use the equations to compute the $\cX_\gamma(\hbar)$ directly.

	\item Finally in \autoref{sec:integral-equations-t3} we write one version of the integral
	equations for the $\cX_\gamma(\hbar)$ of the 
	$T_3$ equation. Here, in order to determine the equations completely,
	one needs to find a closed formula for a 
	certain transformation $\bS_{0,\frac{\pi}{3}}$ relating two different
	branches of $\cX_\gamma(\hbar)$; we formulate this problem carefully
	but do not solve it. We also explain how
	one can approximate $\bS_{0,\frac{\pi}{3}}$ using some integer invariants previously computed in \cite{Hollands:2016kgm} (BPS indices in the Minahan-Nemeschansky $E_6$ theory), and give some numerical evidence
	that this approximation works.
\end{itemize}

All of these analyses just barely scratch the surface; there is much more to do here.

\medskip

A closely related issue is that of the analytic structure of the maximal analytic
continuation of $\cX_\gamma(\hbar)$ from a given initial $\hbar$.
Zeroes, poles, and branch cuts can all occur:

\begin{itemize}
	\item In \autoref{sec:cubic-analytic-cont} we briefly recall the analytic properties of
	the Fock-Goncharov coordinates
	$\cX_\gamma(\hbar)$ for the Schr\"odinger equation with cubic potential. These 
	are relatively simple: the maximal analytic continuation
	is defined on a fivefold cover branched only at $\hbar = 0$, 
	with a concrete monodromy action \eqref{eq:cubic-XAXB-monodromy}.
	The $\cX_\gamma(\hbar)$ can also have poles or zeroes, which come from bound states of the
	Schr\"odinger equation; they occur in infinite discrete families.
	\item In \autoref{sec:mathieu-analytic-cont} we describe the analytic properties of the
	complexified exponentiated Fenchel-Nielsen coordinates $\cX_\gamma(\hbar)$ for the 
	Mathieu equation in the regime $E < -1$.
	These are a bit more complicated: there is infinite-order monodromy \eqref{eq:mathieu-monodromy-1}
	around $\hbar = 0$, and also order-$2$ monodromy \eqref{eq:mathieu-monodromy-2} 
	around an infinite discrete family of other 
	points. The latter points are analytically continued versions of the edges of the bands/gaps in the 
	Mathieu spectrum. In this case we did not explore the positions of poles or zeroes.
	\item In \autoref{sec:t3-analytic-cont} we consider the analytic properties of
	our new coordinates $\cX_\gamma(\hbar)$ for the $T_3$ equation. The picture we find,
	through numerical experimentation, is that
	the maximal analytic continuation lives on a threefold cover, with order-$3$ monodromy around
	$\hbar = 0$, and order-$2$ monodromy around
	$6$ other points. In terms of the coordinate $u' = u / \hbar^3$, the picture is simpler:
	there is only order-2 branching, around the points $u' = \pm u'_*$ where $u'_* \approx 0.041992794$.
	The $\cX_\gamma(\hbar)$ can also have poles or zeroes, which numerically do appear to 
	occur, in infinite discrete families.
\end{itemize}

\subsection{Supersymmetric QFT} \label{sec:physics-intro}

Over the last decade it has turned out that exact WKB is closely connected to
$\cN=2$ supersymmetric quantum field theory in four dimensions. This work was 
motivated by an attempt to understand these connections better.
They arose in two different ways:
\begin{itemize} 
\item On the one hand, \cite{Nekrasov:2009rc} discovered a new connection between 
Nekrasov's $\Omega$-background partition function $\bZ$ in $\cN=2$ 
theories and quantum integrable systems. 
For $\cN=2$ theories of class $S$, the AGT correspondence says $\bZ$ is related to 
Liouville conformal blocks on a Riemann surface $C$ \cite{Alday:2009aq}, while
the quantum 
integrable systems turned out to be spectral problems for Schr\"odinger equations \eqref{eq:schrodinger-intro} on $C$. The investigation of this connection between Schr\"odinger
equations, Liouville conformal blocks and topological strings 
was carried out using WKB methods
beginning in \cite{Mironov2009a,Mironov2009}. This connection has led to a flow
of ideas in both directions. For example, it has been proposed that using
exact WKB one can obtain ``nonperturbative'' information about $\bZ$,
e.g. \cite{Krefl2013,Kashani-Poor2015,Ashok2016}; also,
techniques from the study
of $\bZ$, such as the holomorphic anomaly equations, 
have been imported back to WKB, e.g.
\cite{Codesido2016,Codesido2017}.

\item On the other hand, studying BPS states and supersymmetric defects 
in $\cN=2$ theories of class $S$, \cite{Gaiotto:2009hg,Gaiotto2012} 
were led to develop a version
of exact WKB which applies to a slightly different sort of equation:
instead of the 1-parameter families \eqref{eq:schrodinger-intro} 
or \eqref{eq:deg-3-oper-intro} parameterized
by $\hbar \in \C^\times$, 
\cite{Gaiotto:2009hg,Gaiotto2012} treat
a 2-parameter family of covariant constancy equations for flat 
$\SL(K)$-connections $\nabla_{R,\zeta}$, 
parameterized by $R \in \R_+$ and $\zeta \in \C^\times$:
\footnote{The family of flat connections
 $\nabla_{R,\zeta}$ arises from a solution $(D,\varphi)$ of Hitchin's equations, through the 
 formula
$\nabla_{R,\zeta} = R \zeta^{-1} \varphi + D + R \zeta \varphi^\dagger$.}
\begin{equation} \label{eq:hitchin-flatness-intro}
	\nabla_{R,\zeta} \psi(z) = 0.
\end{equation}

Despite the difference between \eqref{eq:schrodinger-intro} and \eqref{eq:hitchin-flatness-intro}, the 
geometric structures which appear in their exact WKB analysis are the same; in particular
the \ti{Stokes graphs} in exact WKB are the same as the \ti{spectral networks} in \cite{Gaiotto2012}.
A reason for this was conjectured in \cite{Gaiotto2014}, as follows: in the case $K=2$,
taking the scaling limit
$R \to 0$, $\zeta \to 0$ while holding $\hbar = \zeta / R$ fixed reduces the 2-parameter family
of equations \eqref{eq:hitchin-flatness-intro} to the 1-parameter family \eqref{eq:schrodinger-intro}.
For general $K \ge 2$, this scaling limit similarly reduces \eqref{eq:hitchin-flatness-intro} to a
1-parameter family of $\SL(K)$-opers, i.e. order $K$ linear scalar ODEs.
This conjecture was proven in some cases in \cite{Dumitrescu2016}.

\end{itemize}

In this paper we mostly focus on questions internal to exact WKB,
using these developments in physics only as motivation.
However,
in the final section, \autoref{sec:physics}, we return briefly 
to the question of what our computations mean for $\cN=2$ field theory.
We propose that the construction of the functions $\cX_\gamma$ 
provided by exact WKB is related to a
construction of supersymmetric local operators in the field theory
in $\Omega$-background, and comment on the expected relation 
of the $\cX_\gamma$ to the Nekrasov $\Omega$-background partition 
function $\bZ$, motivated by the ideas of \cite{Nekrasov:2011bc}.

\subsection{Some questions}

This project has raised, at least in our minds, many unanswered questions.
Here are some:

\begin{itemize}

\item In our study of the $T_3$ equation we consider only a specific
Stokes graph, the simplest of infinitely many which occur at different
points in the $u'$-plane. Even for this Stokes graph the monodromy
properties of the local WKB solutions turn out to involve a complicated
linear algebra problem, which seems to require real work to solve (in \autoref{sec:enumerating-special-bases}).
What kind of problem will appear at other points of the $u'$-plane?
Is there some systematic way of solving all of them at once?

\item Similarly, what happens in other Minahan-Nemeschansky theories,
like the $E_7$ or $E_8$ theories? Is there a uniform way of describing
the $\cX_\gamma$ and their behavior, or do we have to treat each
example separately?

\item In this paper we reformulate various aspects of exact WKB in the language of abelianization.
One notable exception is the ``P/NP relation'' discussed recently in 
the WKB literature, e.g. \cite{dha,alv1,alv3,Dunne2013,Dunne2014,Codesido2016,Basar2017,Ito2018}.
Does this part of the story have a useful geometric reformulation in the language of abelianization?

\item In our discussions of TBA-type integral equations in \autoref{sec:integral-equations-mathieu}
and \autoref{sec:integral-equations-t3} we make some progress, but do not 
attain the ultimate goal, which would be to completely determine the 
monodromy of the oper in terms of these integral equations. 
It would be very interesting to push this project further. For the $T_3$
equation the main obstruction to doing so is that we have not understood
the coordinate transformation $\bS_{0,\frac{\pi}{3}}$ appearing
in \autoref{sec:integral-equations-t3}. Finding a closed form for 
this transformation would be very interesting in its own right since it
would be equivalent to completely determining the BPS spectrum of 
the $E_6$ Minahan-Nemeschansky theory.

\item In \autoref{sec:t3-analytic-cont} 
we uncover an unexpectedly
interesting analytic structure for the functions $\cX_\gamma$
in the case of the $T_3$ equation.
It is natural to ask what is the physical 
meaning in the $E_6$ Minahan-Nemeschansky theory 
of the nonperturbative monodromy we find around the points
$u' = \pm u'_*$. (A similar question for pure $\cN=2$ supersymmetric
$U(N)$ gauge theory was discussed in
\cite{Jeong2017}, where the relevant physics was proposed to be
the appearance of new massless fields in the theory in $\Omega$-background;
perhaps the monodromy we have found has
a similar meaning.)
It would also be interesting to prove rigorously that there 
is no monodromy around any other points
in the $u'$-plane.

\item In this paper we treat only the case of the $T_3$ equation with unipotent monodromy,
corresponding to the massless Minahan-Nemeschansky
theory.
There is a natural perturbation to consider, taking semisimple monodromy
instead of unipotent, corresponding to the mass-perturbed Minahan-Nemeschansky
theory. It would
be interesting to study this case systematically --- in particular, to see
how the analytic structure of the $\cX_\gamma$ is modified in this case.
(On general grounds we should expect that the structure could be more complicated;
in the massless case the monodromy came ultimately from the fact that there
were $4$ discrete abelianizations of
the $T_3$ equation; in the massive case there are $12$ discrete abelianizations
rather than $4$.)

\item The exact WKB analysis we describe in this paper for equations of
order $K > 2$ is still conjectural, mainly because it has not yet been
proven that the local WKB series are Borel summable. It would be very
interesting to close this gap, perhaps by extending the approach of
Koike-Sch\"afke from the $K=2$ case, 
or by using the integral equations of \cite{Gaiotto:2011tf}.

\item The Darboux coordinates we encounter on moduli of $\SL(3)$-connections over the thrice-punctured
sphere are new as far as we know. It would be interesting to understand explicitly their
relation to other known Darboux coordinate systems on the same space, e.g. the Fock-Goncharov coordinates \cite{MR2233852},
the coordinates introduced
by Goldman \cite{MR1053346}, or the coordinates
obtained from conformal field theory in \cite{Coman2017}.

\item Finally, as we discuss in \autoref{sec:physics}, the exact WKB computations we make
here should have a precise meaning in terms of $\cN=2$ supersymmetric quantum field theories in the
$\Omega$-background. We make some proposals in this direction, but to put these
proposals on a firm footing would seem to require new constructions of supersymmetric
local operators and boundary conditions compatible with the $\Omega$-background. 
It would be very interesting to develop this story further.

\end{itemize}

\subsection*{Acknowledgements}

We thank Dylan Allegretti, Tom Bridgeland, Gerald Dunne,
Dan Freed, Marco Gualtieri, Kohei Iwaki, Saebyeok Jeong, Anton Leykin, Marcos Mari\~{n}o, 
Nikita Nikolaev, Shinji Sasaki, Bernd Sturmfels and Joerg Teschner for useful and enlightening discussions.
LH's work is supported by a Royal Society Dorothy Hodgkin
Fellowship.
AN's work on this paper was supported by NSF grant DMS-1711692 and by a Simons
Fellowship in Mathematics.

\section{Exact WKB for Schr\"odinger equations} \label{sec:wkb-review}

We consider a holomorphic Schr\"odinger equation, of the local form\footnote{In comparing to the ordinary Schr\"odinger equation on the real
line we would have $P = 2(E-V)$.}
\begin{equation} \label{eq:schrodinger}
	\left[ \hbar^{2} \partial_z^2 + P(z,\hbar) \right] \psi(z) = 0.
\end{equation}
The equation \eqref{eq:schrodinger} can be given a global
meaning
on a Riemann surface $C$ equipped with a spin structure
and complex projective structure. 
In that case $\psi(z)$ must be interpreted
as a section of $K_C^{-\half}$, and $P(z,\hbar)$ as a meromorphic 
quadratic differential.
All of our considerations extend to this situation.
Nevertheless, most of the important constructions
can be understood concretely
in a single coordinate patch, and we will write them that way
throughout.

\subsection{WKB solutions} \label{sec:wkb-solutions}

The WKB method is often described in terms of distinguished
local \ti{WKB solutions}.
In this section we briefly recall the construction
of these solutions. (To forestall
confusion we emphasize
that the WKB solutions are exact solutions, not approximate
solutions.)

Suppose we fix a contractible open set $U \subset C$,
a local coordinate $z$ on $U$, and a point $z_0 \in U$.
A WKB solution of \eqref{eq:schrodinger} on $U$ means a solution of the form
\begin{equation} \label{eq:wkb-solutions}
	\psi(z) = \exp \left( \hbar^{-1} \int_{z_0}^z \lambda(z) \, \de z \right).
\end{equation}
For $\psi(z)$ to be a solution of \eqref{eq:schrodinger}, $\lambda$ must
obey the Riccati equation,
\begin{equation} \label{eq:riccati}
	\lambda^2 + P + \hbar \partial_z \lambda = 0.
\end{equation}
The first step in constructing such a $\lambda$ is to build a formal series solution $\lambda^\formal$ of \eqref{eq:riccati} in powers of $\hbar$.
The order-$\hbar^0$ part of \eqref{eq:riccati} is
\begin{equation} \label{eq:riccati-order-0}
	y^2 + p = 0,
\end{equation}
where $y$ (resp. $p$) is the $\hbar^0$ term in $\lambda$ (resp. $P$).\footnote{In the important
special case of $\hbar$-independent $P$, we just have $p = P$.}
Thus we have a two-fold ambiguity, resolved by
choosing one of the two square roots of $-p$.
It will be important to keep careful track of this choice of square root. Thus we introduce the Riemann surface of $\sqrt{-p}$,
\begin{equation} \label{eq:spectral-cover}
\Sigma = \{y^2 + p = 0\}	.
\end{equation}
$\Sigma$ is a branched double cover of $C$.
A sheet of the covering $\Sigma$ corresponds to a choice of $y$
obeying \eqref{eq:riccati-order-0}. We use the generic labels $i,j$
to represent the sheets, and $y_i$, $y_j$ for the corresponding
square roots of $-p$.\footnote{In the WKB literature it is common to write the two square roots simply as $\pm \I \sqrt{p}$, and label the two
sheets as $+$, $-$ instead of $i$, $j$.}

We now choose a sheet $i$, and consider a formal
series solution $\lambda^\formal_i$ of \eqref{eq:riccati}, 
where we choose the $\hbar^0$ term to be $y_i$.
The higher-order expansion of $\lambda^\formal_i$
is then uniquely fixed by \eqref{eq:riccati}, taking the 
form 
\begin{equation} \label{eq:wkb-asymptotics}
	\lambda_i^\formal = y_i + \sum_{n = 1}^\infty \hbar^n \lambda_i^{\formal,n}. 
\end{equation}
For example, if $P$ is $\hbar$-independent, this expansion is
\begin{equation} \label{eq:wkb-asymptotics-concrete}
 	\lambda_i^\formal = y_i  - \hbar \frac{P'}{4P} + \hbar^2 y_i \frac{5 P'^2 - 4 P P''}{32 P^3} + \cdots.
 \end{equation}
Note that although $\lambda_i^\formal$ is a formal solution
of the differential equation \eqref{eq:riccati},
in writing this solution we do not have to do any integrals!

The series $\lambda_i^\formal$ in \eqref{eq:wkb-asymptotics} is 
generally not convergent. Nevertheless, one might
hope that we could interpret $\lambda_i^\formal$ as an 
asymptotic series, and that there would be a unique actual 
solution $\lambda_i$ with $\lambda_i \sim \lambda_i^\formal$
as $\hbar \to 0$.
It turns out that the situation is more complicated. There is
no $\lambda_i$ which has this asymptotic expansion,
if $\hbar$ is allowed to approach $0$ from
an arbitrary direction in the complex plane.
The best one can do in general is to ask for a solution  $\lambda_i^\vartheta$ which
has the expansion  $\lambda_i^\vartheta \sim \lambda_i^\formal$ 
as $\hbar \to 0$ while staying within a closed 
half-plane
\begin{equation}
 \bbH_\vartheta = \{\re (\E^{-\I \vartheta} \hbar) \ge 0\}.
\end{equation}
Such a $\lambda_i^\vartheta$ actually does exist\footnote{This has been a folk-theorem for 
some time, at least for the case of $p$ with
sufficiently generic residues, and a proof has been announced by Koike-Sch\"afke. See \cite{MR3706198} for an account.}, but only 
away from the \ti{$\vartheta$-Stokes curves of type $ij$}, which we 
define next.

For simplicity we assume henceforward 
that $p(z)$ has only simple zeroes.
Then, from each zero of $p(z)$ there emanate three trajectories
along which $\int \E^{-\I \vartheta} \sqrt{-p(z)} \de z$ is purely 
real; we call these $\vartheta$-Stokes curves.
The $\vartheta$-Stokes curves make up the Stokes 
graph $\cW(p, \vartheta)$.
Each Stokes curve is oriented away from the zero, and carries
a label $ij$, determined such that 
$\E^{-\I \vartheta} (y_i - y_j) \de z$ is \ti{positive} 
along the oriented curve.\footnote{Since $y_j = -y_i$ we could also
have just written that $\E^{-\I \vartheta} y_i$ is positive,
and we could have labeled the curve just by the single index
$i$ instead of the ordered pair $ij$.
Our redundant-looking notation is chosen with an eye toward the 
generalization to higher-order equations, in \autoref{sec:wkb-higher} below.} See \autoref{fig:sample-networks-combined} for some examples 
of $\vartheta$-Stokes graphs
in the case where $P(z)$ is a polynomial potential in the plane; 
many other such examples can be found e.g. in 
\cite{MR2182990,Gaiotto:2009hg}.

\insfigscaled{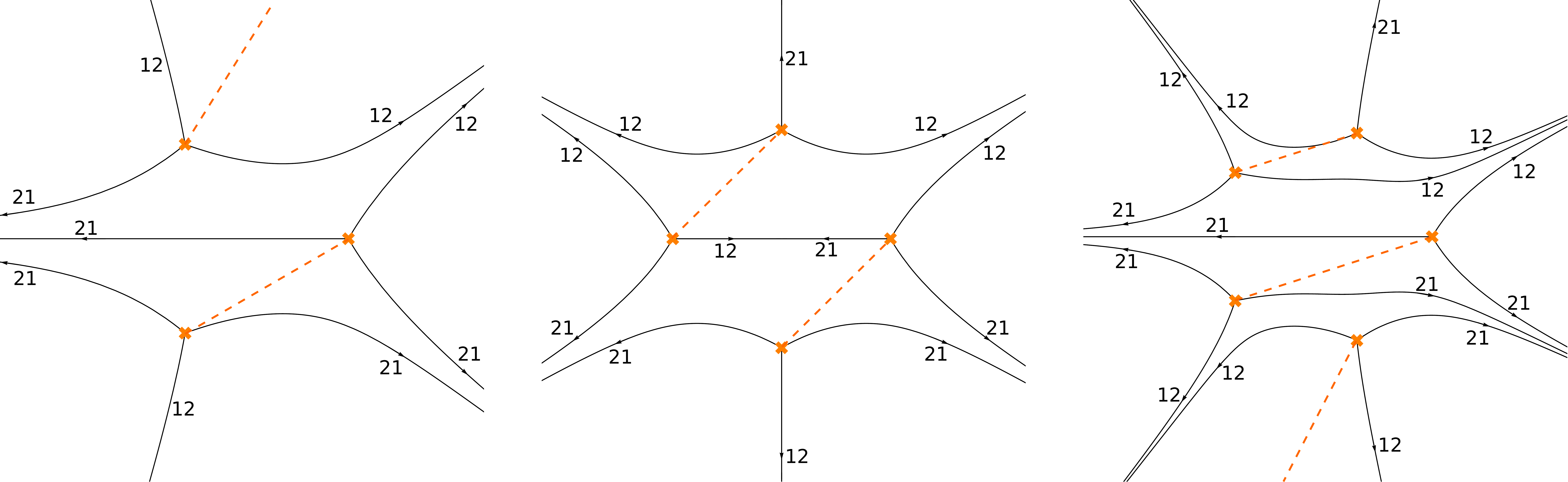}{0.18}{Examples
of $\vartheta$-Stokes graphs at $\vartheta=0$,
with $p(z) = z^n-1$, for $n = 3, 4, 5$. The dashed lines
denote branch cuts of the covering $\Sigma \to C$; the labels
$i = 1,2$ are swapped when we cross a cut.}

As long as the domain $U$ does not contain any $\vartheta$-Stokes curve of type $ij$,
$\lambda_i^\vartheta$ is defined on $U$ and can be integrated to give a WKB solution:
\begin{equation} \label{eq:wkb-solutions-actual}
	\psi_i^\vartheta(z) = \exp \left( \hbar^{-1} \int_{z_0}^z \lambda_i^\vartheta(z) \, \de z \right).
\end{equation}
If $U$ does not contain any $\vartheta$-Stokes curve of either 
type $ij$ or $ji$, then both $\psi_i^\vartheta$ and
$\psi_j^\vartheta$ exist on $U$, and give a basis of solutions
of the Schr\"odinger equation \eqref{eq:schrodinger}.
If $U$ does contain a $\vartheta$-Stokes curve of type $ij$,
then we still get a basis of solutions 
on the complement of the Stokes curve,
but $\psi_i^\vartheta$ jumps by a constant multiple
of $\psi_j^\vartheta$ on crossing the Stokes curve.

\subsection{Abelianization}

The WKB formula \eqref{eq:wkb-solutions-actual} has the awkward
feature that it depends on the arbitrary choice of basepoint
$z_0$. To see the content of \eqref{eq:wkb-solutions-actual}
more clearly, we can observe that it
represents a solution of the \ti{first-order} equation
\begin{equation} \label{eq:wkb-abelian-equation}
	\left( \partial_z - \hbar^{-1} \lambda_i^\vartheta(z) \right) \psi^\vartheta_i(z) = 0.
\end{equation}
\eqref{eq:wkb-abelian-equation} is much simpler than the
original equation \eqref{eq:schrodinger}; a 
lot of the complexity of \eqref{eq:schrodinger} has been swallowed into solving the
Riccati equation to produce $\lambda_i^\vartheta(z)$.

We interpret \eqref{eq:wkb-abelian-equation}
as the condition that $\psi^\vartheta_i (z)$
represents a flat section of a connection $\nabla^{\ab,\vartheta}$ 
in a line bundle $\cL$.
The line bundle $\cL$ lives not over the base $C$ but over the
double cover $\Sigma$, since the 
function $\lambda_i^\vartheta$ depends on the sheet index $i$.
The 1-form $- \hbar^{-1} \lambda_i^\vartheta \de z$ 
represents $\nabla^{\ab,\vartheta}$ relative
to a local trivialization of $\cL$.

\subsection{Gluing across the Stokes graph} \label{sec:stokes-gluing}

Consider a $\vartheta$-Stokes curve of type $ij$.
$\cL$ and $\nabla^{\ab,\vartheta}$ naively do not
extend across the lift of this $\vartheta$-Stokes curve to 
sheet $i$, because the solutions
$\psi_i^\vartheta$ are different on the two sides.
We can nevertheless extend them ``by hand'' by giving a gluing map which takes
$\nabla^{\ab,\vartheta}$-flat sections on one side to $\nabla^{\ab,\vartheta}$-flat sections
on the other, 
i.e. it maps $\psi_i^{\vartheta,L}$ to some constant multiple of $\psi_i^{\vartheta,R}$.
There is a canonical and convenient choice: we glue
$\psi_i^{\vartheta,L}$ to
the unique multiple of $\psi_i^{\vartheta,R}$ which is
of the form $\psi_i^{\vartheta,L} + \beta \psi_j^{\vartheta,L}$,
and glue $\psi_j^{\vartheta,L}$ to $\psi_j^{\vartheta,L}$.
This gluing prescription can be summarized as\footnote{The gluing rule
\eqref{eq:gluing-1} should be regarded as a version of the ``WKB connection formula.''}
\begin{equation} \label{eq:gluing-1}
\begin{pmatrix}
\psi_i^{L} \\
\psi_j^{L} 
\end{pmatrix}
\mapsto 
	\begin{pmatrix} 1 & \beta \\ 0 & 1 \end{pmatrix} \begin{pmatrix}
\psi_i^{L} \\
\psi_j^{L} 
\end{pmatrix}
= 
\begin{pmatrix}
\frac{[\psi_i^L, \psi_j^L]}{[\psi_i^R, \psi_j^L]}
\psi_i^{R} \\
\frac{[\psi_j^L, \psi_i^L]}{[\psi_j^R, \psi_i^L]}
\psi_j^{R}
\end{pmatrix}
\end{equation}
where $[\psi_1, \psi_2]$ means the Wronskian of the two solutions.

An additional subtlety arises if a $\vartheta$-Stokes curve
of type $ij$ coincides with a $\vartheta$-Stokes curve of type $ji$,
as e.g. in the middle of \autoref{fig:sample-networks-combined}.
(This does not occur for generic values of 
$\vartheta$, but it can occur for special $\vartheta$, and in many of
the examples we consider in this paper we take such a special $\vartheta$.) 
In this case we have four distinct solutions
$\psi_i^{\vartheta,L}$, $\psi_j^{\vartheta,L}$, $\psi_i^{\vartheta,R}$, $\psi_j^{\vartheta,R}$ on $U$, and we 
choose a gluing of the form
\begin{equation} \label{eq:gluing-2}
\begin{pmatrix}
\psi_i^{L} \\
\psi_j^{L} 
\end{pmatrix}
\mapsto 
	\begin{pmatrix} \rho & \beta \\ \alpha & \rho \end{pmatrix} \begin{pmatrix}
\psi_i^{L} \\
\psi_j^{L} 
\end{pmatrix}
= 
\begin{pmatrix}
\sqrt{\frac{[\psi_i^L , \psi_j^L]}{[\psi_i^R , \psi_j^R]} \frac{[\psi_i^L , \psi_j^R]}{[\psi_i^R , \psi_j^L]}}
\psi_i^{R} \\
\sqrt{\frac{[\psi_j^L , \psi_i^L]}{[\psi_j^R , \psi_i^R]} \frac{[\psi_j^L , \psi_i^R]}{[\psi_j^R , \psi_i^L]}}\psi_j^{R}
\end{pmatrix}
\end{equation}
where $\rho^2 - \alpha \beta = 1$.

We must make two technical comments about the gluing rule \eqref{eq:gluing-2}:

\begin{itemize}
 \item In writing \eqref{eq:gluing-2} we adopted the choice that the
 two diagonal entries of the gluing matrix should be equal.
We could alternatively have chosen e.g. that the upper left entry of the gluing matrix should be $1$,
or the lower right entry should be $1$. These alternate choices also have their advantages:
they arise naturally if one imagines that the two Stokes curves with labels $ij$
and $ji$ are infinitesimally displaced from one another, 
so that the gluing matrix arises as the product of an upper-triangular and a lower-triangular
matrix.
This infinitesimal displacement was used 
in \cite{Gaiotto2012,Hollands2013} and was called
``resolution'' of the spectral network. 
It appears naturally if we consider an infinitesimal perturbation
of the phase $\vartheta$, either to $\vartheta + \eps$ or $\vartheta - \eps$.
The choice we made in \eqref{eq:gluing-2} is
in some sense an average of these two resolutions, which avoids breaking symmetries.

\item The gluing matrix in \eqref{eq:gluing-2} is determined only
up to an overall sign.
To fix this ambiguity, we need to specify the branches 
of the square roots.
For this purpose (but only for this purpose!) it is convenient to make a definite choice of the
normalization of our solutions $\psi_i^{L/R}$, by choosing the basepoint $z_0$ 
in \eqref{eq:wkb-solutions-actual}
to be on the Stokes curve. Then we choose the principal branch for
both square roots.
The motivation for this choice is that all four Wronskians
appearing under the top square root asymptotically approach $2 \hbar^{-1} y_i$ 
as $\hbar \to 0$, and similarly the four
Wronskians under the bottom square root approach $2 \hbar^{-1} y_j$,
so both ratios approach $1$.

\end{itemize}

After all this gluing, we get a line bundle $\cL$
with flat connection $\nabla^{\ab,\vartheta}$, defined over all of
$\Sigma$ except for the branch points. It remains to consider
the monodromy around the branch points.
By a short calculation (see e.g. \cite{Hollands2013}), using 
the fact that the gluing matrices have determinant $1$, one can
show that $\nabla^{\ab,\vartheta}$ has
monodromy $-1$ on small loops encircling branch points.
We summarize this situation by saying that 
$\nabla^{\ab,\vartheta}$ is an
\ti{almost-flat} connection over $\Sigma$.

\subsection{\texorpdfstring{$\cW$}{W}-framings} \label{sec:w-framings}

The structure we have obtained from WKB
can be encapsulated formally as follows.

The Schr\"odinger equation can be interpreted as
a flat connection $\nabla$ in the $1$-jet 
bundle $J_1(K_C^{-\half})$ over $C$: this is just
the standard maneuver of replacing a second-order equation
by a first-order equation with $2 \times 2$ matrix coefficients,
locally written as
\begin{equation}
\underbrace{\left[\partial_z + \hbar^{-1} \begin{pmatrix} 0 & -P(z) \\ 1 & 0 \end{pmatrix}\right]}_{\nabla} \underbrace{\begin{pmatrix} - \hbar \psi'(z) \\ \psi(z) \end{pmatrix}}_{J(\psi)} = 0.
\end{equation}
Given a flat connection $\nabla$ and a Stokes graph
$\cW$, one can formulate the notion of a \ti{$\cW$-abelianization} 
of $\nabla$, as in \cite{Hollands2013} (see also \cite{Nikolaev2019} for a more
recent and mathematical treatment).
A $\cW$-abelianization consists of:
\begin{itemize}
\item A flat $\SL(2)$-connection $\nabla$ over $C$,
\item An almost-flat $\GL(1)$-connection $\nabla^\ab$ over $\Sigma$,
\item A flat isomorphism $\iota: \pi_* \nabla^\ab \simeq \nabla$ away from the 
walls of $\cW$ (where $\pi: \Sigma \to C$ is the projection),
\end{itemize}
obeying the constraint that, at the walls of $\cW$,
$\iota$ jumps by a unipotent transformation of the 
form \eqref{eq:gluing-1} (for a wall of type $ij$)
or \eqref{eq:gluing-2} (for a wall of type $ij$ and $ji$).

Given the connection $\nabla$, to construct a $\cW$-abelianization of $\nabla$ 
amounts to producing projective bases of $\nabla$-flat sections
in the various domains of $C \setminus \cW$, such that the relations between the bases
in neighboring domains are given by matrices of the form \eqref{eq:gluing-1}
or \eqref{eq:gluing-2}.
This is ultimately a linear algebra problem determined by the combinatorics
of $\cW$ and the monodromy and Stokes data of $\nabla$.
For any particular $\cW$ and $\nabla$, one can ask, how many $\cW$-abelianizations of $\nabla$
are there?
In the examples studied in \cite{Hollands2013}, it turns out that
there are just finitely many of them, and moreover they are in $1-1$
correspondence with some concrete extra data one can attach to 
$\nabla$, called \ti{$\cW$-framings} in \cite{Hollands2013}.
For example,
\begin{itemize}
\item 
Suppose we consider Schr\"odinger equations
on a Riemann surface $C$, taking $P(z)$ meromorphic
with $n$ second-order poles. In this case,
for generic $\vartheta$, 
the $\vartheta$-Stokes graph $\cW$ is 
a ``Fock-Goncharov'' network as described in 
\cite{Gaiotto:2009hg,Hollands2013}. 
A $\cW$-framing in this case is a choice of an eigenline of the 
monodromy around each of the $n$ punctures. For generic
$\nabla$, the monodromy at each puncture has $2$ distinct
eigenlines. Thus $\nabla$ admits $2^n$ distinct
$\cW$-framings.

\item Again, suppose we consider Schr\"odinger equations
on a Riemann surface $C$, taking $P(z)$ meromorphic
with $n$ second-order poles.
For special $\vartheta$, the complement of the $\vartheta$-Stokes graph 
$\cW$ can include regions with the topology of an annulus. 
For such a $\vartheta$, a $\cW$-framing
involves additional data: a choice of an eigenline of the monodromy
around each annulus. Thus $\nabla$ admits $2^{n+m}$ distinct
$\cW$-framings, where $m$ is the number of annuli.
\end{itemize}

Now we come back to WKB.
The discussion of \autoref{sec:wkb-solutions}-\autoref{sec:stokes-gluing} 
above can be rephrased as follows:
when $\nabla$ is the flat $\SL(2)$-connection induced by 
a Schr\"odinger equation \eqref{eq:schrodinger},
and $\cW$ is the Stokes graph with phase $\vartheta = \arg \hbar$,
exact WKB analysis constructs a distinguished $\cW$-abelianization of $\nabla$.
This construction will be developed in more detail in \cite{NNtoappear}.

It is somewhat remarkable that the WKB method automatically
equips $\nabla$ with a distinguished $\cW$-framing.
In the cases above, this boils down to the statement that the local WKB solutions
are automatically eigenvectors
of the relevant monodromies of $\nabla$.

\subsection{Spectral coordinates and their properties} \label{sec:spectral-coordinates}

Starting from the Schr\"odinger equation \eqref{eq:schrodinger} and a choice of phase
$\vartheta$,
we have seen that exact WKB analysis gives rise to 
an almost-flat $\GL(1)$-connection $\nabla^{\ab,\vartheta}$ over
the surface $\Sigma$.
In particular, given any $1$-cycle $\gamma$ on $\Sigma$
there is a corresponding holonomy,
\begin{equation}
	\cX^\vartheta_\gamma = \Hol_{\gamma} \nabla^{\ab, \vartheta} \in \C^\times
\end{equation}
As we have discussed in \autoref{sec:intro-voros},
the quantities $\cX^\vartheta_\gamma$ have various names, among them
\ti{Voros symbols}, \ti{spectral coordinates}, and \ti{quantum periods}.
They turn out to be extremely convenient for the analysis of
the Schr\"odinger equation \eqref{eq:schrodinger}.
Here are a few of their expected properties:
\begin{enumerate}

\item $\cX_\gamma^\vartheta$ admits a complete asymptotic
expansion as $\hbar \to 0$ in $\bbH_\vartheta$, obtained by term-by-term
integration of the formal WKB series \eqref{eq:wkb-asymptotics}:
\begin{equation} \label{eq:X-asymptotics}
	\cX_\gamma^\vartheta \sim \exp \left( \hbar^{-1} \oint_\gamma \lambda^\formal \, \de z \right).
\end{equation}
In particular, assuming $P(z)$ has no term of
order $\hbar$, 
the \ti{leading} asymptotic of $\cX_\gamma^\vartheta$
is controlled by the
classical period: if we define
\begin{equation} \label{eq:period}
	Z_\gamma = \oint_\gamma y \, \de z
\end{equation}
then to leading order
\begin{equation} \label{eq:X-asymptotic-leading}
	\cX^\vartheta_\gamma \sim \pm \exp \left( \hbar^{-1} Z_\gamma \right).
\end{equation}
The sign $\pm$ in \eqref{eq:X-asymptotic-leading} 
is explicitly $\exp \oint_\gamma \frac14 \frac{\de p}{p} = (-1)^{\half w}$,
where $w$ is the number of zeroes of $p(z)$ enclosed by
the projection of $\gamma$, counted with
multiplicity.\footnote{When $C$ is a compact Riemann
surface of genus $g$, to see that $(-1)^{\half w}$ does not
depend on which side we call the ``inside'' of $\gamma$, we use 
the fact that a holomorphic
quadratic differential has $4g-4$ zeroes, which
is divisible by $4$.}

\item $\cX_\gamma^\vartheta$ depends on $\hbar$, on 
the potential $P$, and on the phase $\vartheta$.
As long as the topology of the $\vartheta$-Stokes graph
does not change, the dependence of $\cX_\gamma^\vartheta$ 
on $\vartheta$ is trivial,
while the dependence of $\cX_\gamma^\vartheta$ 
on $\hbar$ and $P$ is holomorphic.
There is a codimension-1 locus in the $(P,\vartheta)$
parameter space where the topology of the $\vartheta$-Stokes graph
does change; we call this the \ti{BPS locus}.
When $(P,\vartheta)$ crosses the BPS locus,
the functions $\cX_\gamma^\vartheta$ jump by a
holomorphic transformation, called \ti{Stokes automorphism} or
\ti{Kontsevich-Soibelman transformation} depending on the context. 
This transformation can be computed
from the Stokes graph at the BPS locus.\footnote{In a generic situation
the Stokes automorphisms which can occur are of the form
$\cX_\mu \to \cX_\mu(1 \pm \cX_\gamma)^{\Omega(\gamma) \IP{\gamma,\mu}}$, 
where $\Omega(\gamma) = +1$ for a ``flip'' of the Stokes graph
and $\Omega(\gamma) = -2$ for a ``juggle'' of the Stokes graph, 
in the terminology of \cite{Gaiotto:2009hg}. The active rays corresponding
to flips are typically isolated in the $\hbar$-plane, 
while juggles occur at the limit of infinite sequences
of flips. A general algorithm
for computing the Stokes automorphism from a Stokes graph at the BPS locus 
is given in \cite{Gaiotto2012}.}

\item 
The asymptotic expansion \eqref{eq:X-asymptotics}
should hold as $\hbar \to 0$ in the half-plane $\bbH_\vartheta$.
If $\hbar$ is exactly in the middle of the 
half-plane $\bbH_\vartheta$, i.e. if
$\vartheta = \arg \hbar$, then we can make a stronger
conjecture, as follows.
If $(P,\vartheta)$ is not on the BPS locus,
$\cX_\gamma^\vartheta$ is the Borel sum of the asymptotic
expansion \eqref{eq:X-asymptotics} along the ray $\E^{\I \vartheta} \R_+$.
If $(P,\vartheta)$ is on the BPS locus, then
\eqref{eq:X-asymptotics} may not be 
Borel summable along the ray $\E^{\I \vartheta} \R_+$,
because of singularities of the Borel transform.
In that case, our conjecture is that $\cX_\gamma^\vartheta$ is obtained 
from \eqref{eq:X-asymptotics} by {\'E}calle's ``median summation''
(in the sense of
\cite{MR1399559,MR1258519}, also reviewed in \cite{MR1704654} page 21.)\footnote{This statement is sensitive to the particular gluing rule 
\eqref{eq:gluing-2} which we chose. Had we chosen a different rule,
as described below \eqref{eq:gluing-2}, we would expect to get 
instead the ``lateral summation'' corresponding to perturbing
$\vartheta$ infinitesimally.}

\end{enumerate}

\subsection{Integral equations} \label{sec:integral-equations}

Finally we come to one of the most interesting properties
of the spectral coordinates of families of Schr\"odinger operators: 
this is the conjecture of 
\cite{Gaiotto2014} which says that they obey
integral equations as functions of $\hbar$.

There is some choice involved in writing down the equations;
one has to first 
choose some function $\vartheta(\arg \hbar)$, subject only to 
the constraint that $\abs{\vartheta(\arg \hbar) - \arg \hbar} \le \frac{\pi}{2}$.
Then one considers the specialization
\begin{equation}
\cX_\gamma^{\RH}(\hbar) = \cX_\gamma^{\vartheta(\arg\hbar)}(\hbar).	
\end{equation}
$\cX_\gamma^{\RH}$ is piecewise analytic in $\hbar$; it jumps
along some rays in the $\hbar$-plane, namely those rays
at which the topology of the Stokes graph $\cW(p, \vartheta(\arg \hbar))$ jumps.
We call these \ti{active rays} and denote them by $r$.
When $\hbar$ lies on 
an active ray $r$,
we let $\cX^{RH,r,\pm}(\hbar)$ denote 
the limit of $\cX^{\RH}(\hbar)$ as $\arg \hbar$ approaches 
the phase of $r$ from the $\pm$ side.
The conjecture of \cite{Gaiotto2014} says
that these functions are the unique solution of a system
of coupled integral equations, of the form
\begin{equation} \label{eq:rh-integral-equation}
	\cX_\gamma^{\RH}(\hbar) = \exp \left[ \frac{Z_\gamma}{\hbar} + \frac{1}{4 \pi \I} \sum_{r \text {  active}} \int_{r} \frac{\de \hbar'}{\hbar'} \frac{\hbar' + \hbar}{\hbar' - \hbar} F_{r,\gamma}(\cX^{RH,r,+}(\hbar')) \right].
\end{equation}
This integral equation is similar to those appearing in the thermodynamic 
Bethe ansatz (TBA), 
and indeed \eqref{eq:rh-integral-equation} 
can be viewed as a generalization of the ``ODE-IM correspondence'' 
as we discussed in the introduction.

We are not aware of a completely rigorous 
proof of \eqref{eq:rh-integral-equation};
morally the idea is that the $\cX_\gamma^\RH(\hbar)$ can
be uniquely characterized in terms of their analytic properties in
the $\hbar$-plane, and a solution of \eqref{eq:rh-integral-equation}
would necessarily have the same analytic properties, so it must
be $\cX_\gamma^\RH(\hbar)$.
One direct argument which derives
\eqref{eq:rh-integral-equation} from reasonable analytic
assumptions is given in \cite{Ito2018}.
In another direction, 
\cite{Gaiotto2014} offers some reasons for optimism based on
identifying \eqref{eq:rh-integral-equation} as the
scaling limit of a better-behaved equation previously considered
in \cite{Gaiotto:2009hg}.
For us, the strongest reason so far to believe \eqref{eq:rh-integral-equation}
is a practical one: it has been checked to high precision in examples.
So far this has been done for various simple potentials,
as reported e.g. in \cite{Dorey2007,Gaiotto2014,Ito2018,dumasneitzke}.

To formulate \eqref{eq:rh-integral-equation} completely, 
as we have explained, one needs
to fix the choice of the function $\vartheta(\arg \hbar)$.
One canonical possibility
is to take
\begin{equation} \label{eq:theta-canonical}
 \vartheta(\arg \hbar) = \arg \hbar.	
\end{equation}
The resulting functions $\cX_\gamma^\RH(\hbar)$ are obtained
by making WKB analysis for each $\hbar$ using the Stokes graph adapted
to the phase $\vartheta = \arg \hbar$.
This choice makes the functions $F_{r,\gamma}$ relatively simple, at the cost that there
may be many active rays (even infinitely many), 
and one has to consider all possible $\vartheta$-Stokes graphs.
See \autoref{fig:uncollapsed-rays}.
\insfig{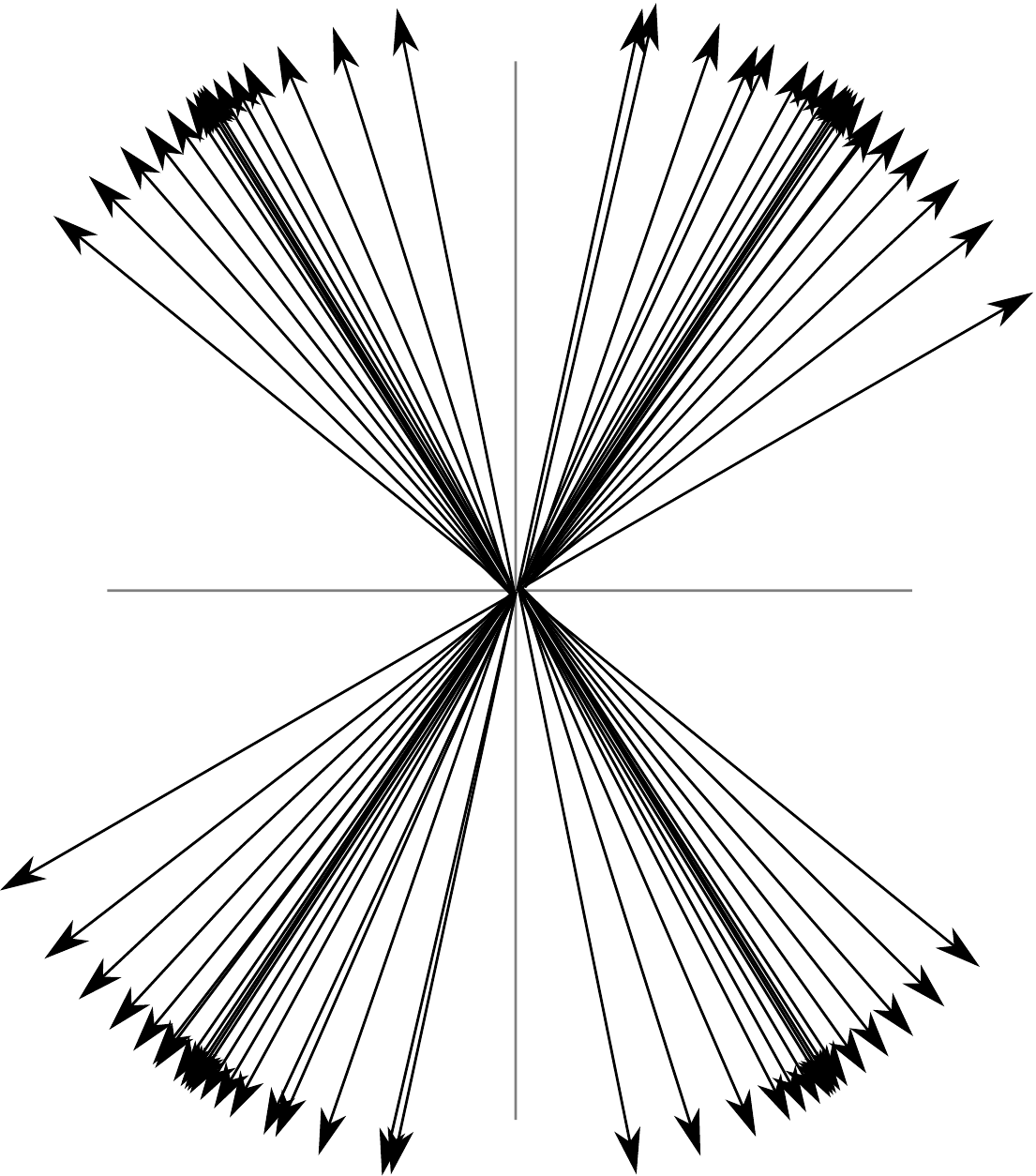}{A sample picture of what the active rays
in the $\hbar$-plane can look like. There are in general infinitely
many such rays, which can accumulate at discrete phases (as shown here)
or even be dense in part or all of the $\hbar$-plane.}

Another natural choice is to take $\vartheta(\arg \hbar)$ 
to be piecewise constant; this has the effect of dividing the
plane into sectors (of opening angle $\le \pi$) 
and collapsing all the active rays in each sector $S_i$
onto a single ``aggregated'' ray $r_i$.
In this case the aggregated 
functions $F_{r_i,\gamma}$ contain equivalent
information to all of the functions $F_{r,\gamma}$ for $r \subset S_i$.

In any case, 
to determine concretely the functions attached to the
active rays, one can use the relation
\begin{equation} \label{eq:f-discontinuity}
	F_{r,\gamma}(\cX_\gamma^{RH,r,+}) = \log \left( \cX_\gamma^{RH,r,+} / \cX_\gamma^{RH,r,-} \right),
\end{equation}
if one knows the 
spectral coordinate systems $\cX_\gamma^\vartheta$
for $\vartheta$ on both sides of the active ray $r$.

The most extreme possibility is to divide the plane into just two sectors,
by fixing a phase $\alpha$ and defining
\begin{equation} \label{eq:theta-sg-type}
	\vartheta(\arg \hbar) = \begin{cases} \alpha & \text{ for } \hbar \in \bbH_\alpha, \\
\alpha+\pi & \text{ for } \hbar \in \bbH_{\alpha+\pi}. \end{cases}
\end{equation}
\insfig{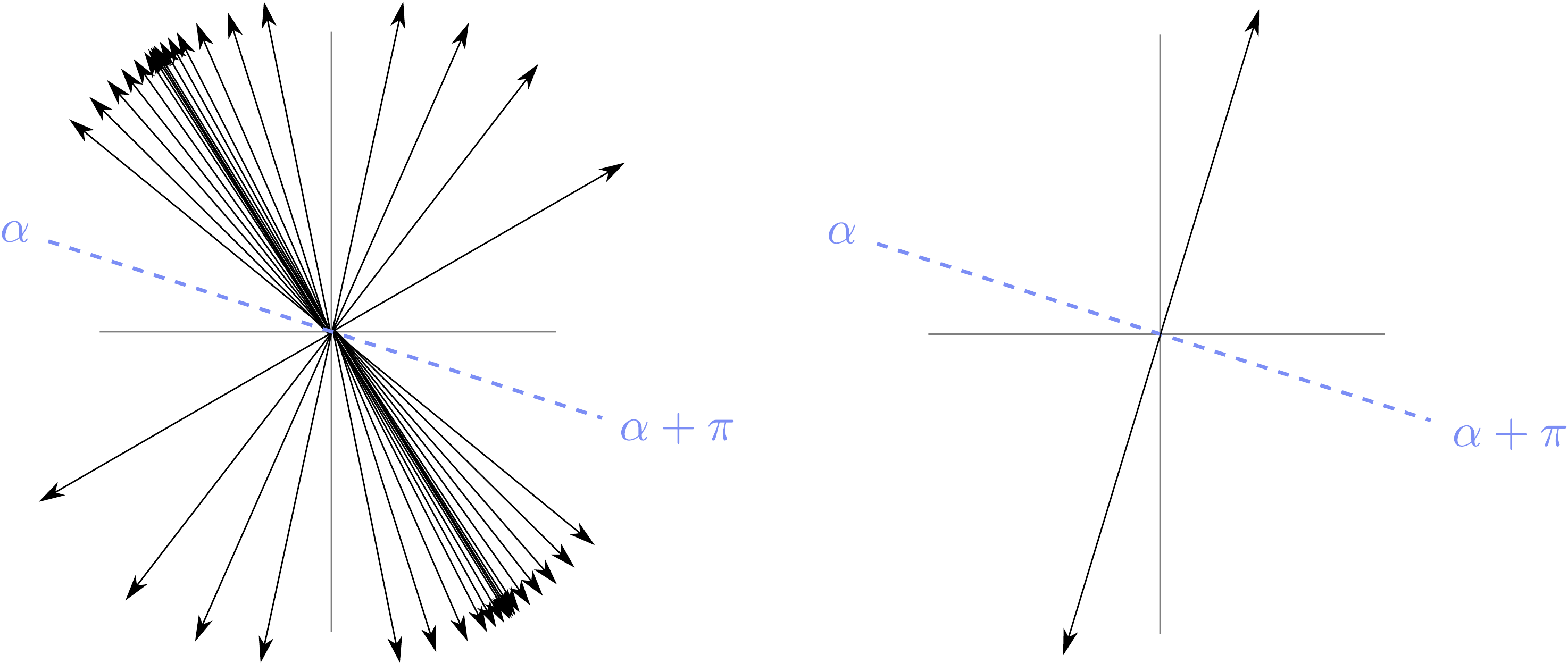}{Collapsing infinitely many active rays down to $2$ by making the choice \eqref{eq:theta-sg-type}. Each active ray on the right carries functions $F_{r,\gamma}$ which should be thought of as containing the same information as all the $F_{r,\gamma}$ in the
corresponding half-plane on the left.}
In this case there are just $2$ active rays $r$, and we only have
to consider two Stokes graphs, the $\alpha$-Stokes graph
and the $(\alpha+\pi)$-Stokes graph.
These two Stokes graphs are moreover identical except for an overall relabeling of all the
Stokes lines, $ij \to ji$.
The function $F_{r,\gamma}$ on each of the $2$ active rays
contains equivalent information to the 
``spectrum generator'' discussed in \cite{Gaiotto:2009hg}.\footnote{In the cluster algebra literature
this object is called the ``Donaldson-Thomas transformation''
or ``DT transformation'' following \cite{Goncharov2016}.} 
When the $\alpha$-Stokes graph is of
``Fock-Goncharov type,''
the spectrum generator has been determined in \cite{Gaiotto:2009hg};
these results were used in \cite{Gaiotto2014} to give several
explicit examples of integral equations \eqref{eq:rh-integral-equation}.

\section{Exact WKB for Schr\"odinger operators with cubic potential} \label{sec:cubic}

The WKB method and exact WKB method have been explored 
rather thoroughly in the case of a Schr\"odinger equation in the 
plane with polynomial potential.
For the WKB method two important references are \cite{MR0209562,MR0486867};
for exact WKB see e.g. the pioneering works \cite{MR553093,MR729194,MR1487700,MR1209700}, 
and \cite{MR2182990} for a clear recent treatment. 

In this section we quickly touch on the very simplest example of this sort, the Schr\"odinger equation
\begin{equation} \label{eq:schrodinger-cubic-potential}
		\left[ \hbar^{2} \partial_z^2 + (z^3-u) \right] \psi(z) = 0,
\end{equation}
for a constant $u \in \C$.
This is an instance of \eqref{eq:schrodinger} with cubic potential
\begin{equation}
	P(z) = z^3 - u.
\end{equation}

\subsection{A Stokes graph}

\insfigscaled{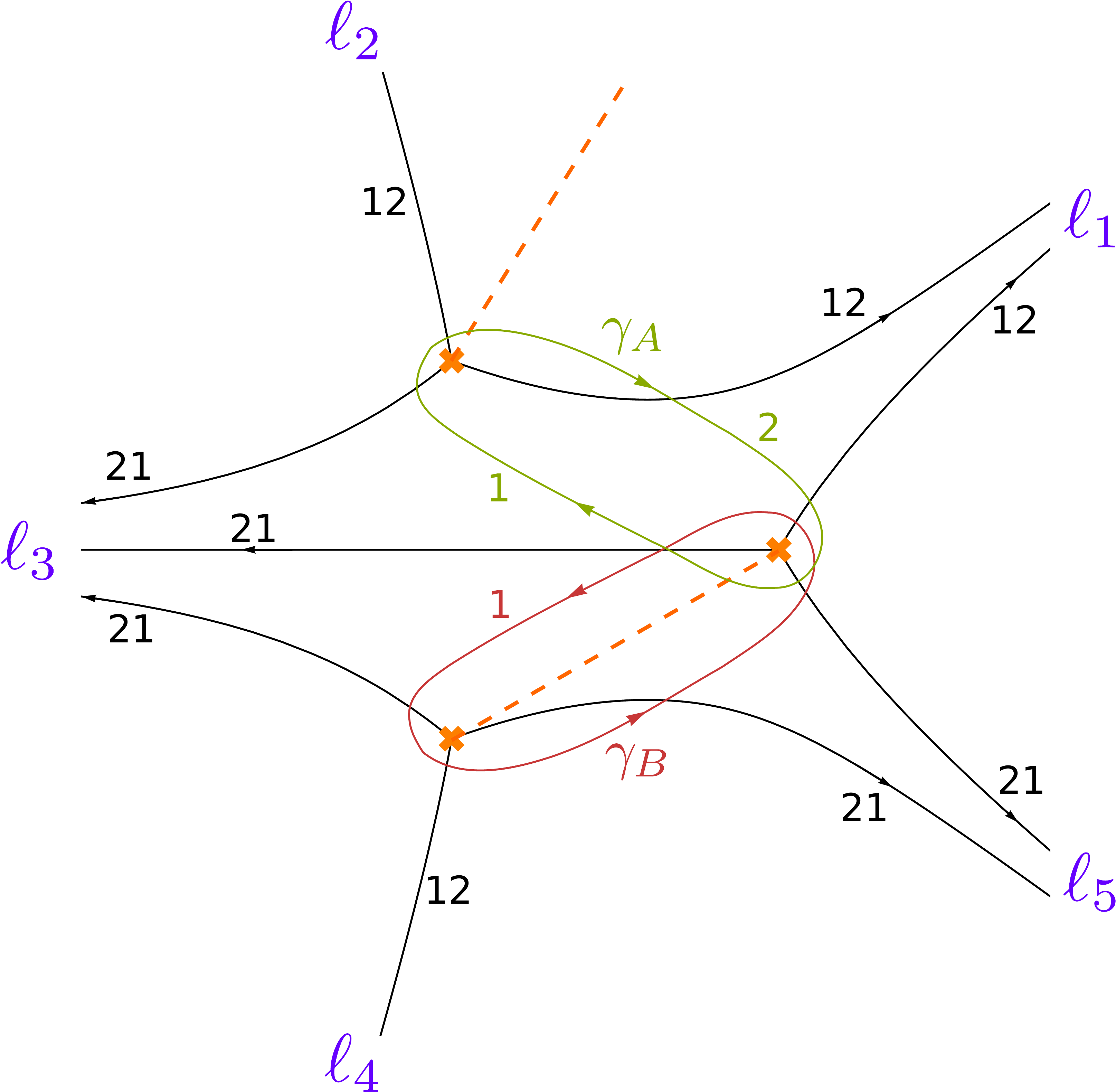}{0.25}{$\vartheta$-Stokes graph for the Schr\"odinger equation with cubic potential \eqref{eq:schrodinger-cubic-potential}, 
at the phase $\vartheta = 0$, and $u = 1$. 
Two 1-cycles $\gamma_A$, $\gamma_B$ on $\Sigma$ are also shown.
Dashed orange segments denote branch cuts; on crossing a cut,
the sheet labels  are exchanged $1 \leftrightarrow 2$. Orange
crosses denote the turning points, zeroes of $p(z) = z^3 - 1$. The 
singularity at $z = \infty$ is not shown.}

Suppose we fix $u = 1$ and $\vartheta = 0$. Then the $\vartheta$-Stokes graph 
is shown in \autoref{fig:cubic-network}.

This graph divides the plane up into $7$ domains.
As we have reviewed in \autoref{sec:wkb-review},
there are canonical local solutions $\lambda^\vartheta_i$ 
of the Riccati equation in each
of these domains, and from these local solutions we can build
local WKB solutions $\psi_i^\vartheta$ of \eqref{eq:schrodinger} ---
or more invariantly, we can build
an almost-flat $\GL(1)$-connection $\nabla^{\ab,\vartheta}$ over
the spectral curve $\Sigma = \{y^2 + z^3 - 1 = 0\}$. 
The connection $\nabla^{\ab,\vartheta}$ abelianizes the $\SL(2)$-connection
$\nabla$ in the $z$-plane associated to \eqref{eq:schrodinger-cubic-potential}.

\subsection{The spectral coordinates} \label{sec:coords-cubic}

Let $\cX_A$ (resp. $\cX_B$) denote the monodromy of 
$\nabla^{\ab,\vartheta}$ along the cycle $\gamma_A$ (resp. $\gamma_B$)
in \autoref{fig:cubic-network}. The Stokes graph of \autoref{fig:cubic-network}
is an example of a \ti{Fock-Goncharov network} in the sense
of \cite{Hollands2013}, and $\cX_A$, $\cX_B$ 
are Fock-Goncharov coordinates
of the flat connection $\nabla$.
Let us explain this more concretely.

We first consider the local WKB solutions in each domain.
These turn out to have a simple and concrete characterization, as follows.

Let $\ell_n$ ($n = 1, \dots, 5$) denote the ray with phase
$\frac{2 \pi}{5}(n + \half)$.
When $\hbar \in \R_+$, for 
each $n$ there exists a solution $\psi^\smsol_n$
such that $\psi^\smsol_n$ 
decays exponentially as $z \to \infty$ along $\ell_n$.
This $\psi^\smsol_n$ is unique up to scalar multiple.
Now let $U$ be one of the domains in the complement of the Stokes
graph. $U$ has two infinite ``ends'' which approach two of the five
rays $\ell_n$.
As $z$ approaches $\ell_n$, the WKB solution $\psi_j^\vartheta$
is exponentially decaying,
where $ij$ is the label on the $\vartheta$-Stokes curves asymptotic
to $\ell_n$.\footnote{This follows from the 
realization of $\lambda_j^\vartheta$ 
as the Borel summation of the WKB series, which
implies that $\lambda_j^\vartheta \de z$ is negative along
$\ell_n$, since every term of the series has this property.}
Thus up to scalar multiple $\psi_j^\vartheta$ is equal to $\psi^\smsol_n$.

Now that we understand the local WKB solutions, we can use them to compute the spectral coordinates. They turn out to be cross-ratios of the
$\psi^\smsol_n$, as follows (see \autoref{app:computations} for the computation):
\begin{equation} \label{eq:spectral-coords-cubic}
	\cX_A = \frac{[\psi_3^\smsol , \psi_2^\smsol]}{[\psi_1^\smsol , \psi_2^\smsol]} \frac{[\psi_1^\smsol , \psi_5^\smsol]}{[\psi_3^\smsol , \psi_5^\smsol]}, \qquad \cX_B = \frac{[\psi_3^\smsol , \psi_4^\smsol]}{[\psi_5^\smsol , \psi_4^\smsol]} \frac{[\psi_5^\smsol , \psi_1^\smsol]}{[\psi_3^\smsol , \psi_1^\smsol]}.
\end{equation}
As we promised above, these 
are Fock-Goncharov coordinates (or ``complexified shear coordinates'') 
of the connection $\nabla$, in the sense of \cite{MR2233852}.

\subsection{Analytic continuation} \label{sec:cubic-analytic-cont}

In our description of $\cX_A$ and $\cX_B$ above we used
the conditions $u=1$ and $\hbar \in \R_+$.
It is interesting to consider the question of 
analytic continuation of these functions in $u$ and $\hbar$.
For this purpose a simple approach is to just start from the final formulas
\eqref{eq:spectral-coords-cubic} and try to continue them
directly. The resulting analytic structure is very simple:

\begin{itemize}
\item First, as we vary $u$, the $z \to \infty$
asymptotic behavior of the equation \eqref{eq:schrodinger-cubic-potential}
does not change; for each $u$ we still have $5$ decaying solutions
$\psi_n^\smsol$, now depending on $u$. Since the equation \eqref{eq:schrodinger-cubic-potential} depends holomorphically on $u$,
so do its decaying solutions.
Thus the formula
\eqref{eq:spectral-coords-cubic} defines single-valued
analytic functions $(\cX_A, \cX_B)$ of $u \in \C$.

These functions may have poles, because
for general $u$ there is nothing preventing $\psi_n^\smsol$ and $\psi_{n'}^\smsol$
from coinciding, as long as $n$ and $n'$ are not consecutive.
Indeed, numerically one finds a discrete sequence of points 
$u = u_1, u_2, \dots$ where $\cX_A$ has a simple pole
($\psi_3^\smsol$ and $\psi_5^\smsol$ become proportional)
and conjugate points
$u = u_1^*, u_2^*, \dots$ where $\cX_B$ has a simple pole
($\psi_3^\smsol$ and $\psi_1^\smsol$ become proportional).
These poles can be thought of as ``bound states'' for the equation
 \eqref{eq:schrodinger-cubic-potential} along a complex contour
asymptotic to $\ell_n$ and $\ell_{n'}$.
The $u_i$ lie on the ray $\arg u = -\frac45 \pi$
(but this is not trivial to see: it was
proven in \cite{Dorey2001}.)

\item Second, we can consider varying $\hbar$. This leads to a slightly
subtler analytic structure. If we vary $\arg \hbar$ by an amount $\beta$, 
the distinguished rays $\ell_n$ 
where we impose the exponential decay condition rotate 
counterclockwise in the plane
by an angle $\frac25 \beta$.
It follows that, when we go clockwise once around the singularity
at $\hbar = 0$, the $\psi_n^\smsol$ are 
permuted by $n \mapsto n+2$ (mod $5$);
this transforms $(\cX_A, \cX_B)$ by
\begin{equation} \label{eq:cubic-XAXB-monodromy}
	(\cX_A, \cX_B) \mapsto \left( \cX_B^{-1} (1 - \cX_A^{-1})^{-1}, \cX_A \right).
\end{equation}
Thus the maximal analytic continuation of the functions $(\cX_A, \cX_B)$
is defined on a 5-fold cover of the punctured plane $\hbar \in \C^\times$.
\end{itemize}

In fact, the continuations in $u$ and $\hbar$ are not unrelated:
the continued functions actually depend only on the combination
$u' = u / \hbar^{\frac65}$, as one sees by dividing
\eqref{eq:schrodinger-cubic-potential} by $\hbar^{\frac65}$
and then taking $z \mapsto \hbar^{\frac25} z$.

Note that the monodromy \eqref{eq:cubic-XAXB-monodromy} acts by a symplectomorphism 
preserving the 
form $\varpi = \de \log \cX_A \wedge \de \log \cX_B$. This is a consistency check of the general
story: $\cX_A$ and $\cX_B$ are local Darboux coordinates on a moduli space of
$\SL(2)$-connections with irregular singularity at $z = \infty$.

We emphasize that the analytic 
continuation of $(\cX_A, \cX_B)$ which we have been discussing 
in this section is not given directly 
by WKB analysis; to make the WKB analysis
directly at a given $(u,\hbar)$ would require us to consider
a different Stokes graph and spectral curve for each $(u, \hbar)$.
This would necessarily lead to single-valued functions 
of $(u, \hbar)$, but ones which are only piecewise analytic,
jumping when the Stokes graph jumps.
These are the functions which we called $\cX_\gamma^\RH$ above;
we will discuss them more in \autoref{sec:integral-equations-cubic}
below.

\subsection{The regular pentagon}

One case worthy of special notice is the case $u = 0$, where
the potential $P(z)$ degenerates to the pure
cubic, $P(z) = z^3$.
At this point the equation \eqref{eq:schrodinger-cubic-potential}
acquires an extra $\Z / 5\Z$ symmetry which acts by
$z \to \E^{2 \pi \I / 5} z$, and thus cyclically permutes the five
rays $\ell_n$. From this symmetry it follows that $(\cX_A, \cX_B)$
is a fixed point of the monodromy \eqref{eq:cubic-XAXB-monodromy}, which
implies
\begin{equation}
	\cX_A = \cX_B = x, \qquad x^2 - x - 1 = 0.
\end{equation}
Numerically we find that the relevant solution of this quadratic is\footnote{The reader might wonder: what about the other solution, where $\cX_A = \cX_B = \frac{1 + \sqrt5}{2}$? That one turns out to be associated to a
Schr\"odinger equation with singular potential, $P(z) = z^3 - \frac34 \frac{\hbar^2}{z^2}$.
The specific coefficient $-\frac34 \hbar^2$ here ensures that the singularity at $z=0$ is
only an ``apparent singularity,'' with trivial monodromy; thus this equation
is still associated to a flat connection $\nabla$ in the plane, and all our
discussion of abelianization applies equally well to this case. Moreover this
equation still has the $\Z / 5\Z$ symmetry (because the two 
terms $z^3$ and $1/z^2$ differ by a factor $z^5$), and numerically one checks 
that it has $\cX_A = \cX_B = \frac{1 + \sqrt5}{2}$. We thank Dylan Allegretti and Tom Bridgeland for several enlightening conversations about Schr\"odinger
equations with apparent singularities.}
\begin{equation} \label{eq:golden-ratio}
	\cX_A = \cX_B = \frac{1 - \sqrt{5}}{2}.
\end{equation}
Since $\cX_A$ and $\cX_B$ depend only on
$u' = u / \hbar^{\frac65}$, it follows that 
this fixed point also governs the $\hbar \to \infty$ behavior
for any constant $u$.

\subsection{Integral equations for spectral coordinates} \label{sec:integral-equations-cubic}

Identifying the cross-ratios \eqref{eq:spectral-coords-cubic} as the 
spectral coordinates coming from WKB
implies that they should have all the properties discussed in
\autoref{sec:spectral-coordinates}-\autoref{sec:integral-equations}.
In particular, when they are extended to functions $\cX_\gamma^\RH(\hbar)$ 
as in \autoref{sec:integral-equations}, they should obey
an integral equation of the form \eqref{eq:rh-integral-equation}.

We make the canonical choice \eqref{eq:theta-canonical}.
Then one direct way to identify the active rays is to 
use a computer to draw the $\vartheta$-Stokes graphs
for various phases $\vartheta$; the active rays are at the phases 
where the $\vartheta$-Stokes graph jumps.
It turns out that there are $6$ such rays, as shown in
\autoref{fig:uncollapsed-rays-cubic}.\footnote{This corresponds to the well known
BPS spectrum of the $(A_1,A_2)$ Argyres-Douglas field theory in its ``maximal chamber,'' discussed
e.g. in \cite{Shapere:1999xr,Gaiotto:2009hg}.}
Each of these rays $r$ has an associated class $\mu \in H_1(\Sigma,\Z)$,
and the function 
$F_{r,\gamma}$ is
\begin{equation}
	F_{r,\gamma}(\cX) = \IP{\gamma, \mu} \log (1 - \cX_\mu)
\end{equation}
where $\IP{\cdot,\cdot}$ is the intersection pairing
on $H_1(\Sigma,\Z)$.

\insfigscaled{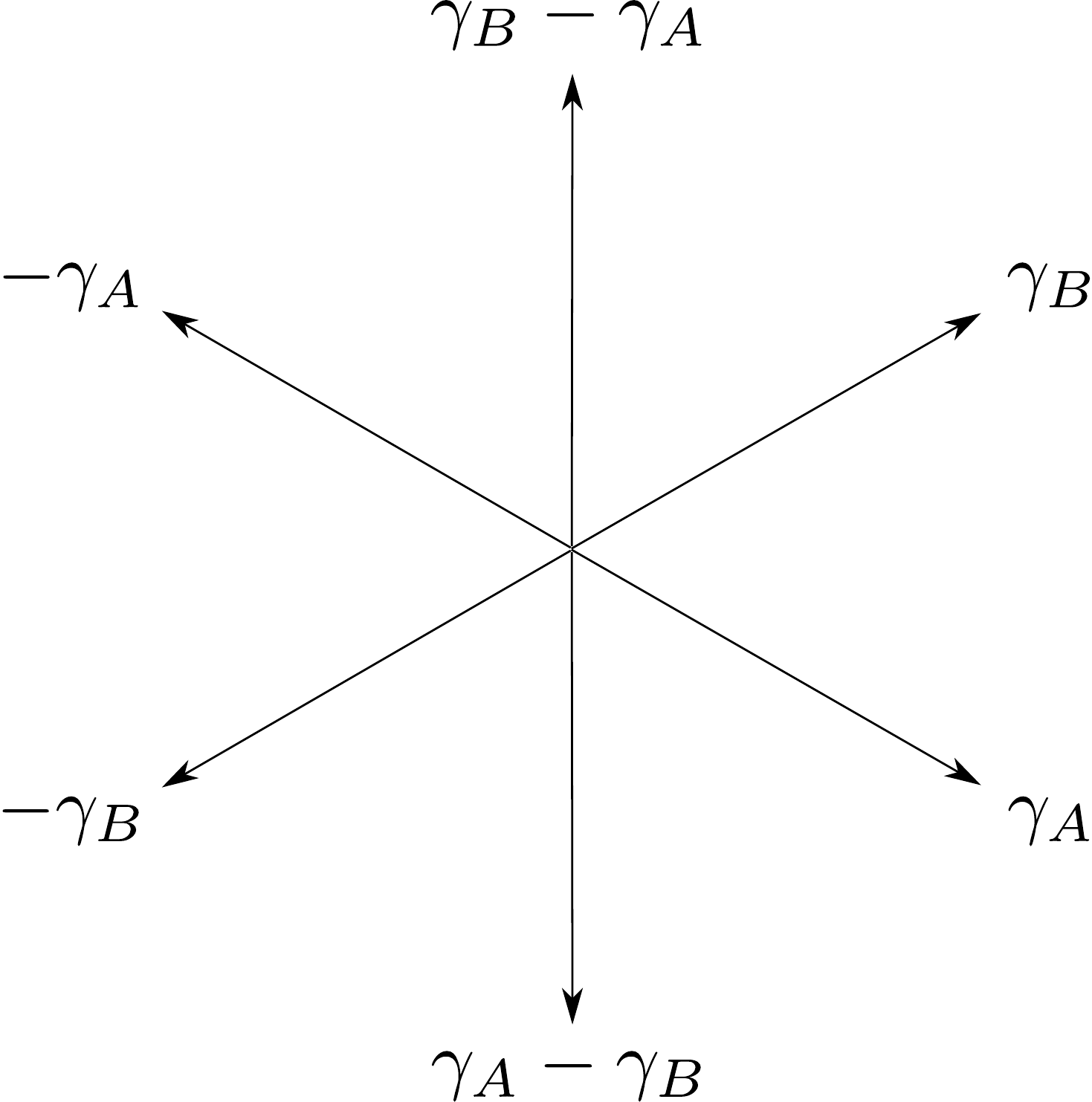}{0.3}{The $6$ active rays in the $\hbar$-plane,
each labeled by its charge $\mu \in H_1(\Sigma,\Z)$. These rays divide the
$\hbar$-plane into $6$ regions. Each region is characterized by a different
topology for the $\vartheta$-Stokes graph, where $\vartheta = \arg \hbar$.}

In this case one can make a direct numerical test of the integral
equation \eqref{eq:rh-integral-equation}. Namely, on the one 
hand we can solve 
\eqref{eq:rh-integral-equation} by numerical iteration,
on the other hand we can determine $\cX_\gamma^\RH(\hbar)$
directly by numerical integration of the Schr\"odinger equation
in the complex plane.
The two computations agree very well.
To give one concrete example, by direct numerical integration we obtain the estimates
\begin{align}
	\cX_A(\hbar = 2 + \I) &\approx -0.230042356-0.324912345\I, \\ \cX_B(\hbar = 2 + \I) &\approx -0.288795812+0.476012574\I,
\end{align}
and each of these agrees with the result obtained from the integral equation \eqref{eq:rh-integral-equation},
to the precision given.
Many similar computations for 
polynomial potentials have been made before, with similarly
good numerical agreement,
e.g. already in \cite{Dorey1998} and more recently
\cite{Gaiotto2014,Ito2018,dumasneitzke}.
The appearance of the fixed point \eqref{eq:golden-ratio}
at the $\hbar \to \infty$ limit was already noticed in
the very early TBA work \cite{Zamolodchikov:1990cf}.

\section{Exact WKB for the Mathieu equation} \label{sec:mathieu}

Now let us recall how exact WKB analysis is applied to the
\ti{Mathieu equation}:
\begin{equation} \label{eq:mathieu}
\left[ - \hbar^2 \partial_x^2 + 2 \cos(x) - 2 E \right] \psi(x) = 0.
\end{equation}
WKB analysis of this equation has been studied extensively;
a review we found particularly helpful is \cite{Dunne2016},
which covers many topics we will not touch here.
For other treatments of exact WKB for this equation see e.g.
\cite{Mironov2009a,He2010,Kashani-Poor2015,Codesido2016},
and more broadly \cite{conuzmar,ZinnJustin:2004ib,ZinnJustin:2004cg}.

\subsection{Exponential coordinate}

Making the replacements
\begin{equation} \label{eq:mathieu-replacements}
 z = \E^{\I x}, \qquad \tilde \psi(z) = \left(\I z\right)^{\half} \psi(x)
\end{equation}
transforms \eqref{eq:mathieu} into an equation defined over
$\CP^1$, with irregular singularities at $z = 0$ and $z = \infty$:
\begin{equation} \label{eq:mathieu-cp1}
	\left[ \hbar^2 \partial_z^2 + P(z) \right] \tilde\psi(z) = 0, \qquad P(z) = \frac{1}{z^3} - \frac{2E-\frac14 \hbar^2}{z^2} + \frac{1}{z}.
\end{equation}
In what follows we will usually use the formulation \eqref{eq:mathieu-cp1}.

\subsection{A simple Stokes graph}

\insfigscaled{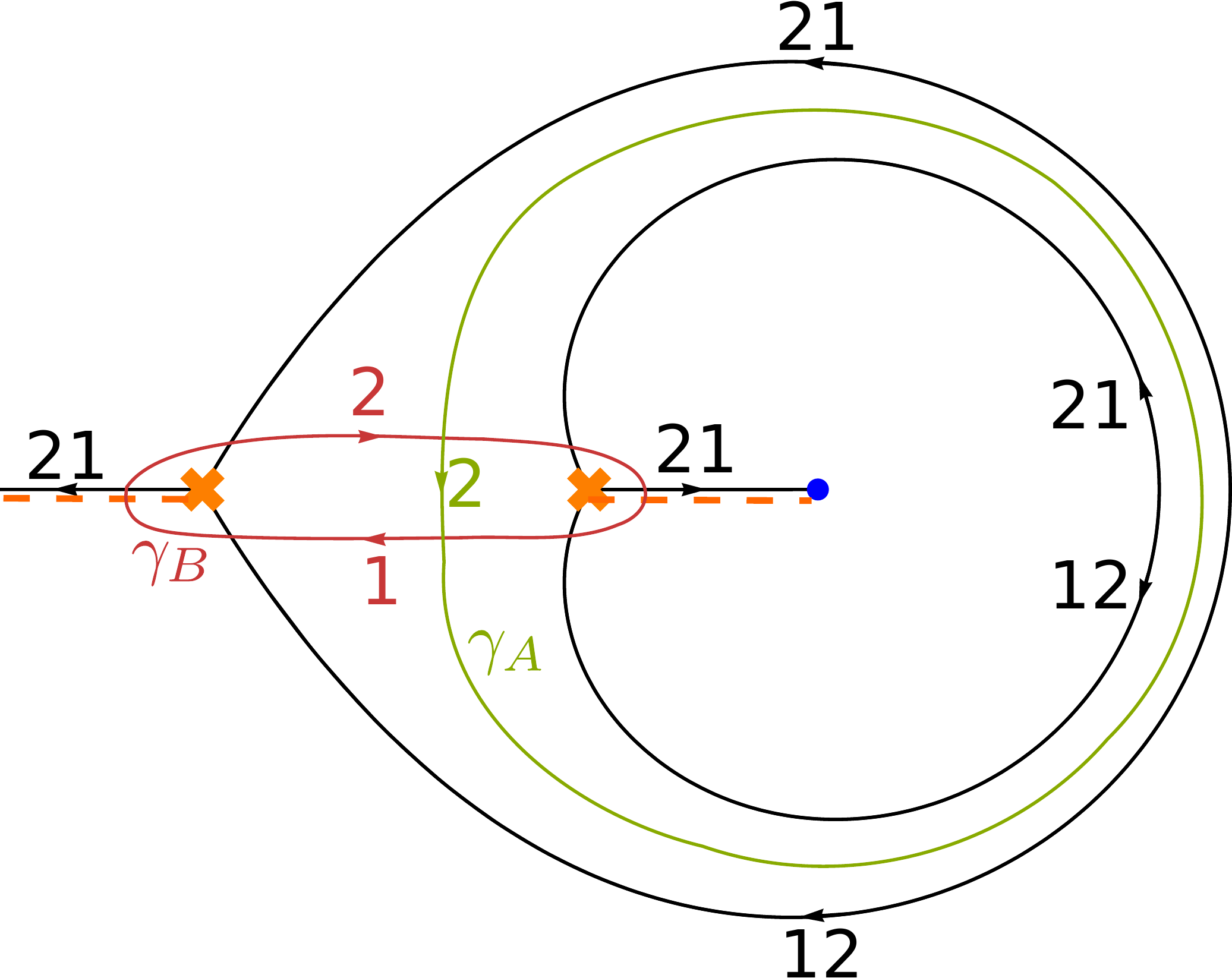}{0.27}{$\vartheta$-Stokes graph for the Mathieu equation, at the phase $\vartheta = 0$, and $E = -\frac98$. 
Two 1-cycles $\gamma_A$, $\gamma_B$ on $\Sigma$ are also shown.
Dashed orange segments denote branch cuts; on crossing a cut,
the sheet labels  are exchanged $1 \leftrightarrow 2$. Orange
crosses denote the turning points, zeroes of $P(z)$. The blue dot
represents the singularity at $z=0$; the singularity at $z = \infty$
is not shown.}

We begin with
real $E < -1$ and $\vartheta = 0$. 
The $\vartheta$-Stokes graph is as shown in \autoref{fig:mathieu-sn-1}. 
The Stokes curves divide the plane into $3$
open domains: a simply connected domain near $z=0$, another near $z=\infty$, and
an annulus containing $z=1$.

\subsection{The spectral coordinates} \label{sec:spectral-coords-mathieu}

Let $\cX_A$ (resp. $\cX_B$) denote the monodromy of 
$\nabla^{\ab,\vartheta}$ along the cycle $\gamma_A$ (resp. $\gamma_B$)
in \autoref{fig:mathieu-sn-1}. This Stokes graph
is an example of a \ti{Fenchel-Nielsen network} in the sense
of \cite{Hollands2013}, and $\cX_A$, $\cX_B$ 
are exponentiated Fenchel-Nielsen coordinates
of the flat connection $\nabla$.
Let us explain this more concretely.

We first consider local WKB solutions in each of the three domains.
\begin{itemize}
\item Let $\psi_i$ be the local WKB solutions near $z=0$.
	$\psi_1$ can be characterized as a solution which exponentially decays as $z \to 0$ along the negative-$z$ ray, similarly to what we saw in \autoref{sec:coords-cubic}.
\item Let $\psi'_i$ be the local WKB solutions near $z = \infty$.
$\psi'_1$ can be characterized as a solution which exponentially decays as $z \to \infty$ 
	along the negative-$z$ ray.
\item Let $\psi''_i$ be the local WKB solutions in some simply connected domain of
the intermediate annulus. These can be characterized as eigenvectors of the counterclockwise monodromy $M$ of $\nabla$.
At $E < -1$ and $\hbar \in \R_+$ the eigenvalues of $M$ are real and negative, and
we let $\mu$ denote the eigenvalue
which has $\abs{\mu} < 1$; then 
$\psi''_1$ is the eigenvector with eigenvalue $\mu^{-1}$, while $\psi''_2$
is the one with eigenvalue $\mu$.
\end{itemize}
With the local WKB solutions understood, we can compute the spectral 
coordinates:
\begin{itemize}
	\item 
	$\cX_A$ is the smaller eigenvalue of 
	monodromy of $\nabla$,
\begin{equation} \label{eq:XA-mathieu}
	\cX_A = \mu.
\end{equation}
Indeed, the representative $\gamma_A$ in \autoref{fig:mathieu-sn-1} does not cross
any Stokes curves, so the eigenvalue of monodromy of $\nabla^\ab$ on sheet $2$ agrees with the 
eigenvalue of monodromy of $\nabla$ acting on $\psi''_2$, which is $\mu$. 
	This is an exponentiated complexified Fenchel-Nielsen 
	length coordinate, in the sense of \cite{Kabaya,Hollands2013}.
	\item $\cX_B$ can be given in terms of Wronskians of the local WKB solutions in the three domains (see \autoref{app:computations} for the computation):
\begin{equation} \label{eq:XB-mathieu}
	 \cX_B = \frac{[\psi_1 , \psi''_2]}{[\psi_1 , \psi''_1]} \frac{[\psi'_1 , \psi''_1]}{[\psi'_1 , \psi''_2]}.
\end{equation}
(In computing these Wronskians we have to evolve all the solutions to a common domain,
which we do along the negative-$z$ ray.) This is an exponentiated complexified 
Fenchel-Nielsen twist coordinate, in the sense of \cite{Kabaya,Hollands2013}.
\end{itemize}

\subsection{Application: bound states} \label{sec:mathieu-bound-states}

Now let us see one application of the spectral coordinates.
We return to the original Mathieu equation 
\eqref{eq:mathieu} and make the substitution $x = \I x' + \pi$ with $x'$ real.
Then $\eqref{eq:mathieu}$ becomes the \ti{modified Mathieu equation},
\begin{equation} \label{eq:mathieu-im}
\left[ - \hbar^2 \partial_{x'}^2 + 2 \cosh(x') + 2 E \right] \psi(x') = 0.
\end{equation}
This is a Schr\"odinger equation
with potential $V(x') = \cosh x'$, for which we can formulate the
usual bound state problem, i.e. we look for $E$ such that
there exists an $L^2$ solution of \eqref{eq:mathieu-im}.
Such a solution exists only for countably many $E = E_1, E_2, \dots$. 
With our sign conventions $E$ is minus the usual energy, so 
all $E_n < -1$.

The condition for existence of a bound
state is that $\psi_1$ is proportional to $\psi'_1$. Substituting this condition
in \eqref{eq:XB-mathieu} gives simply
\begin{equation} \label{eq:mathieu-B-condition}
	\cX_B = 1.
\end{equation}
This is known as the ``exact quantization condition'' for the modified Mathieu
bound states, discussed frequently in the literature, e.g.  \cite{Codesido2017,He2010,Mironov2009a}.

To give some indication of how \eqref{eq:mathieu-B-condition} 
can be used in practice,
let us consider the leading term of the asymptotic expansion,
\begin{equation} \label{eq:mathieu-B-leading}
	\cX_B \approx - \exp(Z_B / \hbar).
\end{equation}
When $E < -1$ we have $Z_B \in \I \R_-$, and recall that $\hbar > 0$;
thus this leading approximation says that
solutions of \eqref{eq:mathieu-B-condition} will be found
when
\begin{equation} \label{eq:mathieu-B-approx-quantization}
	Z_B \approx 2 \pi \I \left(n+\half\right) \hbar.
\end{equation}
To understand \eqref{eq:mathieu-B-approx-quantization} more explicitly, 
we can expand $Z_B$ at large negative $E$: one finds
$Z_B(E) \approx -4 \I \sqrt{-2 E} \log(-E)$.
Thus,
for large negative $E$ and small $\hbar$, the desired
bound states are approximately at
\begin{equation}
\sqrt{-E} \log(-E) \approx \frac{\pi}{2 \sqrt{2}} \left(n+\half\right) \hbar.
\end{equation}
One can improve this estimate by including higher terms --- either
in the WKB expansion of $\cX_B$
in powers of $\hbar$, or in the 
expansion of $Z_B(E)$ in inverse powers of $E$.
We will not explore these improvements here.

\subsection{Analytic continuation} \label{sec:mathieu-analytic-cont}

So far we have considered the spectral coordinates
$\cX_A$ and $\cX_B$ for $E<-1$, $\hbar>0$, built using
the exact WKB method.
It is also interesting to consider the analytic continuation 
of these coordinates to complex parameters. 

To build this analytic continuation, we will 
build a $\cW$-abelianization of $\nabla$
which varies holomorphically with $(E,\hbar)$. Said otherwise, we will build local
solutions $\psi_i, \psi'_i, \psi''_i$ which fit into a $\cW$-abelianization and vary
holomorphically with $(E,\hbar)$. For general $(E,\hbar)$ they 
will not necessarily be given by any kind of WKB analysis.

 The local solutions $\psi''_i$ must be eigenvectors of the monodromy $M$: to decide
which one will be $\psi''_1$ and which will be $\psi''_2$, we just require that our choice is
continuously connected to the choice we got from WKB at $E < -1$, $\hbar > 0$.
This gives a nice analytic continuation along any path in $(E,\hbar)$ space,
except at the codimension-1 locus where the eigenvalues of $M$ coincide. Around this locus
we have an order-2 monodromy exchanging $\psi''_1 \leftrightarrow \psi''_2$.

At our initial locus $(E < -1,\hbar > 0)$, $\psi_1$ can be characterized by the property 
of exponential decay as $z \to 0$ along the negative real axis.
As we vary $(E,\hbar)$ we can define $\psi_1$ by 
a similar condition, except that the negative real axis
has to be replaced by a different path, which asymptotically has 
$z \to 0$ with $\arg z = 2 \arg \hbar + \pi$.
Similar comments apply to $\psi'_1$ except that we use a path with $z \to \infty$
and $\arg z = - 2 \arg \hbar + \pi$.
This gives a nice analytic continuation of $\psi_1$ and $\psi'_1$
along any path in $(E,\hbar)$ space which avoids $\hbar = 0$.
Now we have to consider the possibility of monodromy around
$\hbar = 0$.
As $\arg \hbar$ is continuously increased by $2\pi$, our paths into $z = 0$ and $z = \infty$ 
wind around twice, in opposite directions. The result is that as
we go counterclockwise around $\hbar = 0$ we have an infinite-order monodromy
acting by $\psi_1 \mapsto M^{-2} \psi_1$, $\psi'_1 \mapsto M^{2} \psi'_1$.

(We might also wonder whether the eigenvectors $\psi''_i$ 
of $M$ are exchanged as $\hbar$ goes around $0$;
this cannot occur, since the Mathieu equation depends only on $\hbar^2$, so the monodromy
around $\hbar = 0$ is the square of an order-2 element, hence the identity.)

Using \eqref{eq:XA-mathieu} and \eqref{eq:XB-mathieu}, the analytic structure of $\cX_A$ and $\cX_B$ follows from that
of $\psi_i$, $\psi'_i$, $\psi''_i$; we have unrestricted analytic
continuation in $(E,\hbar)$, except that:
\begin{itemize}
	\item Going around $\hbar = 0$ counterclockwise we have the infinite-order monodromy
\begin{equation} \label{eq:mathieu-monodromy-1}
	(\cX_A, \cX_B) \mapsto (\cX_A, \cX_A^8 \cX_B). 
\end{equation}

	\item Around the locus in $(E,\hbar)$ space where the eigenvalues of $M$ coincide,
	we have the order-$2$ monodromy
\begin{equation} \label{eq:mathieu-monodromy-2}
	 (\cX_A, \cX_B) \mapsto (\cX_A^{-1}, \cX_B^{-1}).
\end{equation}
	
\end{itemize}

Note that both of these monodromies act by symplectomorphisms preserving the 
form $\varpi = \de \log \cX_A \wedge \de \log \cX_B$. This is a consistency check of the general
story: $\cX_A$ and $\cX_B$ are local Darboux coordinates on the moduli space of
$\SL(2)$-connections.

\subsection{Integral equations for spectral coordinates} \label{sec:integral-equations-mathieu}

As we have discussed in \autoref{sec:integral-equations}, one of the
most interesting properties of spectral coordinates for
families of Schr\"odinger equations is that they
conjecturally obey integral equations of the form \eqref{eq:rh-integral-equation}.

In the case of the Mathieu equation, integral equations for spectral coordinates 
were considered in \cite{Gaiotto2014}. There the 
function $\vartheta(\arg \hbar)$ was chosen in the form \eqref{eq:theta-sg-type},
with $\alpha$ a generic phase, collapsing all the active rays onto two aggregated rays. 
In this case the Stokes graphs which appear are of Fock-Goncharov type
in the terminology of \cite{Hollands2013}, and the $\cX_\gamma$ are Fock-Goncharov coordinates. This example is thus qualitatively similar to the one
we considered in \autoref{sec:integral-equations-cubic} 
above, though the details are more intricate.\footnote{In particular, it seems to be harder to find a solution of 
the integral equations \eqref{eq:rh-integral-equation}
directly by iteration in this case. Instead one can
start with a slightly different system of integral equations,
those used in \cite{Gaiotto:2008cd}; these one can solve by iteration;
then one can take the limit
$R \to 0$, $\zeta \to 0$, $\hbar = R / \zeta$, to get solutions of 
\eqref{eq:rh-integral-equation}.}

In this section we try something different: we try to find integral equations 
obeyed by the complexified Fenchel-Nielsen coordinates.
For this purpose we choose the very non-generic phase 
$\alpha = 0$, so that
the aggregated rays are the positive and negative imaginary axes.
See \autoref{fig:collapsing-mathieu}.
\insfigscaled{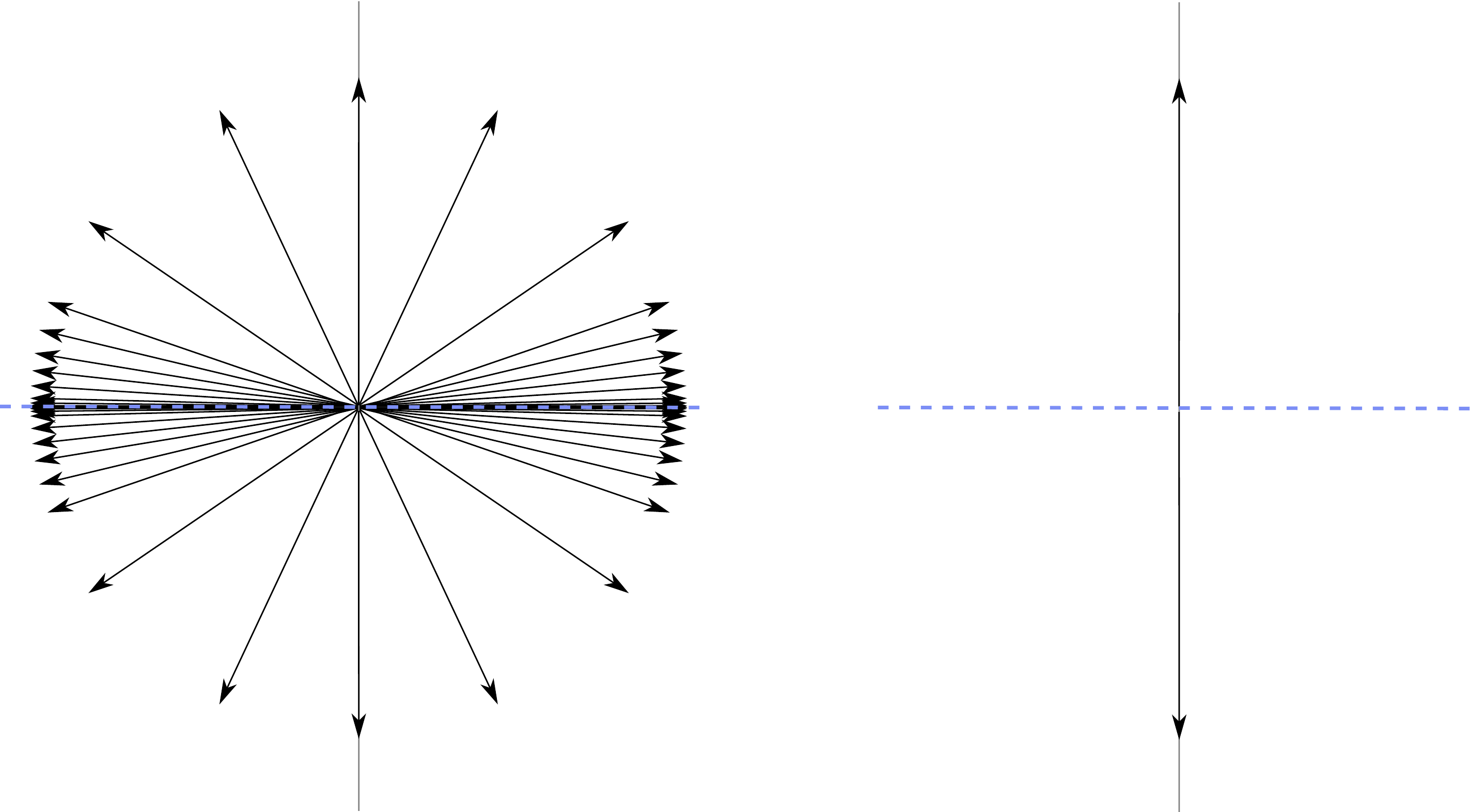}{0.4}{Collapsing the infinitely many active rays down to $2$ by making the choice \eqref{eq:theta-sg-type} with $\alpha = 0$. Each active ray on the right carries functions $F_{r,\gamma}$ which should be thought of as containing the same information as all the $F_{r,\gamma}$ in the
corresponding half-plane on the left. This is a particularly thorny case because the active
rays on the left accumulate at the boundary of the half-planes.}

Then, according to the recipe of \autoref{sec:integral-equations}, the functions $\cX_\gamma^{\RH}$ are:
\begin{equation}
	\cX_{\gamma}^{\RH}(\hbar) = \begin{cases} \cX_\gamma^{\vartheta=0}(\hbar) & \text{for } \re \hbar > 0, \\
	\cX_\gamma^{\vartheta=\pi}(\hbar) & \text{for } \re \hbar < 0. \end{cases}
\end{equation}
Now, to construct the functions $F_{r,\gamma}$ appearing in the 
integral equation \eqref{eq:rh-integral-equation}, 
we need to understand the 
discontinuity of $\cX_\gamma^{\RH}$ across the imaginary axis.
It turns out that this discontinuity has a more complicated form than those
we have previously considered: $\cX_\gamma^{\RH}$ is continuous along some segments
of the axis, and discontinuous along other segments. Correspondingly the functions $F_{r,\gamma}$
must be zero on some segments and nonzero on others, so in particular they cannot be
holomorphic functions of $\cX_\gamma$.
This feature is related to the fact that each $r$ aggregates contributions from
infinitely many rays which accumulate at the boundary of the half-plane,
as shown in \autoref{fig:collapsing-mathieu}.

We can work out the discontinuities of the functions 
$\cX_\gamma$ by keeping track of their symmetry properties.
First, we have
\begin{equation} \label{eq:continuation-rule}
	\cX^{\vartheta=\pi}_\gamma(-\hbar) = \cX_\gamma^{\vartheta =0}(\hbar)^{-1}.
\end{equation}
Second, $\cX_A(\hbar)$
is real for $\hbar > 0$, which implies
the reality property $\cX_A(\bar{\hbar}) = \overline{\cX_A(\hbar)}$.
Combining this with \eqref{eq:continuation-rule} we get
\begin{equation}
	\cX_A^{\RH}(- \bar\hbar) = \overline{\cX_A^{\RH}(\hbar)}^{-1}.
\end{equation}
It follows that the discontinuity of $\cX_A^{\RH}$ at the imaginary 
axis is
\begin{equation}
	\cX_A^{\RH} \mapsto \overline{\cX_A^{\RH}}^{-1} = \cX_A^{\RH} \times \abs{\cX_A^{\RH}}^{-2}.
\end{equation}
For $\cX_B^{\RH}$ it is similar except that 
the reality property has an extra sign,
$\cX_B(\bar{\hbar}) = \overline{\cX_B(\hbar)}^{-1}$,
giving
\begin{equation}
	\cX_B^{\RH}(- \bar\hbar) = \overline{\cX_B^{\RH}(\hbar)}.
\end{equation}
Thus the discontinuity of $\cX_B^{\RH}$ is
\begin{equation}
	\cX_B^{\RH} \mapsto \overline{\cX_B^{\RH}} = \cX_B^{\RH} \times \frac{\overline{\cX_B^{\RH}}}{\cX_B^{\RH}}.
\end{equation}
Substituting these discontinuities into the 
general form
\eqref{eq:rh-integral-equation} using \eqref{eq:f-discontinuity}, 
we get integral equations which 
are most naturally written directly in terms of
$x_\gamma = \log \cX_\gamma$:
\begin{equation} \label{eq:mathieu-integral-equation-1}
	x_A^{\RH}(\hbar) = \frac{Z_A}{\hbar} + \frac{1}{2 \pi \I} \int_0^{\I \infty} \de \hbar' \left( \frac{2\hbar}{\hbar'^2 - \hbar^2} \right) (-2 \re x_A^{\RH}(\hbar'+0)),
\end{equation}
\begin{equation} \label{eq:mathieu-integral-equation-2}
	x_B^{\RH}(\hbar) = \frac{Z_B}{\hbar} + \frac{1}{2 \pi \I} \int_0^{\I \infty} \de \hbar' \left( \frac{2\hbar}{\hbar'^2 - \hbar^2} \right) (-2 \I \im x_B^{\RH}(\hbar'+0)).
\end{equation}
Numerical experimentation gives us some confidence that \eqref{eq:mathieu-integral-equation-1}, \eqref{eq:mathieu-integral-equation-2}
do indeed hold.

These equations by themselves do not fully characterize
$x_A^{\RH}$ and $x_B^{\RH}$; to see this it is enough to 
observe that they admit the ``trivial'' solutions
$x_{\gamma}^{\RH}(\hbar) = \frac{Z_{\gamma}}{\hbar}$.
This is a bit disappointing when we compare to simpler examples like that
of \autoref{sec:integral-equations-cubic}, where it is believed that
the integral equations do characterize the spectral coordinates, and even give
a useful way of computing them.
One hope remains; the actual functions $x_{\gamma}^{\RH}$ obey one more important condition:
for $\hbar \in \pm \I \R$, the quantity $x_A^{\RH} \pm 2 x_B^{\RH}$ is always either
real or pure imaginary. It would be interesting to know whether this property
together with \eqref{eq:mathieu-integral-equation-1}, \eqref{eq:mathieu-integral-equation-2}
is enough to determine the functions $x^{\RH}_{\gamma}$.

\subsection{Another Stokes graph}

To get good information about the region $E > -1$ from WKB, we switch to
considering the $\vartheta$-Stokes graphs relevant for that region.
There are two possibilities, depending on whether $E \in (-1,1)$ or $E>1$.
Here we will just discuss $E > 1$. Then the $\vartheta$-Stokes graph for
$\vartheta = 0$ is shown in \autoref{fig:mathieu-sn-3}.

\insfigscaled{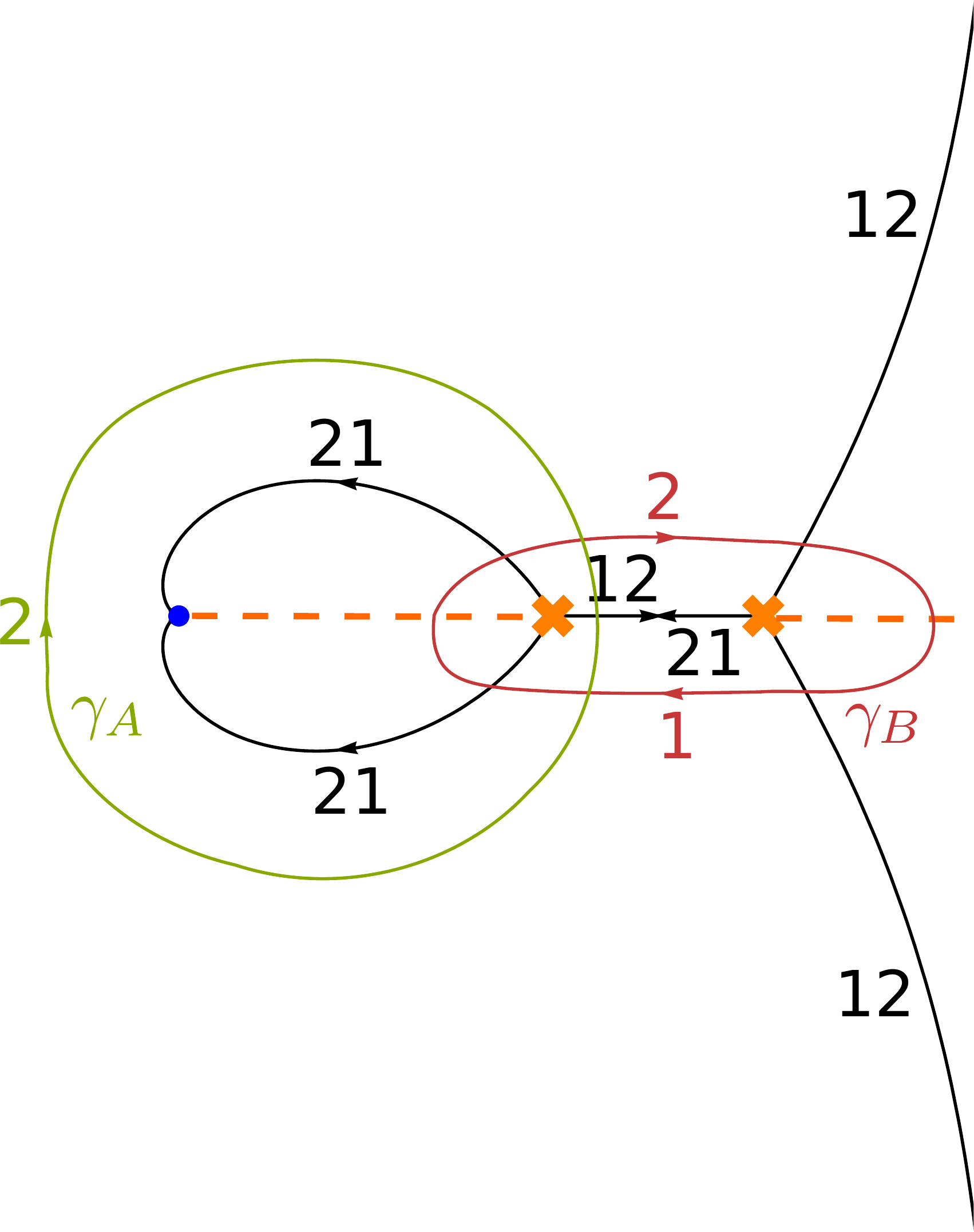}{0.30}{$\vartheta$-Stokes graph for the Mathieu equation, at the phase $\vartheta = 0$, and $E = \frac{41}{40}$. All notation is as in \autoref{fig:mathieu-sn-1}.
}

\subsection{The spectral coordinates}

Let $\psi$ denote the unique solution 
of \eqref{eq:mathieu-cp1} which decays exponentially as $z \to 0$
along the negative real axis, $\psi'$ the 
unique solution which decays exponentially as $z \to \infty$
along the negative real axis, and $M$ the operator of counterclockwise
monodromy around $z = 0$.
Then we have (see \autoref{app:computations})
\begin{equation} \label{eq:mathieu-coordinates-eg1}
	\cX_A = \pm \sqrt{\frac{[\psi , M \psi']}{[M \psi , \psi']}}, \qquad
 \cX_B = \frac{[\psi , M \psi][\psi' , M \psi']}{[\psi , \psi']^2}.	
\end{equation}

In particular, unlike
\autoref{sec:spectral-coords-mathieu},
here there is no spectral coordinate which is equal
to an eigenvalue of $M$. 
Nevertheless, we can express
the trace of the monodromy in terms of spectral coordinates:
\begin{equation} \label{eq:mathieu-trace-formula}
	\Tr M = (\cX_A + \cX_A^{-1}) \sqrt{1 - \cX_B}.
\end{equation}
One quick way to see \eqref{eq:mathieu-trace-formula} is to write $M$
relative to the basis $(\psi,\psi')$ as a matrix $\begin{pmatrix} a & b \\ c & d \end{pmatrix}$;
then \eqref{eq:mathieu-trace-formula} becomes
\begin{equation}
	a + d  = \pm \left(\sqrt{\frac{d}{a}} + \sqrt{\frac{a}{d}}\right) (\sqrt{1 + bc})
\end{equation}
which indeed holds, using the fact that $ad - bc = 1$.
To fix the sign we use the facts that, at small $\hbar$,
$\sqrt{1 - \cX_B}$ is exponentially
close to $1$, and $\cX_A$ is exponentially close
to an eigenvalue of $M$.

\subsection{Application: quasiperiodic solutions} \label{sec:mathieu-quasiperiodic}

Now we consider the application of these spectral coordinates
to another classical spectral problem.
If we consider $x$ to be a real variable, then \eqref{eq:mathieu} is
a Schr\"odinger equation with periodic potential, $V(x) = \cos x$.
The standard analysis of such equations involves
fixing $\nu \in \R / 2 \pi \Z$ (quasimomentum)
and looking at solutions obeying the quasiperiodic boundary condition
\begin{equation} \label{eq:quasiperiodic}
	\psi (x + 2 \pi) = \E^{\I \nu} \psi (x).
\end{equation}
For fixed $\nu$,
solutions of \eqref{eq:mathieu}, \eqref{eq:quasiperiodic}
exist only at a countable
set of energies $E = E_1, E_2, \dots$, with all $E_n > -1$.
These can be thought of as analogues of the bound state energies
for a confining potential on the real line.

Using \eqref{eq:mathieu-trace-formula} we can rewrite the quasiperiodicity condition \eqref{eq:quasiperiodic} 
as\footnote{The minus sign on the right side 
in \eqref{eq:mathieu-A-rewritten}
arises because of the square-root cut in the transformation
\eqref{eq:mathieu-replacements}.}
\begin{equation} \label{eq:mathieu-A-rewritten}
	(\cX_A + \cX_A^{-1}) \sqrt{1 - \cX_B} = -2 \cos \nu.
\end{equation}
This is another example of an ``exact quantization condition'' in the terminology
of exact WKB (however, we have not found precisely \eqref{eq:mathieu-A-rewritten}
in the literature.)

When $E>1$ we have $Z_B \in \R_-$, and the
leading WKB asymptotic $\cX_B \approx -\exp(Z_B / \hbar)$,
so the factor $\sqrt{1 - \cX_B}$ in \eqref{eq:mathieu-A-rewritten} 
gives an exponentially small correction.
As a first approach we could try
neglecting this correction. Then \eqref{eq:mathieu-A-rewritten}
reduces to
\begin{equation} \label{eq:mathieu-A-simplified}
	\cX_A \approx -\E^{\pm \I \nu}.
\end{equation}
To derive concrete predictions from \eqref{eq:mathieu-A-simplified}
we can use the WKB series for
$\cX_A$.
For example, suppose we take the leading asymptotic
$\cX_A \approx -\exp(Z_A / \hbar)$, and further take large $E$,
so that $Z_A \approx 2 \pi \I \sqrt{2 E}$: then we get
\begin{equation}
	\E^{2 \pi \I \sqrt{2 E} / \hbar} \approx \E^{\pm \I \nu},
\end{equation}
i.e.
\begin{equation} \label{eq:mathieu-free}
	E \approx \frac{\hbar^2}{2} \left(n \pm \frac{\nu}{2\pi}\right)^2.
\end{equation}
This is indeed the leading behavior of the energies at large $E$ and small $\hbar$;
in fact, in this limit we can approximate the quasiperiodic solutions with given $\nu$
simply by the free-particle wavefunctions, $\psi(x) \approx \E^{\I (\pm n + \nu / 2\pi) x}$.

To improve the accuracy
one could include subleading terms in the WKB series of
$\cX_A$; this gives perturbative corrections in a power series in $\hbar$. Likewise one could 
take more terms in the expansion of $Z_A$ around large $E$.
This would modify the relation between $E$ and $(n,\nu)$ 
but preserve the basic feature that for every $E$ there
is some corresponding $(n,\nu)$ with $\nu$ real.
Indeed, even if we used the \ti{exact} $\cX_A$ in \eqref{eq:mathieu-A-simplified}, 
we would still find
that for every $E$ there is a corresponding $(n,\nu)$ with $\nu$ real;
this follows from the fact that $\abs{\cX_A} = 1$ for all large enough real $E$,
a consequence of \eqref{eq:mathieu-coordinates-eg1}.

Now, let us consider the nonperturbative correction
$\sqrt{1 - \cX_B}$ in \eqref{eq:mathieu-A-rewritten}.
This has a qualitatively new effect: when $\cX_A(E)$ is
close to $\pm 1$,
the LHS of \eqref{eq:mathieu-A-rewritten} can have absolute value
larger than $2$. For such an $E$ 
there is no solution to \eqref{eq:mathieu-A-rewritten}
for any real $\nu$; the eigenvalues of the monodromy become complex.
This is the well-known phenomenon of ``gaps'' 
in the Mathieu spectrum.

It is known that the width of the gaps is exponentially suppressed 
by $\frac12 Z_B / \hbar$; see e.g. \cite{Dunne2016} for discussion and 
references on this point.\footnote{In this context the 
quantity $\frac12 Z_B$ might be called a ``$1$-instanton action''
since it corresponds to the change in the exponent of a WKB solution upon integrating along
a one-way path from one branch point to another, as opposed to $Z_B$ which is the
integral over the round-trip path $\gamma_B$.} Let us see how to 
recover this fact from
the exact quantization condition \eqref{eq:mathieu-A-rewritten}.
Taking $\cos \nu = -1$, expanding $\cX_A = 1 + \delta \cX_A$
and taking $\cX_B$ small, \eqref{eq:mathieu-A-rewritten} gives
\begin{equation}
	\left(2 + (\delta \cX_A)^2 \right) \left(1 - \frac12 \cX_B\right) \approx 2,
\end{equation}
i.e. the leading-order displacement of $\cX_A$ from the gap center is
\begin{equation}
	\delta \cX_A \approx \pm \sqrt{\cX_B},
\end{equation}
and thus the leading-order displacement of $E$ from the gap center is
\begin{equation}
	\delta E \approx \pm \frac{\sqrt{\cX_B}}{\partial \cX_A / \partial E}.
\end{equation}
If we further take the leading $\hbar \to 0$
asymptotics of $\cX_A$ and $\cX_B$, this becomes\footnote{As a check against blunders,
we numerically computed the width of a few of the gaps and obtained reasonable
agreement: for example, when $\hbar = 0.2$, there is a
gap extending from $E_- \approx 1.3836418$ to $E_+ \approx 1.3838946$, which thus has
$\delta E = \frac12(E_+ - E_-) \approx 0.0001264$, while the estimate \eqref{eq:gap-width-estimate} gives
$\delta E \approx 0.0001278$.}
\begin{equation} \label{eq:gap-width-estimate}
	\delta E \approx \pm \hbar \left( \I \frac{\partial Z_A}{\partial E} \right)^{-1} \exp\left(\frac1{2\hbar} Z_B\right).
\end{equation}
One could try to go beyond this leading-order estimate 
using the full $\hbar$ expansions of $\cX_A$ and $\cX_B$.
It would be interesting to know whether in this way one can recover the more detailed results
on the gap widths explained in \cite{Dunne2016}.

\section{Exact WKB for higher order equations} \label{sec:wkb-higher}

So far we have been discussing order $2$ differential equations
\eqref{eq:schrodinger}.
We now move to the case of order $3$ equations, involving
two meromorphic ``potentials'' $P_2$ and $P_3$:
\begin{equation} \label{eq:deg-3-oper}
	\left[ \partial_z^3 + \hbar^{-2} P_2 \partial_z + (\hbar^{-3} P_3 + \frac12 \hbar^{-2} P'_2) \right] \psi(z) = 0.
\end{equation}
The equation \eqref{eq:deg-3-oper} can be given a global meaning on a Riemann surface $C$ with a complex projective structure,
as with \eqref{eq:schrodinger} above; 
in this case $\psi(z)$ is a section of $K_C^{-1}$, $P_2(z)$ is a meromorphic quadratic
differential, and $P_3(z)$ is a meromorphic cubic differential.

In this section we explain how the exact WKB method is 
expected to extend to equations of the form \eqref{eq:deg-3-oper}.
In this situation there are no rigorous
results yet,
but there is a reasonable conjectural picture.
(The same picture is expected to work for equations of any order $K \ge 2$;
we stick to $K=3$ to be concrete, and because our main example has $K=3$.)
Some numerical evidence supporting this conjectural picture 
in special cases has been obtained in \cite{Hollands2017,dumasneitzke}.
We will give more numerical evidence
in the case of the $T_3$ equation in \autoref{sec:t3-numerics} and
\autoref{sec:integral-equations-t3} below.

All the formal structures in the story are parallel to the 
order $2$ case, so this section is organized in parallel
to \autoref{sec:wkb-review}, and we will be very brief.

\subsection{WKB solutions}

WKB solutions of \eqref{eq:deg-3-oper} are solutions
of the form
\begin{equation} \label{eq:wkb-solutions-higher}
	\psi(z) = \exp \left(\hbar^{-1} \int_{z_0}^z \lambda(z) \, \de z \right),
\end{equation}
where now $\lambda$ must obey a higher analogue of the Riccati
equation \eqref{eq:riccati},
\begin{equation} \label{eq:riccati-higher}
	\lambda^3 + 3 \hbar \lambda \partial_z \lambda + \hbar^2 \partial_z^2 \lambda + P_2 \lambda + P_3 + \half \hbar P_2' = 0.
\end{equation}
One again constructs WKB solutions $\lambda_i^\formal$ as power
series in $\hbar$. As before, one meets an ambiguity at order
$\hbar^0$; this ambiguity is resolved by choosing a sheet $i$ of the
3-fold covering
\begin{equation}
	\Sigma = \{ y^3 + p_2(z) y + p_3(z) = 0 \}.
\end{equation}
Now the conjectural picture is that, as in the order $2$ case, there exist actual solutions 
of \eqref{eq:riccati-higher} which have
the asymptotic behavior $\lambda^\vartheta_i \sim \lambda_i^\formal$
in the half-plane $\bbH_\vartheta$, away from $\vartheta$-Stokes curves.

The $\vartheta$-Stokes curves carry labels $ij$.
Along a $\vartheta$-Stokes curve of type $ij$,
$\E^{-\I \vartheta} (y_i - y_j) \de z$ is real and positive.
We make the simplifying assumption that all branch points of $\Sigma$
are simple branch points, i.e. only two $y_i$ collide at a time.
For the construction of the $\vartheta$-Stokes graph in this case
see \cite{Gaiotto2012}.
One key new feature of the higher-order case, first
discovered in \cite{berk:988} and further investigated in e.g. 
\cite{akt-wkb,MR2132714,Gaiotto2012},
is that Stokes curves of type
$ik$ can be born from intersections of Stokes curves of
types $ij$ and $jk$. See \autoref{fig:sample-sn-higher}
for an example.

The local solution $\lambda^\vartheta_i$ of \eqref{eq:riccati-higher} 
is supposed to exist away from $\vartheta$-Stokes curves of type $ij$, 
as in the order $2$ case.
On crossing a $\vartheta$-Stokes curve of type $ij$, we conjecture
that the local
WKB solution $\psi_i^\vartheta$ jumps by a constant multiple of $\psi_j^\vartheta$.\footnote{Some evidence
for this conjecture has been given in \cite{MR2020659}. We thank Kohei Iwaki for pointing out this
reference.}

\insfigscaled{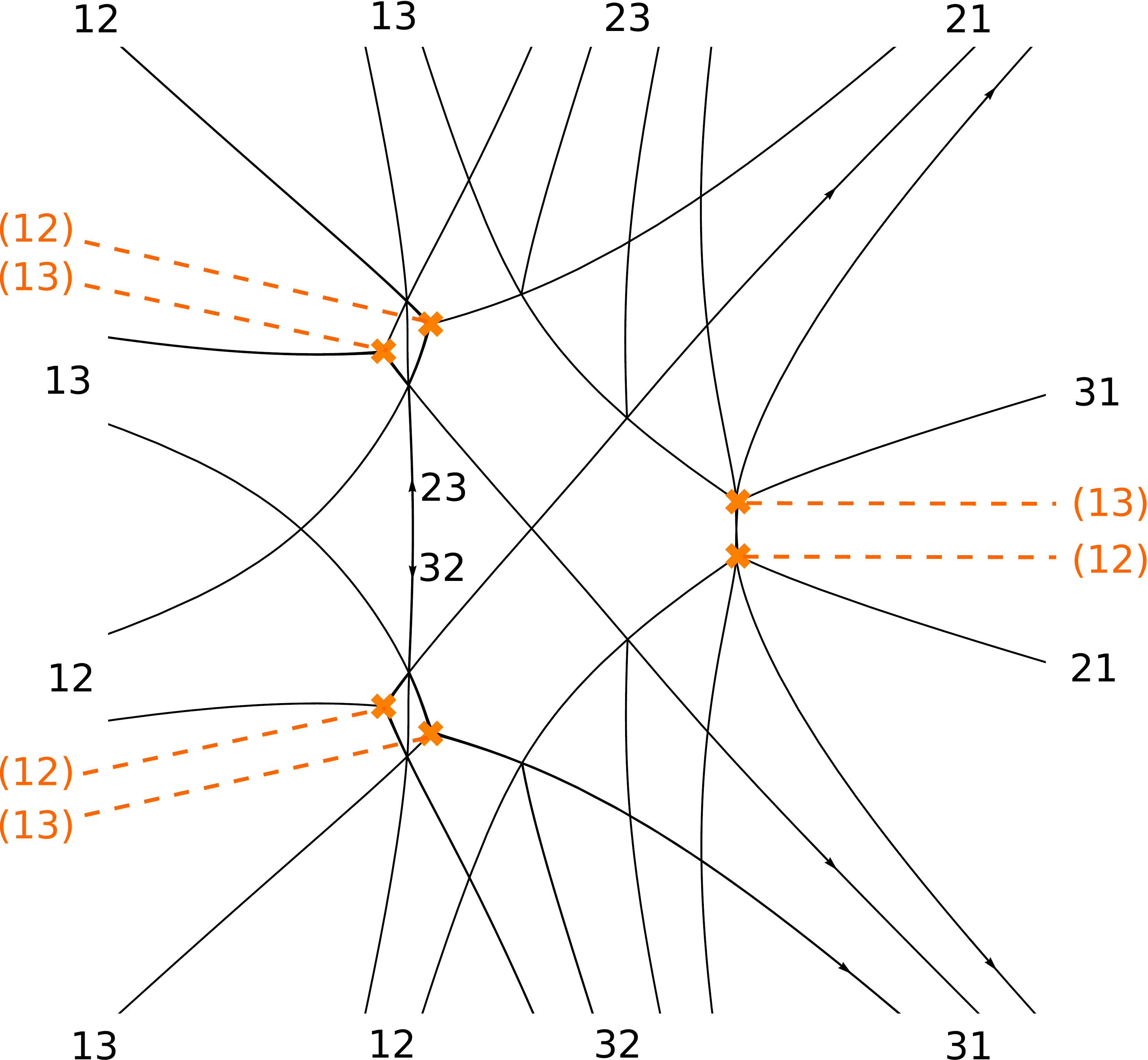}{0.24}{An example
of a $\vartheta$-Stokes graph at $\vartheta=0$,
with $p_2(z) = 1$ and $p_3(z) = z^3-1$.}

\subsection{Abelianization}

As in the order $2$ case, 
the WKB solutions of \eqref{eq:deg-3-oper} 
can be thought of as solutions of a first-order equation
over $\Sigma$, built using the $\lambda_i^\vartheta$.
Thus, as before, 
exact WKB analysis of \eqref{eq:deg-3-oper} 
leads to a line bundle $\cL$ with almost-flat
connection $\nabla^{\ab,\vartheta}$ over $\Sigma$,
away from the $\vartheta$-Stokes curves.

\subsection{Gluing across the Stokes graph}

Also as before, we can glue $\cL$ and $\nabla^{\ab,\vartheta}$ across the 
$\vartheta$-Stokes curves. 
At a $\vartheta$-Stokes curve of type $ij$ the gluing takes the form
(cf. \eqref{eq:gluing-1})
\begin{equation} \label{eq:gluing-1-rank3}
\begin{pmatrix}
\psi_i^{L} \\
\psi_j^{L}  \\
\psi_k^{L}
\end{pmatrix}
\mapsto 
	\begin{pmatrix} 1 & \beta & 0 \\ 0 & 1 & 0 \\ 0 & 0 & 1 \end{pmatrix} \begin{pmatrix}
\psi_i^{L} \\
\psi_j^{L} \\
\psi_k^{L} 
\end{pmatrix}
= 
\begin{pmatrix}
\frac{[\psi_i^L , \psi_j^L , \psi_k^L]}{[\psi_i^R , \psi_j^L , \psi_k^L]}
\psi_i^{R} \\
\frac{[\psi_j^L, \psi_k^L, \psi_i^L]}{[\psi_j^R, \psi_k^L, \psi_i^L]}  \psi_j^{R} \\
\frac{[\psi_k^L, \psi_i^L, \psi_j^L]}{[\psi_k^R, \psi_i^L, \psi_j^L]}  \psi_k^{R}

\end{pmatrix}.
\end{equation}
If $\vartheta$-Stokes curves of type $ij$ and $ji$ coincide, we choose a gluing
of the form (cf. \eqref{eq:gluing-2})\footnote{As in the order $2$ case 
(see \autoref{sec:stokes-gluing}) 
this is not the only possible choice, but
it is the most invariant choice.}
\begin{equation} \label{eq:gluing-2-rank3}
\begin{pmatrix}
\psi_i^{L} \\
\psi_j^{L} \\
\psi_k^{L} 
\end{pmatrix}
\mapsto 
	\begin{pmatrix} \rho & \beta & 0 \\ \alpha & \rho & 0 \\ 0 & 0 & 1 \end{pmatrix} \begin{pmatrix}
\psi_i^{L} \\
\psi_j^{L} \\
\psi_k^{L}
\end{pmatrix}
= 
\begin{pmatrix}
\sqrt{\frac{[\psi_i^L , \psi_j^L , \psi_k^L]}{[\psi_i^R , \psi_j^R , \psi_k^L]} \frac{[\psi_i^L , \psi_j^R , \psi_k^L]}{[\psi_i^R , \psi_j^L , \psi_k^L]}}
\psi_i^{R} \\
\sqrt{\frac{[\psi_j^L , \psi_i^L , \psi_k^L]}{[\psi_j^R , \psi_i^R , \psi_k^L]} \frac{[\psi_j^L , \psi_i^R , \psi_k^L]}{[\psi_j^R , \psi_i^L , \psi_k^L]}}\psi_j^{R} 
\\ \frac{[\psi_k^L, \psi_i^L, \psi_j^L]}{[\psi_k^R, \psi_i^L, \psi_j^L]}  \psi_k^{R}
\end{pmatrix},
\end{equation}
with $\rho^2 - \alpha \beta = 1$. (The branches of the square roots are fixed as was done above in the $K=2$ case.)
By this process we obtain a line bundle $\cL$ with almost-flat
connection $\nabla^{\ab,\vartheta}$ over $\Sigma$.

\subsection{Spectral coordinates}

Finally we can introduce higher-order versions of the spectral
coordinates: as before, these are defined by
\begin{equation}
	\cX_\gamma^\vartheta = \Hol_\gamma \nabla^{\ab,\vartheta} \in \C^\times.
\end{equation}
The $\cX_\gamma^\vartheta$ 
are expected to have all the same formal properties as in the order $2$
case, 
discussed in \autoref{sec:spectral-coordinates}-\autoref{sec:integral-equations};
we will not repeat those here.

\section{Exact WKB for the \texorpdfstring{$T_3$}{T3} equation} \label{sec:t3}

Now we consider
a specific instance of \eqref{eq:deg-3-oper}, a 
third-order ODE over $\CP^1$ with three regular 
singularities. By convention we place the singularities at $\{1,\omega,\omega^2\}$ where $\omega = \E^{2 \pi \I / 3}$:\footnote{Our conventions
here differ from those of \cite{Hollands:2016kgm} by the replacement $u \to -u$. Sorry.}
\begin{equation} \label{eq:t3-equation}
	\left[ \partial_z^3 + \hbar^{-2} P_2 \partial_z + (\hbar^{-3} P_3+\frac12 \hbar^{-2} P'_2) \right] \psi(z) = 0, \qquad
	P_2 = \frac{9 \hbar^2 z}{(z^3 - 1)^2}, \quad P_3 = \frac{u}{(z^3-1)^2}.
\end{equation}
We call \eqref{eq:t3-equation} the $T_3$ equation.
This equation actually does not depend on $u$ and $\hbar$ separately,
but only on the combination $u' = u / \hbar^3 \in \C$.

\subsection{A simple Stokes graph}

The $\vartheta$-Stokes graphs for the $T_3$ equation were
investigated in \cite{Hollands:2016kgm}. It was found there that the
topology of the $\vartheta$-Stokes graph depends on the phase of the quantity
$w = \E^{-3 \I \vartheta} u$.
For a generic phase of $w$, it seems likely that 
the Stokes graph is ``wild'' --- in particular, that it is dense at least 
in some parts of $\CP^1$. 
WKB analysis involving such a wild Stokes graph may 
ultimately be very interesting, but we
are not brave enough to try it today.\footnote{In the order $2$ case, some of the necessary
analytic technology for dealing with wild Stokes graphs is developed in \cite{Fenyes2015a}. It would
be exciting to develop the higher-rank analogue of this.} 
Instead, we focus on the
non-generic situation where the Stokes graph is compact; this happens for a countable
set of phases of $w$. We will not rederive the form of the Stokes graphs
here, but simply lift them from \cite{Hollands:2016kgm}.

The simplest Stokes graph arises when $w$ is real; to be completely concrete, we
take $u > 0$ and $\vartheta = 0$.
See \autoref{fig:circle-network}.
\insfigscaled{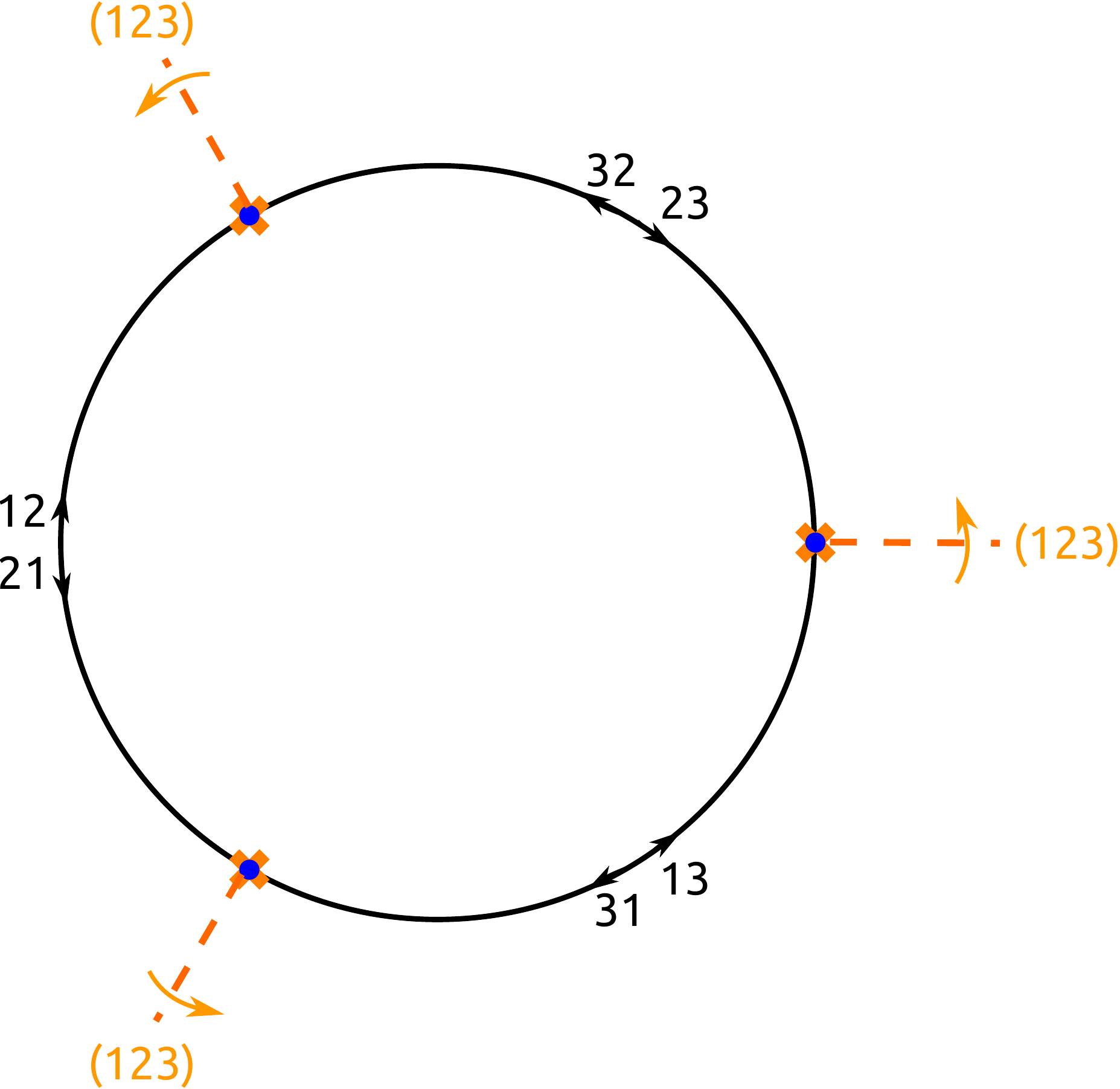}{0.31}{The $\vartheta$-Stokes graph for
the $T_3$ equation, in case $u > 0$ and $\vartheta = 0$. The three
branch cuts emanating from the singularities meet at $z = \infty$.}
Applying the higher-order exact WKB method is expected 
to produce a $\cW$-abelianization of the $T_3$ equation.
Thus, we should begin by understanding concretely
what this means.

We explained in \autoref{sec:w-framings} that 
$\cW$-abelianizations of a meromorphic
Schr\"odinger equation with second-order poles
are in 1-1 correspondence with discrete data called 
$\cW$-framings, 
and the choice of a $\cW$-framing amounts to
choosing one of the two eigenvectors of the monodromy
around each singularity and each cylinder.
In the case of the $T_3$ equation, we will have a formally similar
story, except that the linear-algebra problem one has to solve to find
$\cW$-abelianizations is more intricate: it does not just correspond
to choosing eigenvectors of monodromy matrices.

\subsection{The abelianization problem for the \texorpdfstring{$T_3$}{T3} equation}

The local solutions of \eqref{eq:t3-equation} in a neighborhood of $z=0$
form a $3$-dimensional vector space $V$.
In \autoref{fig:base-cycles} we show three cycles $A, B, C$ on $\CP^1 \setminus \{1,\omega,\omega^2\}$, beginning and ending at $z = 0$.
\insfigscaled{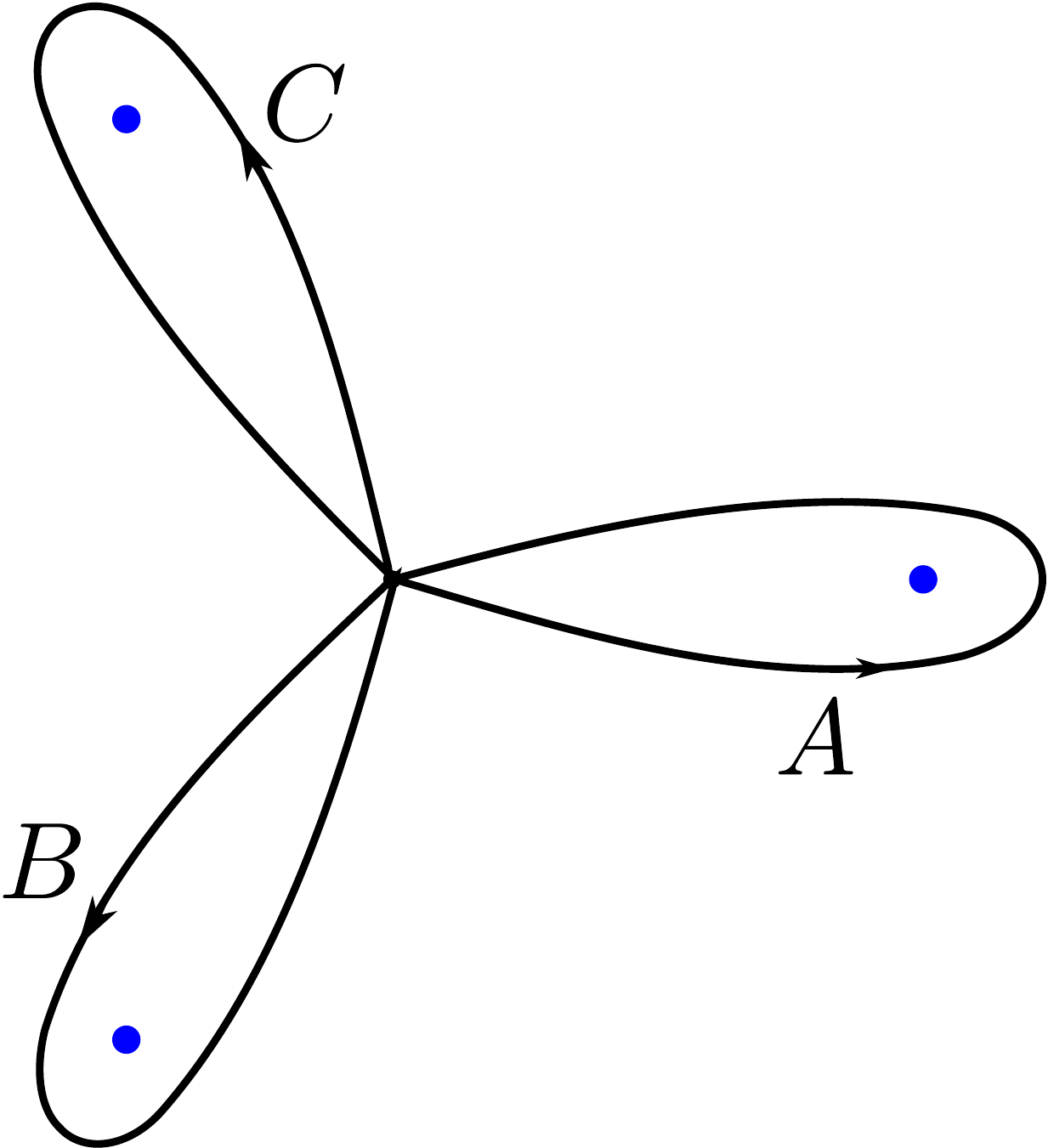}{0.3}{Three cycles on $\CP^1 \setminus \{1,\omega,\omega^2\}$.}
Let $\bA, \bB, \bC$ denote the maps $V \to V$ induced by monodromy
of \eqref{eq:t3-equation} around
these three cycles. Note they satisfy
\begin{equation} \label{eq:abc}
	\bA \bB \bC = 1.
\end{equation}
We say a basis $(\psi_1, \psi_2, \psi_3)$ of $V$ is in \ti{special position} if
the following conditions are satisfied:
\begin{subequations} \label{eq:planarity-relations}
\begin{gather}
\bC \psi_1, \bB^{-1} \psi_2 \in \IP{\psi_1, \psi_2}, \label{eq:planarity-2} \\
\bA \psi_2, \bC^{-1} \psi_3 \in \IP{\psi_2, \psi_3}, \label{eq:planarity-1} \\
\bB \psi_3, \bA^{-1} \psi_1 \in \IP{\psi_3, \psi_1}. \label{eq:planarity-3}
\end{gather}
\end{subequations}
A concrete way to think about the special-position constraint 
is that relative to the basis
$(\psi_1,\psi_2,\psi_3)$ the monodromy 
endomorphisms must have zeroes in specific places:
\begin{subequations} \label{eq:monodromy-constraints}
\begin{alignat}{3}
	\bA &= \begin{pmatrix} * & 0 & * \\ * & * & * \\ * & * & * \end{pmatrix}, \quad & 
	\bB &= \begin{pmatrix} * & * & * \\ * & * & 0 \\ * & * & * \end{pmatrix}, &
	\bC &= \begin{pmatrix} * & * & * \\ * & * & * \\ 0 & * & * \end{pmatrix}, \\
	\bA^{-1} &= \begin{pmatrix} * & * & * \\ 0 & * & * \\ * & * & * \end{pmatrix}, &
	\bB^{-1} &= \begin{pmatrix} * & * & * \\ * & * & * \\ * & 0 & * \end{pmatrix}, \quad &
	\bC^{-1} &= \begin{pmatrix} * & * & 0 \\ * & * & * \\ * & * & * \end{pmatrix}.
\end{alignat}
\end{subequations}

The special-position constraint 
is invariant under rescalings of the vectors
$(\psi_1, \psi_2, \psi_3)$: it depends only
on the \ti{projective} basis of $V$, which we can view as
a 3-tuple of points in the projective space
${\bbP}(V) \simeq \CP^2$.

The point of this definition is the following,
proven in \autoref{app:computations}: 
\ti{$\cW$-abelianizations for the $T_3$
equation are in 1-1 correspondence with projective bases
$(\psi_1, \psi_2, \psi_3)$ of $V$ in special position.}

Now the question arises:
how can we enumerate the possible
projective bases of $V$ in special position? 
Note that \eqref{eq:monodromy-constraints} imposes 
$6$ conditions on the basis, so a naive dimension count
would suggest that bases obeying these constraints 
should occur discretely. To enumerate them precisely is a problem
of algebraic geometry, which we address in \autoref{sec:enumerating-special-bases} below.
The outcome is that when
$\bA$, $\bB$, $\bC$ are unipotent and generic enough 
there are ``$4+\infty$'' projective bases
in special position: $4$ occurring discretely plus
a 1-parameter family.

\subsection{Projective bases in special position} \label{sec:enumerating-special-bases}

In this section we consider the following question. Suppose given unipotent endomorphisms 
$\bA$, $\bB$, $\bC$ of a $3$-dimensional complex vector space
$V$, obeying $\bA \bB \bC = {\mathbf 1}$.
Assume that $\bA$, $\bB$, $\bC$ are in general position; concretely this means that each of $\bA$, $\bB$, $\bC$ preserves
a unique complete flag, and these flags are in general position. 
How do we enumerate the
projective bases of $V$ in special position?

We begin with an observation. Let $\IP{e_{\bA}}$ denote the eigenline of $\bA$
and similarly for $\bB, \bC$.
Suppose that $(\psi_1,\psi_2,\psi_3)$
is a projective basis in special position. 
Assume that $\IP{\psi_1} \neq \IP{e_\bC}$.
Then $\IP{\psi_1, \bC \psi_1}$ is a plane, and \eqref{eq:planarity-2}
says this plane contains both $\psi_2$ and $\bB^{-1} \psi_2$.
Equivalently, we have
\begin{equation}
	\psi_2 \in \IP{\psi_1, \bC \psi_1}, \qquad \psi_2 \in \bB \IP{\psi_1, \bC \psi_1}.
\end{equation}
Now, these two planes are not equal (if they were, then \eqref{eq:planarity-3} would 
show that this plane also contains $\psi_3$, contradicting the linear independence
of the $\psi_i$.)
Since both contain
$\psi_2$, their intersection must
be precisely $\IP{\psi_2}$:
\begin{equation} \label{eq:psi2-determined}
	\IP{\psi_2} = \IP{\psi_1, \bC \psi_1} \cap \bB  \IP{\psi_1, \bC \psi_1}.
\end{equation}
Let $X = \bbP(V) \simeq \CP^2$.
The relation \eqref{eq:psi2-determined} can be expressed as
\begin{equation}
\psi_2 = \Phi_{\bB,\bC}(\psi_1)
\end{equation}
where $\Phi_{\bB,\bC}: X \dashrightarrow X$ is 
the product of two ``quadratic transformations''\footnote{A useful reference
on quadratic transformations is \cite{cremona}.}
\begin{equation}
	\Phi_{\bB,\bC} = \Phi_{\bB^*} \circ \Phi_\bC,
\end{equation}
with $\Phi_\bC: X \dashrightarrow X^*$
the quadratic transformation taking the line $\IP{\psi}$ to the 
plane $\IP{\psi,\bC \psi}$,
and $\Phi_{\bB^*}: X^* \dashrightarrow X$
the dual quadratic transformation taking a plane $p$
to the line $p \cap \bB p$.
Thus $\Phi_{\bB,\bC}$ is a
birational map (Cremona transformation) of degree $4$, i.e.
defined by three homogeneous degree $4$ polynomials.

Thus $\psi_2$ is determined by $\psi_1$.
Repeating this process using \eqref{eq:planarity-3}, \eqref{eq:planarity-1} shows $\psi_3$ is determined by $\psi_2$,
and $\psi_1$ is determined by $\psi_3$:
\begin{equation} \label{eq:psi3-determined}
\psi_3 = \Phi_{\bC,\bA}(\psi_2), \qquad \psi_1 = \Phi_{\bA,\bB}(\psi_3)
\end{equation}
Altogether, this means $\psi_1$ is constrained to obey
\begin{equation}
	\psi_1 = \crem (\psi_1)
\end{equation}
where $\crem: X \dashrightarrow X$ is a degree $64$
birational map
\begin{equation}
	 \crem = \Phi_{\bA,\bB} \circ \Phi_{\bC,\bA} \circ \Phi_{\bB,\bC}.
\end{equation}
Thus, whenever $(\psi_1,\psi_2,\psi_3)$ is a projective basis
in special position, $\IP{\psi_1} \in X$ is a fixed point of
$\crem$, and \eqref{eq:psi3-determined} then determines
the rest of the basis.
This translates the job of finding projective bases in special position
to the job of finding the fixed locus of $\crem$.

This problem is simplified by the observation that
$\crem$ preserves the ratio of two cubic forms.
Indeed, suppose we define a cubic form on $V$ by
\begin{equation}
F_{M,M'}(\psi) = [\psi, M \psi, M' \psi],
\end{equation}
and dually on $V^*$
\begin{equation}
F^*_{M,M'}(\eta) = [\eta, M^T \eta, M'^T \eta].
\end{equation}
Then we have an identity of sextic forms on $V$,\footnote{We have no great
insight into why this identity is true, although we have checked it in Mathematica; 
it is a specialization of a ``remarkable identity'' 
originally due to Zagier, given as equation 14 in \cite{arxiv-1202.4161}.}
\begin{equation} \label{eq:sextic-identity}
	F^*_{M,M'}(\Phi_{M'}(\psi)) = F_{M,M'}(\psi) F_{M'^{-1},M}(\psi).
\end{equation}
Now we consider the ratio of cubic forms
\begin{equation}
	r(\psi) = \frac{F_{\bC,\bA^{-1}}(\psi)}{F_{\bC^{-1},\bA}(\psi)}.
\end{equation}
Using \eqref{eq:sextic-identity} six times
we obtain the desired invariance:
\begin{equation}\label{eq:preserved-cubic}
		r(\crem(\psi)) = r(\psi).
\end{equation}
\eqref{eq:preserved-cubic} is equivalent to saying that $\crem$ preserves
a one-parameter family (pencil) of cubic curves $E_t \subset X$,
\begin{equation} \label{eq:cubic-pencil}
	E_t = \{F_{\bC,\bA^{-1}} (\psi) + t F_{\bC^{-1},\bA} (\psi) = 0\} \subset X.
\end{equation}

There are three points of $X$ which are common to all of the $E_t$, or
said otherwise, this pencil of cubic curves has a base locus supported at three points of $X$. Two of the base points are easy to spot:
if $\psi = e_\bA$ or $\psi = e_\bC$
then $F_{\bC,\bA^{-1}}(\psi) = F_{\bC^{-1},\bA}(\psi) = 0$ and so $\psi$ lies on every $E_t$.
The last base point is trickier: it is $p_\bB \cap \bC^{-1} p_\bB$
where $p_\bB$ is the unique plane fixed by $\bB$. (Indeed if $\psi \in p_\bB \cap \bC^{-1} p_\bB$
then $\psi$, $\bC \psi$ and $\bA^{-1} \psi$ all lie in $p_\bB$, showing that
$F_{\bC,\bA^{-1}}(\psi) = 0$; similarly $F_{\bC^{-1},\bA}(\psi) = 0$.) These three base points lie on a line $\ell \subset X$. In fact the line $\ell$
(with multiplicity $3$) is equal to $E_t$ for some $t = t_*$.
Any point of $\ell$ is a fixed point of $\crem$ (with the exception of the three base points, where $\crem$ is not defined). This gives a $1$-parameter family of projective bases in special position.

Now we want to see if there are any other fixed points.
For this purpose
the fact that $\crem$ is not defined everywhere is technically inconvenient. To resolve its indeterminacies we blow up the base 
locus.
This results in a singular surface, but by further
blowing up the singular points, 
we obtain a smooth rational elliptic
surface $\tX$. See \autoref{fig:rational-surface}. 
\insfigscaled{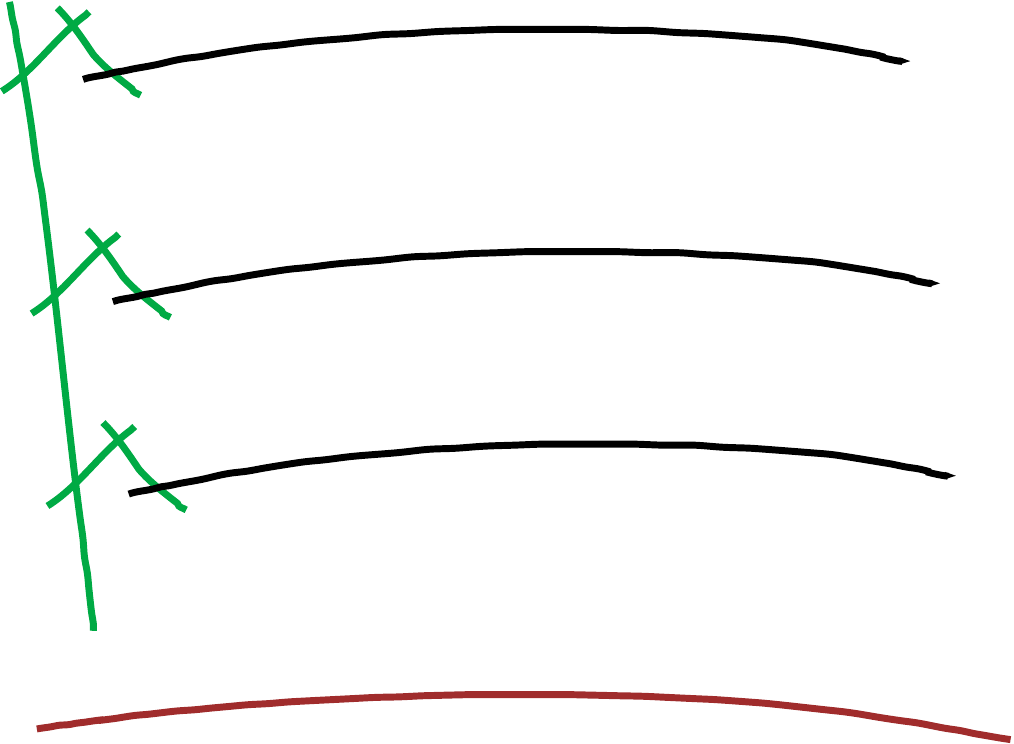}{0.4}{A neighborhood of the $IV^*$
fiber (green) in the rational elliptic surface $\tX$. The preimage
of each point of the base locus is a chain of three
rational curves; two of the three are in the $IV^*$ fiber (green),
while the last is a section of the elliptic fibration (black).
There are generically $4$ other singular fibers elsewhere in $\tX$ 
(not shown).
}
$\tX$ has one fiber of Kodaira type $IV^*$
(affine $E_6$ configuration), which maps to the line $E_{t_*}$
through the base points in $X$.
This fiber has Euler characteristic $8$.
A smooth rational elliptic surface has Euler characteristic
$12$, and the smooth fibers do not contribute to
the Euler characteristic, so there must be some other singular fibers
in $\tX$;
the most generic possibility is to have $4$ more singular fibers, 
each of type $I_1$ (nodal torus), so that altogether
$\tX$ has singular fibers $IV^* + 4 I_1$.
For some special $\bA$, $\bB$, $\bC$ it may happen that
some of the $I_1$ fibers collide.
In particular, in \autoref{sec:t3-analytic-cont} below we will meet
a phenomenon where two $I_1$ fibers collide to make an $II$ fiber
(cusp), so that $\tX$ has singular fibers $IV^* + II + 2 I_1$.

The birational automorphism $\crem$ of $X$ 
lifts to a regular map $\tX \to X$, 
so in particular $\crem$ acts by an honest automorphism
of each fiber except for the $IV^*$ fiber.
Since these fibers are (smooth or nodal) elliptic curves, their
automorphism groups are easy to understand, and indeed
by direct computations near a base point
one can show that $\crem$ is not trivial and not an inversion;
thus it must act by a nontrivial translation on each fiber.
It follows that the only place a fixed point can occur
is at a singularity of a fiber; in particular, in the generic
$IV^* + 4 I_1$ case, the fixed points are exactly the $4$
nodes.
Combining these with the continuous family we found before,
we conclude finally that $\crem$ has ``$4 + \infty$''
fixed points, and thus there are ``$4 + \infty$''
projective bases in special position, as we claimed above.

This description of the projective bases 
in special position gives some small insight into their nature
and their number, but more importantly for us, it 
is efficient enough to be used
for numerical computations: starting from $\bA$, $\bB$,
$\bC$, we use Mathematica to solve the polynomial system determining
the singularities of the cubic curves \eqref{eq:cubic-pencil};\footnote{In particular this seems to be much more efficient than trying to solve the coplanarity constraints \eqref{eq:planarity-relations} directly.}
these give the desired basis elements $\psi_1$;
then we determine $\psi_2$ and $\psi_3$ using
\eqref{eq:psi2-determined}, \eqref{eq:psi3-determined}.

Finally let us comment on the case of $\bA$, $\bB$, $\bC$
semisimple instead of unipotent.
(This case would arise if, instead of the conformally invariant 
Minahan-Nemeschansky theory, we considered its mass deformation.)
In this case the analysis is very similar
to the above,
except that the rational elliptic surface $\tX$ which appears is a bit
different: it arises by blowing up $9$ distinct 
points of $X$ (lying on a cubic
curve), instead of $3$ with multiplicity.
The result is that instead of singular fibers of type $IV^* + 4 I_1$
one generically gets $12 I_1$, and so
instead of ``$4+\infty$'' $\cW$-abelianizations
there are generically just $12$ $\cW$-abelianizations.

\subsection{The spectral coordinates}

Now we are in a position to decribe the spectral coordinates
concretely.

\insfigscaled{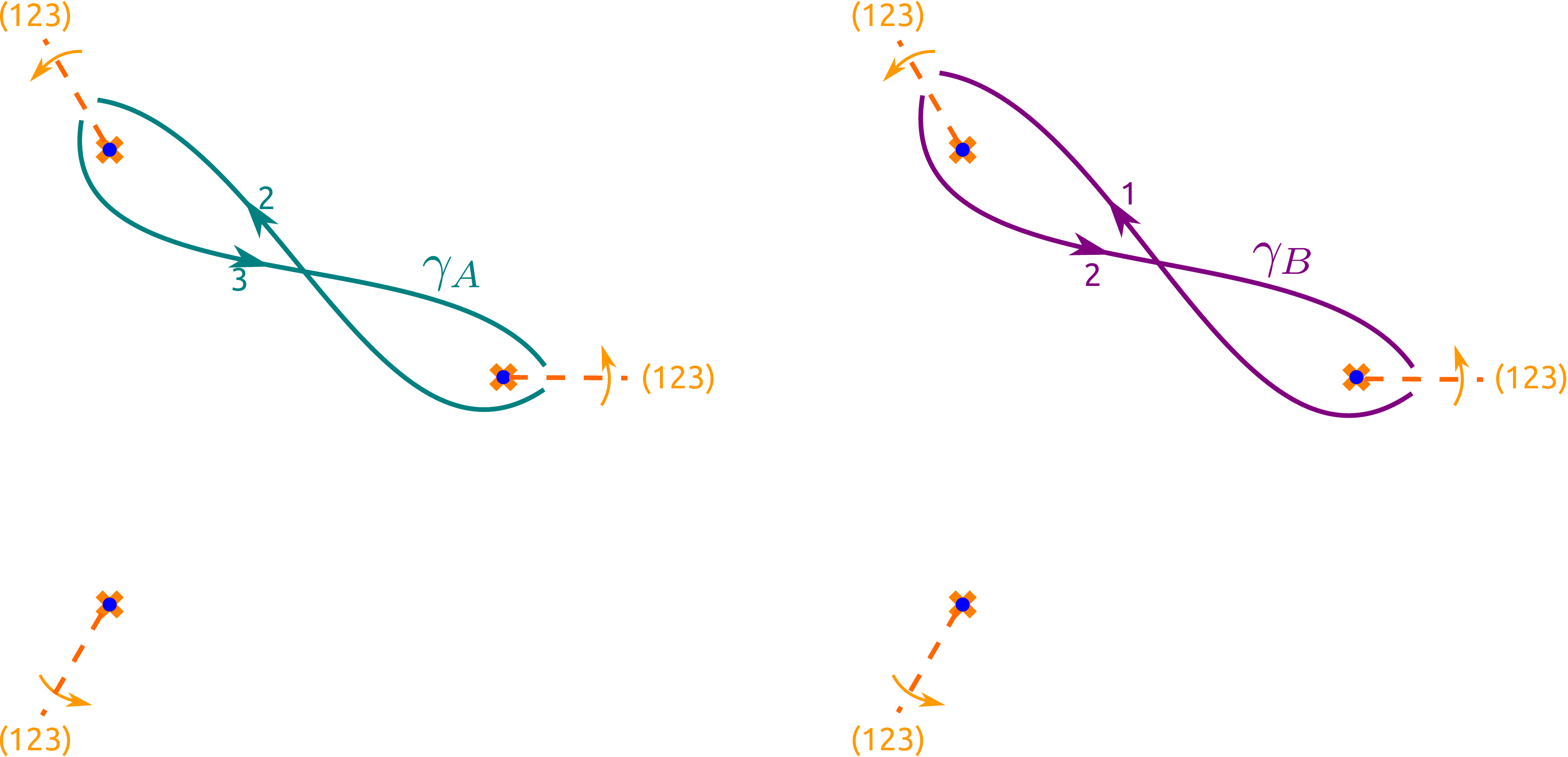}{0.25}{Cycles $\gamma_A$ and $\gamma_B$ on the 3-fold branched cover $\Sigma$.}

Let $\gamma_A$, $\gamma_B$ be the cycles on $\Sigma$ shown in \autoref{fig:circle-network-cycles}.
Fix an $\SL(3)$-connection $\nabla$ over $C$, with unipotent holonomy, and fix a
$\cW$-abelianization of $\nabla$.
Let $(\psi_1, \psi_2, \psi_3)$ be the corresponding basis of solutions near
$z = 0$. As we have explained above,
$(\psi_1, \psi_2, \psi_3)$ are in special position.
Then, the spectral coordinates are (see \autoref{app:computations})
\begin{subequations} \label{eq:spectral-coords-t3}
\begin{align}
\cX_A &= \frac{[\psi_2 , \psi_3 , \psi_1]}{[\bC^{-1} \psi_3 , \bA \psi_2 , \psi_1]}, \label{eq:XA-t3} \\
\cX_B &= \sqrt{- \frac{[\bC \psi_1 , \bB^{-1} \psi_2 , \psi_3][\bC \psi_1 , \psi_1 , \psi_3] [\psi_2 , \bA \psi_2 , \psi_1] [\bB \psi_3 , \bA^{-1} \psi_1 , \psi_2] [\bB \psi_3 , \psi_3 , \psi_2]}{[\psi_2 , \bB^{-1} \psi_2 , \psi_3][\bC^{-1} \psi_3 , \bA \psi_2 , \psi_1] [\bC^{-1} \psi_3 , \psi_3 , \psi_1] [\psi_1 , \psi_3 , \psi_2] [\psi_1 , \bA^{-1} \psi_1 , \psi_2]} }. \label{eq:XB-t3}
\end{align}
\end{subequations}

\subsection{Spectral coordinates for the continuous family of abelianizations}
\label{sec:t3-fg-coords}

In this section we record an interesting curiosity, not
required for the rest of the paper.

Recall that there is a continuous family of $\cW$-abelianizations,
with the property that $\psi_1$ is a linear combination of
the eigenvectors of $\bA$ and $\bC$, and similarly
for $\psi_2$, $\psi_3$.
It turns out that the 
spectral coordinates $\cX_A$ and $\cX_B$ are independent of which 
member of the continuous family we take, so all of these
$\cW$-abelianizations are actually isomorphic, and in some sense
they should be considered as just \ti{one} abelianization.
Moreover, these spectral coordinates are \ti{Fock-Goncharov coordinates} associated
to an ideal triangulation of $\CP^1$.\footnote{The triangulation is
made up of $2$ triangles, whose interiors are $\{\abs{z} < 1\}$
and $\{\abs{z} > 1\}$.}
Indeed, let $a_1$ be an eigenvector for $\bA$, and $a_2$
another vector such that $\IP{a_1,a_2}$ is the unique plane
preserved by $\bA$; likewise define $b_1, b_2$ and $c_1, c_2$,
and $d_1, d_2$ associated to the operator $\bD = \bC^{-1} \bB \bC$.
Then the \ti{triple ratio} and \ti{edge coordinate} from \cite{MR2233852} are
\begin{equation}
	t = \frac{[a_1 , a_2 , b_1] [b_1 , b_2 , c_1] [c_1 , c_2 , a_1]}{[a_1 , a_2 , c_1] [b_1 , b_2 , a_1] [c_1 , c_2 , b_1]}, \qquad
	e = \frac{[b_1 , c_1 , a_1][d_1 , a_2 , a_1]}{[a_2 , c_1 , a_1][b_1 , d_1 , a_1]}.
\end{equation}
These coordinates turn out to be related to the spectral coordinates for the
continuous family of $\cW$-abelianizations, by
\begin{equation}
	\cX_A = e, \qquad \cX_B = \frac{t}{\sqrt{e}}.
\end{equation}

It is not clear to us why the Fock-Goncharov coordinates appear
as spectral coordinates for the Stokes graph $\cW$. 
In \cite{Gaiotto:2012db} it was shown that Fock-Goncharov 
coordinates do
appear as spectral coordinates for a specific sort of 
spectral network associated to a triangulation, but that is a different
spectral network from $\cW$. It would be interesting to understand
this better.

At any rate, the Fock-Goncharov coordinates will not play much role
in the rest of the paper; most of our attention will be focused 
instead on the $4$ discrete abelianizations, since these are the ones 
which turn out to be directly related to WKB
for the $T_3$ equation.

\subsection{The monodromy matrices}

Relative to the projective basis $(\psi_1, \psi_2, \psi_3)$
we can write the monodromy explicitly. Its form depends
on which $\cW$-abelianization we take.
For the $4$ discrete $\cW$-abelianizations, it is (up to a diagonal
gauge transformation)
\begin{subequations} \label{eq:t3-monodromies}
\begin{align}
 \bA &= \begin{pmatrix} -f(\cX_A) \cX_A & 0 & \cX_A \cX_B^{-1} \sqrt{1 + f(\cX_A)^2 \cX_A} \\ (1 + f(\cX_A)^2 \cX_A) \cX_B & f(\cX_A) & -f(\cX_A) \cX_A \sqrt{1 + f(\cX_A)^2 \cX_A} \\ f(\cX_A) \cX_B \sqrt{1 + f(\cX_A)^2 \cX_A} & \cX_A^{-1} \sqrt{1 + f(\cX_A)^2 \cX_A} & - f(\cX_A)^2 \cX_A \end{pmatrix}, \\
 \bB &= \begin{pmatrix} -f(\cX_A)^2 \cX_A & f(\cX_A) \sqrt{1 + f(\cX_A)^2 \cX_A} & \cX_B^{-1} \sqrt{1 + f(\cX_A)^2 \cX_A} \\ \cX_A \sqrt{1 + f(\cX_A)^2 \cX_A} & -f(\cX_A) \cX_A & 0 \\ -f(\cX_A) \cX_B \sqrt{1 + f(\cX_A)^2 \cX_A} & (1 + f(\cX_A)^2 \cX_A) \cX_B \cX_A^{-1} & f(\cX_A) \end{pmatrix}, \\
 \bC &= \begin{pmatrix} f(\cX_A) & -f(\cX_A)\sqrt{1 + f(\cX_A)^2 \cX_A} & 1 + f(\cX_A)^2 \cX_A \\ \sqrt{1 + f(\cX_A)^2 \cX_A} & - f(\cX_A)^2 \cX_A & f(\cX_A) \cX_A \sqrt{1 + f(\cX_A)^2 \cX_A} \\ 0 & \sqrt{1 + f(\cX_A)^2 \cX_A} & -f(\cX_A) \cX_A \end{pmatrix},
 \end{align}
 where
\begin{equation} \label{eq:t3-f}
	f(\cX_A) = \frac{1 - \cX_A \pm \sqrt{1 - 14 \cX_A + \cX_A^2}}{2 \cX_A}.
\end{equation}
\end{subequations}
The formulas \eqref{eq:t3-monodromies} can be obtained directly by ``nonabelianization:''
we begin with $\nabla^\ab$ and reconstruct $\nabla$ from it, using only the constraint
that the gluing matrices across Stokes curves are of the block form
\eqref{eq:gluing-2-rank3}.

It we choose the $-$ sign in \eqref{eq:t3-f}, then 
$f(\cX_A)$ is regular at $\cX_A = 0$, with an expansion of the form
\begin{equation}
	f(\cX_A) = 3 \cX_A + 12 \cX_A^2 + \cdots
\end{equation}
This expansion played an important role in the analysis of BPS particles
of the Minahan-Nemeschansky $E_6$ theory in \cite{Hollands:2016kgm}; its coefficients
count BPS solitons in the Minahan-Nemeschansky theory coupled to a certain \half-BPS
surface defect.

The $-$ branch of the square root is also the one which appears 
for the $\cW$-abelianization coming from exact WKB:
when we take $u > 0$ and $\hbar \to 0$ with
$\arg \hbar = 0$, the WKB abelianization
has $\cX_A$ exponentially small, and likewise $f(\cX_A)$ exponentially
small.
On the other hand, when $\cX_A$ is not small, there is in general
no canonical choice of branch in \eqref{eq:t3-f}; 
both possibilities are possible. 
This suggests that we should pay attention to the
locus where the branches collide: this occurs when $1 - 14 \cX_A + \cX_A^2 = 0$
ie $\cX_A = 7 \pm 4 \sqrt{3}$. Indeed this locus will turn out
to be important below.

\subsection{Testing the predictions of WKB} \label{sec:t3-numerics}

As we have described, when $u > 0$ and $\hbar > 0$,
we conjecture that the higher-rank exact WKB method
with $\vartheta = 0$ furnishes a
$\cW$-abelianization of the $\SL(3)$-connection
associated to the $T_3$ equation.

In fact we can go a bit further:
since the $T_3$ equation depends only on $u' = u / \hbar^3$,
we could equally well study it by using exact WKB
with $u > 0$ but $\vartheta = \arg \hbar = \frac{2 \pi}{3}$,
or $\vartheta = \arg \hbar = \frac{4 \pi}{3}$.
The corresponding Stokes graphs $\cW^\vartheta$ are not equal to $\cW = \cW^{\vartheta=0}$,
but differ from $\cW$ only by cyclic permutations of the
sheet labels $(123)$. Thus the $\cW^\vartheta$-abelianization provided
by exact WKB can be converted
to a $\cW$-abelianization, by cyclically permuting the projective basis 
$(\psi_1, \psi_2, \psi_3)$. In this way exact WKB should produce two additional
$\cW$-abelianizations $A_\vartheta$.

Altogether then, we expect that for $u' \gg 0$,
among the $\cW$-abelianizations of the $T_3$ equation we should
find three coming from exact WKB, $A_\vartheta$
($\vartheta = 0, \frac{2\pi}{3}, \frac{4\pi}{3}$). The spectral coordinates
associated to these three $\cW$-abelianizations should have the 
small-$\hbar$ asymptotic behavior
\begin{equation}
	\cX_{\gamma} \approx \exp\left(Z_{\gamma}(u) / \hbar\right), \qquad \arg \hbar = \vartheta.
\end{equation}
In fact, these asymptotics should hold not only for 
$\arg \hbar = \vartheta$ but more generally for $\arg \hbar \in (\vartheta-\frac{\pi}{2}, \vartheta+\frac{\pi}{2})$.

It is convenient to rewrite these asymptotics in terms of the invariant parameter
$u'$, using the explicit formulas for the periods:
\begin{equation} \label{eq:T3-periods}
	Z_{A} = -M u^{\frac13}, \qquad
	Z_{B} = -\E^{-\frac{2 \pi \I}{3}} M u^{\frac13},
\end{equation}
where
\begin{equation}
M = 2^{-\frac23} \pi^{-\frac12} \Gamma\left(\frac13\right) \Gamma\left(\frac16\right) \approx 5.2999\,.
\end{equation} 
Then the prediction is
\begin{equation} \label{eq:t3-concrete-prediction}
	\cX_{A} \approx \exp(-M \E^{- \I \vartheta} u'^{\frac13}), \qquad \cX_{B} \approx \exp(-M \E^{- \I (\vartheta + \frac{2 \pi}{3})} u'^{\frac13}),
\end{equation}
This should hold for $u' \gg 0$ but also more generally when $u'$
is analytically continued; in fact, since changing $\arg \hbar$ by
 $\frac{\pi}{2}$ changes $\arg u'$ by $\frac{3\pi}{2}$, the prediction
 \eqref{eq:t3-concrete-prediction} can be analytically continued
 to give the asymptotics as $u' \to \infty$ along an arbitrary ray.

We can test this prediction experimentally as follows:
\begin{itemize} 
\item Numerically compute the monodromy matrices 
$\bA$, $\bB$, $\bC$ for the $T_3$ equation,
for various values of $u'$.
\item Use the method of \autoref{sec:enumerating-special-bases}
to determine the $\cW$-abelianizations for each $u'$.
\item Use the formulas \eqref{eq:XA-t3}, \eqref{eq:XB-t3} 
to compute the spectral
coordinates $\cX_A$ and $\cX_B$ for each abelianization.
\item Check that $3$ of the $\cW$-abelianizations have
the behavior
\eqref{eq:t3-concrete-prediction}
when $\abs{u'} \to \infty$.
\end{itemize}

Experimentally this indeed works; for a sample of the
numerical evidence, see \autoref{fig:X-numerics}.
\insfigscaled{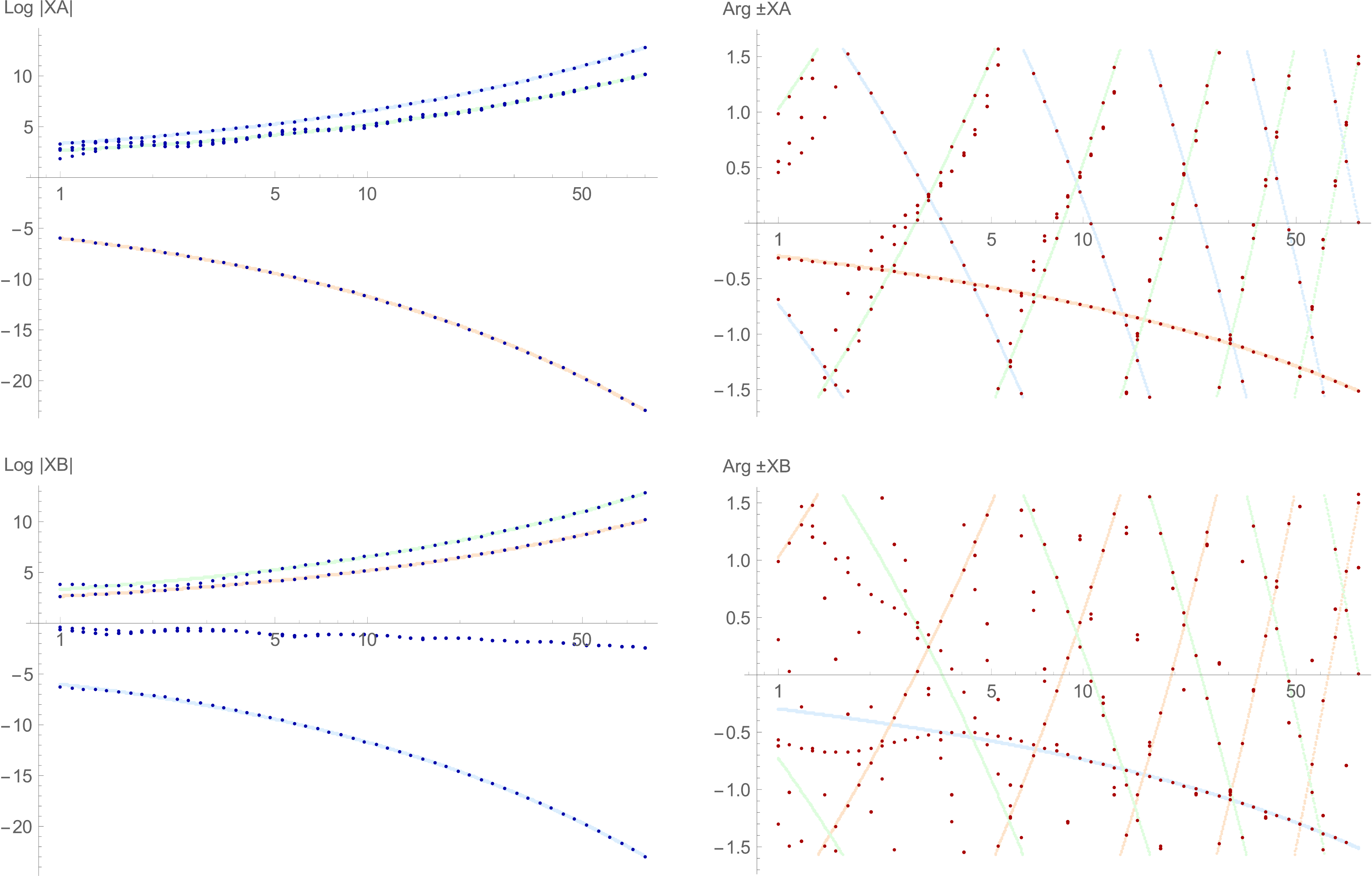}{0.43}{A numerical study of $\cX_A(u')$ and $\cX_B(u')$
for $\arg u' = 0.2$ and 
$1<\abs{u'}<80$. For each value of $u'$, the values of $\cX_A$ and $\cX_B$ 
for all of the $\cW$-abelianizations are plotted. The 3 WKB asymptotic formulas
are also plotted, with $\vartheta = 0$ (orange), $\vartheta = -\frac{2\pi}{3}$ (blue), $\vartheta = \frac{2\pi}{3}$ (green). In each case the curve plotted
is the sum of the first three terms of the WKB
asymptotic series.}

Finally we consider what happens for $-u' \gg 0$.
We can reach this situation by taking $u > 0$ and $\hbar < 0$.
The resulting Stokes graph $\cW^{\vartheta = \pi}$ 
is identical to $\cW$, except that the sheet labels are reversed.
Because all walls of $\cW$ are double,
the notion of $\cW$-abelianization is actually unaffected by this reversal of the sheet labels;
a $\cW^{\vartheta = \pi}$-abelianization is the same thing as a $\cW$-abelianization.
Then, in parallel to $u' \gg 0$, exact WKB at the three phases 
$\vartheta = \arg \hbar = \pi, \frac{5\pi}{3}, \frac{\pi}{3}$
gives three $\cW$-abelianizations $A_\vartheta$ of the $T_3$ equation with $-u' \gg 0$.

\subsection{Analytic continuation} \label{sec:t3-analytic-cont}

Now let us consider the analytic continuation of the
spectral coordinates $\cX_\gamma$ in $u$ and $\hbar$.
The $\cX_\gamma$ are really defined on the 
$4$-fold cover given by the discrete $\cW$-abelianizations; 
thus studying their monodromy is equivalent to studying 
the monodromy of the $\cW$-abelianizations. Since the
$T_3$ equation depends only on $u' = u / \hbar^3$ this 
reduces to working out the monodromy in the $u'$-plane.
We have not found an analytic way of computing this monodromy, but
we have studied it numerically, by tracking the spectral coordinates
$\cX_A$ and $\cX_B$ directly as functions of $u'$.

Let us begin with large $\abs{u'}$.
As we have discussed above, at either $u' \gg 0$
or $u' \ll 0$ we have three $\cW$-abelianizations
$A_\vartheta$ coming from WKB.
As we continue counterclockwise from one side to the other,
these three $\cW$-abelianizations continue as 
$A_\vartheta \to A_{\vartheta + \frac{\pi}{3}}$;
thus, going counterclockwise around a large circle in the $u'$-plane
induces the order-$3$ 
monodromy $A_\vartheta \to A_{\vartheta + \frac{2 \pi}{3}}$.
The behavior of $\cX_A$ as we go around 
the circle $\abs{u'}=25$ is shown in \autoref{fig:t3-x-u=25}.
\insfigscaled{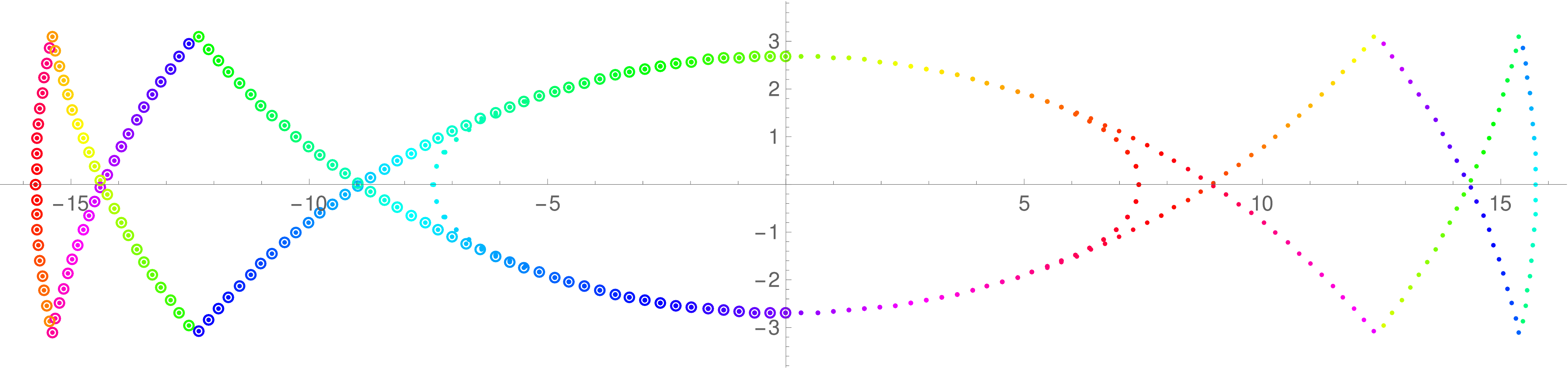}{0.37}{The coordinate $\log \cX_A(u')$, plotted in
$\C / 2 \pi \I \Z$, for $\abs{u'} = 25$.
The hue indicates the phase $\arg u'$. For each value of $u'$, there are
$4$ solid points on the plot, 
representing the values of $\cX_A(u')$ for the $4$ discrete
$\cW$-abelianizations. $3$ of these points lie on a large loop,
while the fourth point lies on a smaller loop; the two loops come
very close to one another. 
As $\arg u'$ advances by $2 \pi$, $\cX_A(u')$ moves 
one-third of the way around the large loop, or all the way around
the small loop. This reflects the
fact that the monodromy permutes $3$ of the discrete 
$\cW$-abelianizations while leaving the 
fourth one invariant. The hollow circles on the plot show the WKB asymptotic
formula for $\cX_A(u')$, analytically continued from $\arg u' = 0$ to 
the region $-\frac{3\pi}{2} < \arg u' < \frac{3\pi}{2}$;
the fact that these points track closely with one of the $4$ $\cW$-abelianizations 
in this range confirms the prediction of WKB.}

Now we can ask what happens in the interior of the $u'$-plane.
By numerical exploration
we found monodromy around just two points,
located at $u' = \pm u'_*$, where $u'_* \approx 0.041992794$.
Coming in from $u' \gg 0$, we find that the two
$\cW$-abelianizations which we called $A_{\frac{2\pi}{3}}$
and $A_{\frac{4\pi}{3}}$ above
collide at $u' = u'_*$. When they collide they have
$\cX_A = 7 + 4 \sqrt{3}$ and $\abs{\cX_B}^{-2} = \cX_A$. 
Traveling around a small 
loop around $u'_*$, these two $\cW$-abelianizations are
exchanged.
Similarly, coming in from $u' \ll 0$,
we find that the two $\cW$-abelianizations we called
$A_{\frac{\pi}{3}}$ and $A_{\frac{5\pi}{3}}$ are
exchanged around
$u' = -u'_*$, with $\cX_A = 7 - 4 \sqrt{3}$
there.

By numerical experimentation we have not 
found monodromy anywhere else in the 
$u'$-plane. Thus we conjecture that the only monodromy
is around $\pm u'_*$.
It is straightforward to check that this 
gives a consistent global picture: the order-$3$ 
monodromy we found at large $\abs{u'}$
can be factorized into the two order-$2$
monodromies around $\pm u'_*$.

It is interesting to compare the monodromy of the $\cX_\gamma$ with that
of the periods $Z_\gamma$. At large $\abs{u'}$ the two monodromies
agree. At small $\abs{u'}$ the $Z_\gamma$ have a single singularity
at $u' = 0$, while the $\cX_\gamma$ have two singularities at $\pm u'_*$.
Since $\cX_\gamma \sim \exp (Z_\gamma / \hbar)$ one might wonder whether
one can globally take logs, to obtain a deformation
$\widetilde Z_\gamma = \hbar \log \cX_\gamma(\hbar)$.
Were this possible, we would just have two holomorphic 
functions $\widetilde Z_\gamma$
in the $u'$-plane, transforming linearly under monodromy around the
two points $\pm u'_*$.
Then it would be tempting to try to realize
the $\widetilde Z_\gamma$ directly as periods of a globally defined $1$-form 
on a family of deformed spectral
curves. The real situation is more delicate, because the analytically
continued functions $\cX_\gamma$ may have zeroes or poles at some 
values of $u'$; upon analytic continuation around such a $u'$,
$\widetilde Z_\gamma$ has an additive shift by $\pm 2 \pi \I \hbar$.
To see examples of this kind of singularity concretely, we plot
the spectral coordinates for all abelianizations on the line $u'>0$:
see \autoref{fig:t3-x-ureal}.
\insfigscaled{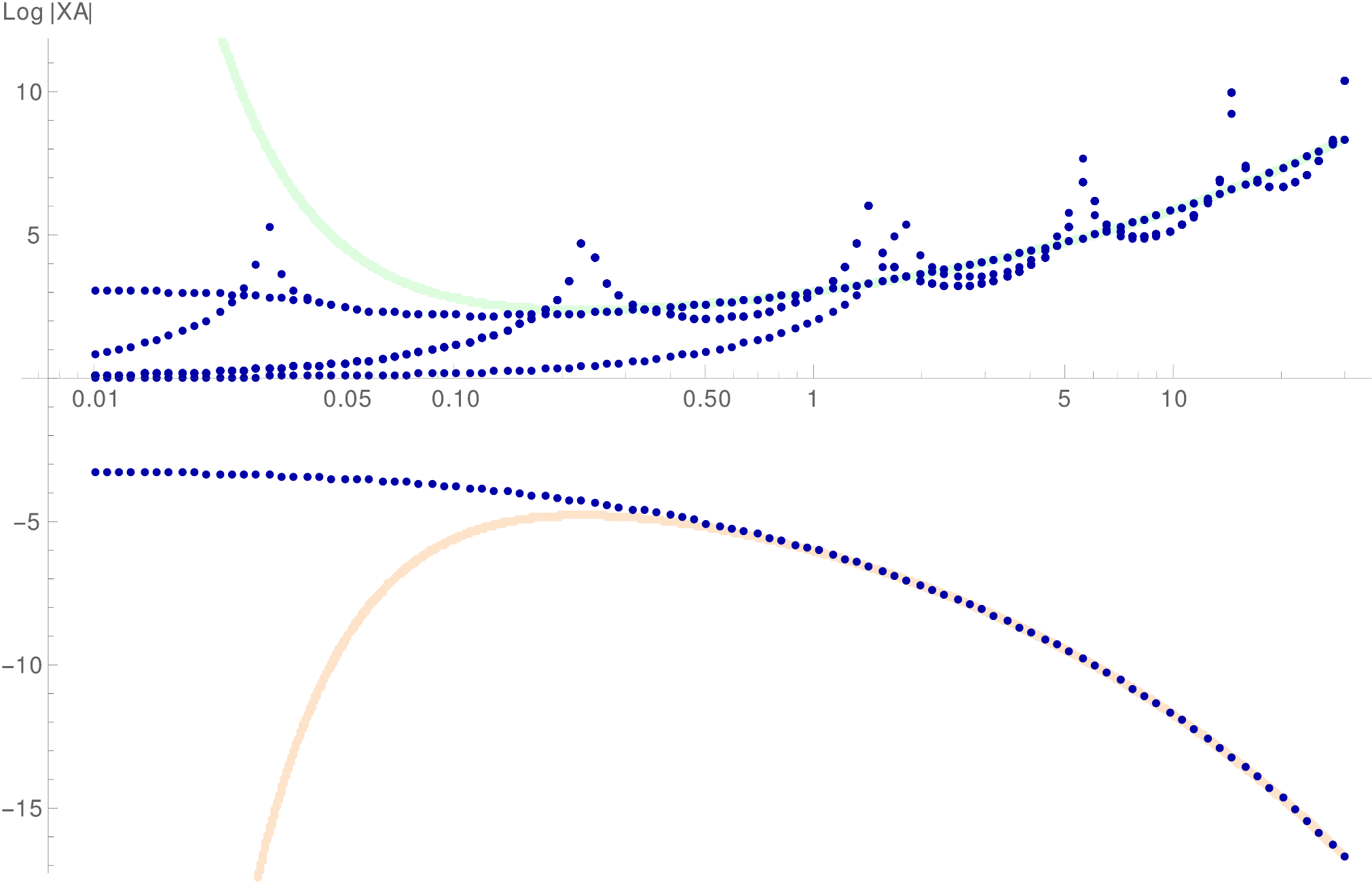}{0.5}{A numerical study of the spectral coordinate $\cX_A(u')$
for $0.01 < u' < 30$. Notation is as in \autoref{fig:X-numerics}.}
Along the ray $u' > 0$ there appear to be infinitely many 
such singularities, with the first few at
$u' \approx 0.03013837, 0.23370955, 1.75819973, \dots$.
Similarly along the ray $u' \in \I \R_+$ there are
singularities which occur at 
$u' \approx 0.4595\I, \dots$

So far we have been discussing the $\cW$-abelianizations which occur discretely.
For the $\cW$-abelianization which occurs in a continuous family,
the situation is simpler: there is no monodromy mixing
it with the other $\cW$-abelianizations. This matches with the fact
from \autoref{sec:t3-fg-coords}
that the corresponding spectral coordinates are the Fock-Goncharov
coordinates, which are uniquely determined by the connection $\nabla$
as long as each of $\bA$, $\bB$, $\bC$ preserves a unique flag.

\subsection{The uniformization point}

It is also interesting to ask what happens at $u' = 0$.
This point is a singularity for the periods $Z_\gamma$, but 
the $T_3$ equation
at $u' = 0$ is perfectly regular. Indeed, its
monodromy representation can be described
explicitly, because it
has a simple interpretation: it is the image of the uniformization
representation of the 3-punctured sphere, 
$\pi_1(C) \to \Gamma_0(2) \subset \SL(2,\Z)$, under the 
symmetric square $\Sym^2: \SL(2,\Z) \to \SL(3,\Z)$.
Thus it can be represented explicitly by the matrices
\begin{align}
\bA &= \Sym^2 \begin{pmatrix} 1 & 2 \\ 0 & 1 \end{pmatrix} = \begin{pmatrix} 1 & 2 & 4 \\ 0 & 1 & 4 \\ 0 & 0 & 1 \end{pmatrix}, \\ \qquad \bB &= \Sym^2 \begin{pmatrix} 1 & 0 \\ -2 & 1 \end{pmatrix} = \begin{pmatrix} 1 & 0 & 0 \\ -4 & 1 & 0 \\ 4 & -2 & 1 \end{pmatrix} , \\ \qquad \bC &= (\bA \bB)^{-1} = \begin{pmatrix} 1 & -2 & 4 \\ 4 & -7 & 12 \\ 4 & -6 & 9 \end{pmatrix}.
\end{align}
The integrality of these matrices implies that the spectral coordinates
$\cX_A$, $\cX_B$ are algebraic.
Indeed, we have computed them explicitly:
at $u' = 0$ two of the discrete $\cW$-abelianizations have
\begin{equation}
	(\cX_A, \cX_B) = \left( \frac15 (59 \pm 24\sqrt{6}), \sqrt{ \frac15 (59 \mp 24\sqrt{6}) } \right)
\end{equation}
and the other two have the coincident value
\begin{equation}
	(\cX_A, \cX_B) = \left( -1, 1 \right),
\end{equation}
while the continuous family of $\cW$-abelianizations have
\begin{equation}
	(\cX_A, \cX_B) = \left( 1, 1 \right).
\end{equation}
If we approach $u' = 0$ starting from $u' \gg 0$,
the WKB abelianization $A_0$ smoothly approaches
the one with
$\cX_A = \frac15(59 - 24 \sqrt{6}) \approx 0.0424492$
(see the bottom curve in \autoref{fig:t3-x-ureal}).

\subsection{Integral equations} \label{sec:integral-equations-t3}

Now let us consider the construction of an integral equation
\eqref{eq:rh-integral-equation} obeyed by
the spectral coordinates, following
the scheme of \autoref{sec:integral-equations}.
For concreteness, we fix
$u > 0$ (it is easy to restore more general $u$ dependence if needed.)

In the scheme of
\autoref{sec:integral-equations} we have to choose
a function $\vartheta(\arg \hbar)$. It would be inconvenient in this
example to choose $\vartheta(\arg \hbar) = \arg \hbar$; the results of 
\cite{Hollands:2016kgm} imply that there
are infinitely many active rays, and indeed the active rays are everywhere dense.
We pick instead
\begin{equation}
\vartheta(\arg \hbar) = n \frac{\pi}{3} \qquad \text{ for } \qquad \arg \hbar \in \left(n \frac{\pi}{3}-\frac{\pi}{6}, n \frac{\pi}{3}+\frac{\pi}{6} \right).
\end{equation}
This choice has the effect of collapsing the infinitely many active rays 
down to $6$ rays $r_n$ with phases $\frac{\pi}{6} + n \frac{\pi}{3}$. 
To write the integral equation \eqref{eq:rh-integral-equation} we need to
determine the functions $F_{r_n,\gamma}$ attached to those $6$ rays.
According to \eqref{eq:f-discontinuity}, this amounts to determining
the coordinate transformation which relates the spectral coordinates
$\cX^{\vartheta = n \frac{\pi}{3}}$ to the $\cX^{\vartheta = (n+1) \frac{\pi}{3}}$.

To be concrete let us focus on the ray $r_0$, with phase $\frac{\pi}{6}$; 
the others are essentially the same.
The functions $x_\gamma = \cX_\gamma^{\vartheta = 0}(\hbar)$ for $\arg \hbar = 0$,
and the functions $y_\gamma = \cX_\gamma^{\vartheta = \frac{\pi}{3}}(\hbar)$ for
$\arg \hbar = \frac{\pi}{3}$, are associated to a $\cW$-abelianization
and a $\cW^{\vartheta = \frac{\pi}{3}}$-abelianization respectively.
We analytically continue $x$ and $y$ to a common sector
$\arg \hbar \in (-\eps, \frac{\pi}{3} + \eps)$, which in particular contains
$r_0$.
In this sector $x_\gamma$ and $y_\gamma$ have the same
asymptotics as $\hbar \to 0$, 
but they are not the same; the ``nonperturbative'' difference between them,
$F_{r_0,\gamma} = y_\gamma / x_\gamma$, is what we are after.

We can describe this difference a bit more concretely.
Just as in \autoref{sec:t3-numerics}, note that a
$\cW^{\vartheta = \frac{\pi}{3}}$-abelianization also induces 
a $\cW$-abelianization.
In fact, the $x_\gamma$ are the spectral coordinates for the
$\cW$-abelianization $A_0$,
while the $y_\gamma$ are obtained by applying a cyclic
permutation of the basis cycles, $(A,B,-A-B) \to (B,-A-B,A)$,
to the spectral coordinates
for the $\cW$-abelianization $A_{-\frac{2\pi}{3}}$.
(For example, $x_B$ is given by the points along the orange curve in the lower
part of \autoref{fig:X-numerics}, while $y_B$ is given by the points along the
green curve in the upper part of that figure.)

We do not have a closed formula for the coordinate transformation $y = \bS_{0,\frac{\pi}{3}}(x)$
giving the $y_\gamma$ as a function of the $x_\gamma$.
However, we do have some partial information.
As we vary $\vartheta$ from $0$ to $\frac{\pi}{3}$, 
the $\vartheta$-Stokes graph jumps at a countable dense set of phases, 
and correspondingly $\bS_{0,\frac{\pi}{3}}$ admits a factorization into 
a countable product of Stokes automorphisms, of the form \cite{ks1,Gaiotto:2008cd,Gaiotto2012}
\begin{equation} \label{eq:s-factorization}
	\bS_{0,\frac{\pi}{3}} = \bT_{\frac{\pi}{3}}^{\frac12} \circ \left( \prod^\cwarrow_{\vartheta \in (0,\frac{\pi}{3})} \bT_\vartheta \right) \circ \bT_0^{\frac12}.
\end{equation}
In \eqref{eq:s-factorization} the product over $\vartheta$ is taken in decreasing order, $\bT_\vartheta$ is a coordinate transformation of the form
\begin{equation} \label{eq:ks-trans}
	\bT_\vartheta = \prod_{\gamma: \arg(-Z_\gamma) = \vartheta} \cK_\gamma^{\Omega(\gamma)}, \quad
	\cK_\gamma^*\cX_\mu = \cX_\mu (1 - \sigma(\gamma) \cX_\gamma)^{\IP{\mu,\gamma}},
\end{equation}
$\sigma: H_1(\Sigma,\Z) \to \{\pm 1\}$ is
\begin{equation}
	\sigma(a \gamma_A + b \gamma_B) = (-1)^{a + b + ab},
\end{equation}
and most crucially, there appear some integers $\Omega(\gamma) \in \Z$, determined
by the jumping of the Stokes graphs. In the relation
to $\cN=2$ supersymmetric field theory, $\Omega(\gamma)$ is a helicity 
supertrace counting BPS particles with charge $\gamma$.
Note that $\bT_\vartheta = 1$ except for countably many phases $\vartheta$,\footnote{The 
results of \cite{Hollands:2016kgm} show that $\Omega(\gamma) \neq 0$ for every primitive charge
$\gamma$, so all of the countably many 
phases $\vartheta$ which could give nontrivial $\bT_\vartheta$ indeed do.} and 
for each such phase $\bT_\vartheta$ is a countable product,
so altogether the product in \eqref{eq:s-factorization} involves
a countably infinite number of $\cK_\gamma$.

The effect of the transformation $\cK_\gamma$ is to multiply each $\cX_\mu$ by some power of
$(1 \pm \cX_\gamma)$. For the $\cK_\gamma$ which contribute 
to $\bS_{0,\frac{\pi}{3}}$ we have $\arg(-Z_\gamma) \in [0,\frac{\pi}{3}]$.
When $\arg \hbar \in (-\eps, \frac{\pi}{3} + \eps)$, these $\cX_\gamma$ are exponentially
suppressed like $\exp(Z_\gamma / \hbar)$ 
as $\hbar \to 0$, and thus $\cK_\gamma$ acts by an exponentially
small transformation on the coordinates.
In particular, if $\arg \hbar = \frac{\pi}{6}$, 
then for $\gamma = (a,b) = a \gamma_A + b \gamma_B$, 
$\re(-Z_\gamma / \hbar)$ is proportional to $a-b$; all $(a,b)$
which contribute have $a-b > 0$, and of those 
the least suppressed $\cK_{a,b}$ are the ones
with $a-b = 1$, next are the ones with $a-b = 2$, and so on.

We do not know all of the $\Omega(\gamma)$, but we do know some of them, by the results of
\cite{Hollands:2016kgm}; in particular we know all of the $\Omega(a,b)$
with $a-b \le 3$; see \autoref{fig:t3-bps-counts}.
\insfigscaled{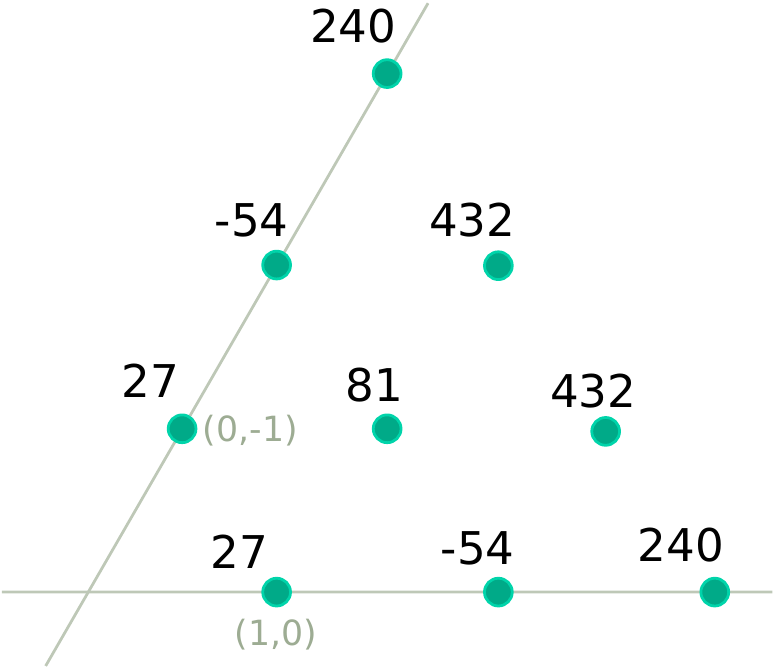}{0.55}{Some degeneracies of BPS particles 
in the $T_3$ theory with $u > 0$. Each green dot represents a charge
$\gamma = (a,b) = a \gamma_A + b \gamma_B$, and is plotted at the point $-Z_\gamma \in \C$,
and decorated by the BPS count $\Omega(\gamma) \in \Z$.
The charges shown are the ones with $\arg (-Z_\gamma) \in [0, \frac{\pi}{3}]$, and with the smallest values of $\re(-Z_\gamma / \hbar)$ when $\arg \hbar \approx \frac{\pi}{6}$.}
Thus we can try approximating $\bS_{0,\frac{\pi}{3}}$ by
just the contributions from these least-suppressed $\Omega(a,b)$;
this gives a sequence of approximations,
\begin{align}
	\bS_{0,\frac{\pi}{3}}^{(1)} &= \cK_{0,-1}^{\frac12 27} \cK_{1,0}^{\frac12 27} ,\\
	\bS_{0,\frac{\pi}{3}}^{(2)} &= \cK_{0,-1}^{\frac12 27} \cK_{0,-2}^{-\frac12 54} \cK_{1,-1}^{81} \cK_{1,0}^{\frac12 27} \cK_{2,0}^{-\frac12 54}, \\
	\bS_{0,\frac{\pi}{3}}^{(3)} &= \cK_{0,-1}^{\frac12 27} \cK_{0,-2}^{-\frac12 54} \cK_{0,-3}^{\frac12 240} \cK_{1,-2}^{432} \cK_{1,-1}^{81} \cK_{2,-1}^{432} \cK_{1,0}^{\frac12 27} \cK_{2,0}^{-\frac12 54} \cK_{3,0}^{\frac12 240},
\end{align}
and so on.
To write the next approximation $\bS_{0,\frac{\pi}{3}}^{(4)}$ 
would require us to know the BPS count $\Omega(3,-1)$, which was 
not computed in \cite{Hollands:2016kgm}, so for now we stop here.

We have tested these approximations numerically; for example,
at $\hbar = \E^{\frac{\pi \I}{6}}$ and $u=1$,
we find:
\sisetup{retain-zero-exponent = true}
\sisetup{group-digits = false}
\begin{center}
\begin{tabular}{|c|c|c|} \hline
    & $A$ & $B$ \\ \hline
$x$                              & $\num{-3.81327+4.08339i E-3}$ & $\num{-1.207491+1.440995i E2} $\\
$\bS_{0,\frac{\pi}{3}}^{(1)} x$  & $\num{-3.40103+4.07226i E-3}$ & $\num{-1.220866+1.303183i E2} $ \\
$\bS_{0,\frac{\pi}{3}}^{(2)} x$  & $\num{-3.41706+4.07711i E-3}$ & $\num{-1.221625+1.308395i E2} $ \\
$\bS_{0,\frac{\pi}{3}}^{(3)} x$  & $\num{-3.41628+4.07696i E-3}$ & $\num{-1.221619+1.308141i E2} $ \\ \hline
$y$                              & $\num{-3.41630+4.07694i E-3}$ & $\num{-1.221611+1.308147i E2} $ \\ \hline
\end{tabular}
\end{center}
As expected, the $\bS_{0,\frac{\pi}{3}}^{(k)}x$ are converging
to $y$ as $k$ increases. Also as expected, the speed of convergence increases
as we increase $\abs{u}$; for example, at $\hbar = \E^{\frac{\pi \I}{6}}$ and $u=10$, we find:
\begin{center}
\begin{tabular}{|c|c|c|} \hline
    & $A$ & $B$ \\ \hline
$x$                              & $\num{2.86472-2.57616i E-5}$ & $\num{1.929843-1.734237i E4}$\\
$\bS_{0,\frac{\pi}{3}}^{(1)} x$  & $\num{2.86673-2.57616i E-5}$ & $\num{1.929986-1.735579i E4}$ \\
$\bS_{0,\frac{\pi}{3}}^{(2)} x$  & $\num{2.86673-2.57616i E-5}$ & $\num{1.929986-1.735579i E4}$ \\
$\bS_{0,\frac{\pi}{3}}^{(3)} x$  & $\num{2.86673-2.57616i E-5}$ & $\num{1.929986-1.735579i E4}$ \\ \hline
$y$                              & $\num{2.86673-2.57616i E-5}$ & $\num{1.929986-1.735579i E4}$ \\ \hline
\end{tabular}
\end{center}
We regard these results as strong evidence for the consistency of the whole story.

We could also run this program in reverse: since we can
compute $x$ and $y$ numerically for any given $u$ and in particular for
large $\abs{u}$,
we could try to determine the BPS counts $\Omega(\gamma)$ from the condition
that $\bS_{0,\frac{\pi}{3}} x = y$. It is easy in this way to ``discover'' the fact that
$\Omega(1,0) = \Omega(0,-1) = 27$, and in principle one could iteratively 
determine the higher $\Omega(a,b)$ by the same strategy. As $a-b$ increases, 
so does the needed precision in the numerical computations of $x$ and $y$.

With our confidence thus bolstered, we tried writing down 
approximate versions of the desired integral equation 
\eqref{eq:rh-integral-equation}, taking $\vartheta(\arg \hbar) = \arg \hbar$,
but truncating as follows:
we fix some $k$, and then include only the $\Omega(a,b)$
shown in \autoref{fig:t3-bps-counts} with $a-b \le k$, 
together with their images under the obvious $\Z_6$ symmetry.
It is not clear \ti{a priori} whether the resulting approximate
equations have any right to work; nevertheless,
we tried solving them numerically anyway, with the following results:
\begin{center} $u=5$: \qquad
\begin{tabular}{|c|S[table-format = 2.5e+1, table-number-alignment=center]|c|} \hline
    & $A$ & $B$ \\ \hline
$\cX^{(0)}$  & 11.59062E-5 & $\num{0.005041+0.92884i E2}$ \\
$\cX^{(1)}$  & 8.42628E-5 & $\num{0.308629+1.04475i E2}$ \\
$\cX^{(2)}$  & 8.00913E-5 & $\num{0.362395+1.05700i E2}$ \\
$\cX^{(3)}$  & 7.87397E-5 & $\num{0.380959+1.06060i E2}$ \\ \hline
$\cX$        & 7.77949E-5 & $\num{0.394281+1.06300i E2}$ \\ \hline
\end{tabular}
\end{center}

\begin{center} $u=1$: \qquad
\begin{tabular}{|c|c|c|} \hline
    & $A$ & $B$ \\ \hline
$\cX^{(0)}$  & $\num{4.99201E-3}$ & $\num{0.17298+1.40473i E1}$ \\
$\cX^{(1)}$  & $\num{2.93480E-3}$ & $\num{1.07747+1.49881i E1}$ \\
$\cX^{(2)}$  & $\num{2.68345E-3}$ & $\num{1.23050+1.48742i E1}$ \\
$\cX^{(3)}$  & $\num{2.60470E-3}$ & $\num{1.28429+1.47980i E1}$ \\ \hline
$\cX$        & $\num{2.55054E-3}$ & $\num{1.32318+1.47307i E1}$ \\ \hline
\end{tabular}
\end{center}
\begin{center} $u=0.01$: \qquad
\begin{tabular}{|c|S[table-format = 2.5e+1, table-number-alignment=center]|c|} \hline
    & $A$ & $B$ \\ \hline
$\cX^{(0)}$  & 31.92335E-2 & $\num{0.97281+1.47856i E0}$ \\
$\cX^{(1)}$  &  7.03803E-2 & $\num{3.62200+1.04386i E0}$ \\
$\cX^{(2)}$  &  5.00813E-2 & $\num{4.35361+1.00679i E0}$ \\ 
$\cX^{(3)}$  &  4.38490E-2 & $\num{4.67643+0.96774i E0}$ \\ \hline
$\cX$        &  3.91347E-2 & $\num{4.98408+0.84366i E0}$ \\ \hline
\end{tabular}
\end{center}
In each of these tables, $\cX^{(k)}_\gamma$ is the value computed numerically 
from the $k$-th truncated integral equation,
and $\cX_\gamma$ is the value computed numerically from the monodromy of 
the $T_3$ equation.
These results offer some support for the
conjecture that $\lim_{k \to \infty} \cX^{(k)}_\gamma = \cX_\gamma$.

Rather than studying these successive approximations,
what would be really desirable would be to give a closed
formula for $\bS_{0,\frac{\pi}{3}}$; then we could write
down a version of the integral equation \eqref{eq:rh-integral-equation} which
would compute the exact $\cX_\gamma$. This remains as a problem
for the future.

\subsection{Spectral problem}

Finally, we briefly consider a spectral problem for the $T_3$ equation,
analogous to those we considered for the Mathieu equation
in \autoref{sec:mathieu-bound-states} and \autoref{sec:mathieu-quasiperiodic}:
we search for those $u'$ such that the $T_3$ equation
admits a discrete $\cW$-abelianization with 
\begin{equation} \label{eq:t3-spectral-A}
\cX_A = 1.
\end{equation}
We recall that for large $u'$ the asymptotics of $3$ of the $4$ discrete $\cW$-abelianizations
are given by \eqref{eq:t3-concrete-prediction}.
Thus a natural first place to look for solutions of \eqref{eq:t3-spectral-A} 
at large $u'$ is at the $u'$ satisfying
\begin{equation}
	1 = \cX_A \approx \exp(-M u'^{\frac13}),
\end{equation}
where $u'^{\frac13}$ is allowed to be any of the three cube roots.
This leads to potential solutions at
\begin{equation} \label{eq:t3-spectral-A-approximate-solutions}
	u' \approx 8 \pi^3 \I n^3 / M^3 = \pm 1.666221\I, \pm 13.32977\I, \pm 44.9880\I, \pm 106.6381\I, \dots
\end{equation}
By numerical experimentation we find actual solutions at
\begin{equation}
	u' \approx \pm 0.0610186\I, \pm 2.148003\I, \pm 14.24769\I, \pm 46.3655\I, \pm 108.4752\I, \dots
\end{equation}
which asymptotically indeed appear to approach the
values \eqref{eq:t3-spectral-A-approximate-solutions}.

The reader might find our choice of spectral problem a little unmotivated,
since its very formulation involves the spectral coordinates $\cX_A$.
It might be some comfort to know that the solutions of \eqref{eq:t3-spectral-A} 
can be alternatively described as points $u'$ for which
\begin{equation} \label{eq:spectral-problem-t3-1}
	\Tr \bA \bB^{-1} - \Tr \bB \bA^{-1} = \pm 12 \sqrt{3} \I,
\end{equation}
as one sees by substituting \eqref{eq:t3-spectral-A} into 
the monodromy matrices \eqref{eq:t3-monodromies}.
In the parlance of exact WKB, one would say 
\eqref{eq:t3-spectral-A} is the ``exact quantization condition''
for the solutions of \eqref{eq:spectral-problem-t3-1}.
One could also go the other way, starting with one's favorite condition
on the matrices $\bA$, $\bB$, $\bC$ and finding the corresponding
exact quantization condition in terms of the spectral coordinates
$\cX_A$, $\cX_B$; we have not explored in this direction.

\section{Supersymmetric field theory} \label{sec:physics}

In the main part of this paper we have been exploring the 
exact WKB method for certain differential equations
(opers) of order $2$ and $3$.
In this final section we consider the relation of
our constructions to $\cN=2$ supersymmetric quantum field theories
of class $S$ in four spacetime dimensions.
Our discussion here is somewhat open-ended;
we hope to return to these questions in the future.

\subsection{Opers and QFT of class \texorpdfstring{$S$}{S}}

Fixing a Lie algebra $\fg$
and a punctured Riemann surface $C$ with singularity
data at the punctures
determines an $\cN=2$ theory $\fX(\fg,C)$ of class $S$.
It has been known for some time that there is a connection
between the theory $\fX(\fg,C)$ and the space of $\fg$-opers on $C$;
see e.g. \cite{Nekrasov:2009rc,Nekrasov:2010ka,Hollands2017,Jeong2018} for various aspects of this connection.
In this section we describe a slightly different version of the connection.

The Coulomb branch of the theory $\fX(\fg,C)$ is the base $\cB_0(\fg,C)$
of the Hitchin integrable system. The algebra $\cA_0$
of chiral local operators in theory $\fX(\fg,C)$ is canonically
identified with the space of holomorphic functions on $\cB_0(\fg,C)$.
Following \cite{Nekrasov:2009rc}, suppose we deform the theory
by turning on the ``$\half \Omega$-background'' associated
to a rotation in the $x_2$-$x_3$ plane, with parameter $\varepsilon = \hbar$.
This modification deforms $\cA_0$ into a new algebra $\cA_\hbar$,
consisting of supersymmetric local operators inserted at the
origin of the $x_2$-$x_3$ plane, still free to move in the $x_0$ and $x_1$
directions.
$\cA_\hbar$ can be thought of as the algebra of functions on a deformation
$\cB_\hbar(\fg,C)$ of $\cB_0(\fg,C)$.
By studying the Hilbert space of the theory on $S^3$ and using the 
state-operator map, together with known facts about how $S$-duality
acts in the theory reduced on $S^1$,
one can show that the deformed space $\cB_\hbar(\fg,C)$ is canonically 
isomorphic to the space of $\fg$-opers on $C$.\footnote{We thank David Ben-Zvi
for explaining this point to us.}
So, in short, turning on the $\half \Omega$-background deforms
the Coulomb branch into the space of opers.

This deformation might sound a bit trivial since, when considered simply as complex
manifolds, the Coulomb branch and the space of opers are isomorphic; however,
the two spaces come equipped with natural presentations in terms of 
holomorphic functions, which are different in the two cases, 
as we will discuss below.

The three spaces of opers we considered in this paper 
correspond in this way to familiar quantum 
field theories:
\begin{center}
\begin{tabular}{|c|c|c|c|} 
\hline
opers & $\fg$ & $C$ & theory $\fX(\fg,C)$ \\
\hline
cubic potential  \eqref{eq:schrodinger-cubic-potential} & $A_1$ & $\CP^1$, irregular puncture & $(A_1,A_2)$ Argyres-Douglas theory  \\
Mathieu \eqref{eq:mathieu} & $A_1$ & $\CP^1$, $2$ irregular punctures & $\cN=2$ Yang-Mills, $G=\SU(2)$ \\
$T_3$ equation \eqref{eq:t3-equation} & $A_2$ & $\CP^1$, $3$ regular punctures & $E_6$ Minahan-Nemeschansky \\ \hline
\end{tabular}
\end{center}

\subsection{Spectral coordinates as vevs}

The stars of this paper are the spectral coordinate functions $\cX_\gamma(\hbar)$
on $\cB_\hbar(\fg,C)$.  What is their meaning in the theory $\fX(\fg,C)$?

The function 
\begin{equation}
\widetilde Z_{\gamma}(\hbar) = \hbar \log \cX_\gamma(\hbar)
\end{equation}
is a deformation of the function $Z_\gamma$ on $\cB_0(\fg,C)$
(if we momentarily ignore the multivaluedness of the $\log$).
Since $Z_\gamma$ is the vev of the vector multiplet scalar $a_\gamma$,
we suspect that
$\widetilde Z_\gamma(\hbar)$ is likewise the vacuum expectation value
of an operator $\widetilde a_\gamma(\hbar)$. 
The operator $\widetilde a_\gamma(\hbar)$ should be a deformation of $a_\gamma$
which preserves supersymmetry in the $\half \Omega$-background.
Such a deformation might not be simple to construct; nevertheless,
\ti{a posteriori}, the WKB expansion \eqref{eq:X-asymptotics} of 
$\cX_\gamma(\hbar)$ suggests that there is a universal 
$\widetilde a_\gamma(\hbar)$ to all orders in $\hbar$.

What about going beyond series in $\hbar$?
We have seen that the $\cX_\gamma(\hbar)$ can be defined beyond perturbation theory
in various ways, corresponding to the different choices of spectral network.
One particularly interesting nonperturbative definition is the function we called
$\cX_\gamma^\RH(\hbar)$ in \S\ref{sec:integral-equations}, with the canonical
choice \eqref{eq:theta-canonical}.
Thus we conjecture that this canonical choice corresponds to a
canonical nonperturbative definition of $\widetilde a_\gamma(\hbar)$.

This canonical $\widetilde a_\gamma(\hbar)$ must have some
new features compared to $a_\gamma$:
\begin{itemize}
\item $\widetilde a_\gamma(\hbar)$ 
should suffer from a nonperturbative discontinuity as a function 
of $\hbar$ whenever there
exists a BPS state whose central charge is aligned with $\hbar$,
corresponding to the fact that the functions 
$\cX^\RH_\gamma(\hbar)$ jump at the active rays.
We might interpret this as saying that the operators 
$\widetilde a_\gamma(\hbar)$ are defined only in the IR (like the $a_\gamma$),
and the scale below which this IR description is appropriate goes to zero
as $\hbar$ approaches an active ray.

\item
$\widetilde a_\gamma(\hbar)$ should also suffer 
from an additive ambiguity, because $\widetilde Z_\gamma(\hbar)$
has an ambiguity by shifts by $2 \pi \I \hbar$. This ambiguity presumably
comes from the possibility of shifting
by a local operator built from background supergravity fields.
(After dimensional reduction to $\cN=(2,2)$ theory in 
the $x^0$-$x^1$ plane, the rotation in the $x^2$-$x^3$ plane becomes a global symmetry;
then $\hbar$ can be interpreted
as a complex twisted mass for this global symmetry, and the ambiguity we are after would
come from shifting by the scalar in the background vector multiplet.)
\end{itemize}

It would be very interesting to give a direct construction
of the operator $\widetilde a_\gamma(\hbar)$ and to understand
more precisely why it has the above features.

\subsection{Scaling line defects}

Although we do not have a direct construction of 
the operators $\widetilde a_\gamma(\hbar)$ in hand,
we can at least propose a construction which should
yield the operators $\exp(\widetilde a_\gamma(\hbar) / \hbar)$,
as follows.

We recall that in an
$\cN=2$ theory one has families of 
$\half$-BPS line defects $L(\zeta)$ labeled by a parameter
$\zeta \in \C^\times$.
It was argued in \cite{Gaiotto:2010be} 
that in the low-energy limit of the theory
there exist distinguished $\half$-BPS ``IR line defects'' $L_\gamma$.
The vacuum expectation values of these line defects on $\R^3 \times S^1$
are functions $\hat\cX_\gamma(R,\zeta)$ which are 
close analogues of the functions $\cX_\gamma^\RH(\hbar)$;
the precise relation was proposed in \cite{Gaiotto2014},
\begin{equation}
\lim_{R \to 0} \hat\cX_\gamma(R,\zeta = \hbar R) = \cX_\gamma^\RH(\hbar).
\end{equation}
So far this is only a relation on the level of functions; can we promote it to the level of 
operators?

Here is a possible approach.
After the $\Omega$-background deformation in the $x_2$-$x_3$ plane
we expect that, for any $R>0$, 
$L_\gamma$ can be wrapped supersymmetrically 
around the circle $(x_2)^2 + (x_3)^2 = R^2$.\footnote{Here is a heuristic
way to understand why $L_\gamma$ can be wrapped supersymmetrically 
around the circle. Suppose $\hbar$ is real. We imagine lifting
the 4-dimensional theory to a 5-dimensional theory on an $\R^4$ bundle over $S^1$, where
the $S^1$ base has length $\rho$, and the $x_2$-$x_3$ plane in the 
fiber is rotated by an angle $\rho \hbar$ as we 
go around the $S^1$ base. In the limit $\rho \to 0$ this gives rise to an effectively
4-dimensional theory, which can be identified with the $\Omega$-background
deformation of the original theory. On the other hand, 
this 5-dimensional background is locally
Euclidean space, and in the 5-dimensional theory, we can put the
line defect $L_\gamma$ supersymmetrically on any straight line.
We choose a straight line in the $x_4$ direction, beginning
at some point $(x_0,x_1,x_2,x_3,x_4 = 0)$.
After going around the $S^1$ fiber this line will return to $(x_0,x_1,x'_2,x'_3,x_4 = 0)$
where $(x'_2,x'_3)$ is the image of $(x_2,x_3)$ under rotation by an angle $\rho \hbar$. If $\rho \hbar = \frac{2\pi}{N}$, then after going around $N$ times, the line closes up to a loop, which pierces
the $\R^4$ fiber in $N$ points arranged around a circle in the $x_2$-$x_3$ plane. In the limit as $\rho \to 0$
ie $N \to \infty$, these $N$ points just look like a line wrapped around the circle.}
Taking the limit $R \to 0$ then gives a supersymmetric local operator 
placed at the origin of the $x_2$-$x_3$ plane,
which we propose to identify with $\exp (\widetilde a_\gamma(\hbar) / \hbar)$.

To get a different viewpoint on this construction, following 
\cite{Nekrasov:2010ka}, we can deform the $x_2$-$x_3$ plane
to a ``cigar'' metric and then compactify on the radial circle. 
The result is a 3-dimensional theory on a half-space, with a boundary condition
corresponding to the origin of the $x_2$-$x_3$ plane. At low energies the 3-dimensional
theory is described by a sigma model into a moduli space $\cM(\fg,C,\hbar)$ of
flat $\fg$-connections on $C$, and it was proposed in
\cite{Nekrasov:2010ka} that the boundary condition we get corresponds
to a Lagrangian subspace $\cL_\oper \subset \cM(\fg,C,\hbar)$,
whose points are the opers.\footnote{By a change of variable
introduced in \cite{Nekrasov:2010ka}, $\cM(\fg,C,\hbar)$ can be identified with the
moduli space of the theory \ti{without} $\Omega$-background, compactified on a circle
of radius $R = \abs{\hbar}^{-1}$. This moduli space is \hk, with complex structures
labeled by $\zeta \in \CP^1$; the boundary condition we get
preserves the subalgebra labeled by $\zeta = \frac{\hbar}{\abs{\hbar}}$.}
This is consistent with our proposal, as follows.
Wrapping $L_\gamma$ around the compactification circle gives a local
operator $O_\gamma$ in the sigma model. As we approach the boundary 
the radius of the compactification circle shrinks to zero,
so at the boundary our proposal says $O_\gamma$ should become identified with
$\exp (\widetilde a_\gamma(\hbar) / \hbar)$.
This is what the $\cL_\oper$ 
boundary condition enforces: it requires that the
$O_\gamma$ obey the same relations as the
$\exp (\widetilde a_\gamma(\hbar) / \hbar)$.

\subsection{Opers and instanton counting}

Concretely, what are the relations obeyed by the local operators $\widetilde a_\gamma(\hbar)$,
or by their vevs $\widetilde Z_\gamma(\hbar)$?

The functions $Z_\gamma$ on $\cB_0(\fg,C)$ obey well-known relations:
choosing a symplectic basis $\{A_1, \dots, A_r, B^1, \dots, B^r\}$ for 
the charge lattice $\Gamma$, the $Z_B$ are determined by the $Z_A$, via the formula
\begin{equation}
	Z_{B^I} = \partial \cF(Z_{A_1}, \dots, Z_{A_r}) / \partial Z_{A_I},
\end{equation}
for a locally defined holomorphic function $\cF$ called ``prepotential.''
The existence of such an $\cF$ reflects the fact that $Z$ gives
a local Lagrangian embedding of the Coulomb branch $\cB_0$ into the symplectic
vector space $\Gamma^* \otimes \C$.
Physically, $\cF$ gives a Lagrangian description of the 
$\cN=2$ theory on its Coulomb branch.

At $\hbar \neq 0$ there is a very similar picture: any log spectral coordinate system
$\widetilde Z_\gamma$ gives local Darboux coordinates on the moduli space $\cM(\fg,C,\hbar)$,
and the fact that $\cL_\oper$ is a Lagrangian subspace means that there is a locally
defined $\widetilde \cF$ for which $\cL_\oper$ is 
given by the equations\footnote{In conformal theories $\widetilde \cF$ depends only on the 
$\widetilde Z_{A_i}$ and not on $\hbar$. In non-conformal theories there are complex parameters $m_i$
with the dimension of mass, and then $\widetilde \cF$ depends on $\hbar$ through the combinations $m_i / \hbar$.}
\begin{equation}
	\widetilde Z_{B^I} = \partial {\widetilde \cF}(\widetilde Z_{A_1}, \dots, \widetilde Z_{A_r}, \hbar) / \partial \widetilde Z_{A_I}.
\end{equation}
Now it is natural to ask: what 
is the meaning of $\widetilde \cF$ in the language of supersymmetric
field theory?
In \cite{Nekrasov:2011bc} this question was 
considered in the special case where $\fg = A_1$ and the $\widetilde{Z}_\gamma$
are complexified Fenchel-Nielsen coordinates, like those we considered in 
\autoref{sec:spectral-coords-mathieu} above.
In this case (as long as $C$ has only regular punctures), the theory $\fX(\fg,C)$ is a supersymmetric
gauge theory \cite{Gaiotto:2009we}, and so one can formulate the Nekrasov instanton partition function
$\bZ(\varepsilon_1, \varepsilon_2; a)$ \cite{Nekrasov:2002qd,Nekrasov:2003rj}.
The proposal of \cite{Nekrasov:2011bc} is that $\widetilde \cF$ is the $\varepsilon_2 \to 0$
limit of $\bZ$, or more precisely,
\begin{equation} \label{eq:nrs-proposal}
\widetilde \cF \left({\widetilde Z}_A, \hbar = \varepsilon_1 \right) = \frac{1}{\varepsilon_1} \lim_{\varepsilon_2 \to 0} \varepsilon_2 \log \bZ(\varepsilon_1,\varepsilon_2 ; a = \varepsilon_1 {\widetilde Z}_A).
\end{equation}
The formula \eqref{eq:nrs-proposal} is a direct link between two very different-looking
objects: on the LHS the monodromy of $\SL(2)$-opers on the Riemann surface $C$, on the RHS equivariant 
integrals over moduli of instantons in $\R^4$.
It has been extended in \cite{Hollands2017,Jeong2018} to a broader class of Lagrangian field theories
of class $S$; in those cases the LHS involves monodromy of $\SL(N)$-opers on $C$, 
expressed in terms of $\widetilde Z_A$
which are higher-rank analogues of complexified Fenchel-Nielsen coordinates.

It is difficult to check \eqref{eq:nrs-proposal} directly.
Nevertheless, in \cite{Nekrasov:2011bc,Hollands2017} evidence 
for \eqref{eq:nrs-proposal} has been given, and in \cite{Jeong2018} a proof in many cases. 
The strategy is as follows. In Lagrangian field theories of class $S$ one always
has parameters $q_i$ which can be varied: from the field theory point of view these are gauge couplings,
while from the point of view of $C$ they are moduli of the complex structure.
One considers a degeneration limit ``$q_i \to 0$'': in field theory this is a weak-coupling limit, and 
in the complex moduli space of $C$ it is a limit where $C$ maximally degenerates to a chain of three-punctured
spheres. Expanding both sides of \eqref{eq:nrs-proposal} in powers of the 
$q_i$, each term is a well-defined nonperturbative
function of $\varepsilon_1$. Thus the statement \eqref{eq:nrs-proposal} is sensitive to the
precise nonperturbative definition of $\widetilde Z_\gamma$, and 
as is shown in 
\cite{Nekrasov:2011bc,Hollands2017,Jeong2018}, it holds only when one takes the
$\widetilde Z_\gamma$ to be complexified Fenchel-Nielsen coordinates (or their 
higher-rank analogues).

In \autoref{sec:t3} of this paper, we have been exploring a specific coordinate system
$\widetilde Z_\gamma$ which arose naturally from the exact WKB analysis of 
the locus of opers associated to the $\half \Omega$-deformed $E_6$ Minahan-Nemeschansky theory.
One might ask whether some analogue of \eqref{eq:nrs-proposal}
holds in this setting.
To formulate this question sharply would require us 
to understand precisely how to define $\bZ$ in the non-Lagrangian
Minahan-Nemeschansky theory. We suspect that the proper formulation of $\bZ$ in a general
non-Lagrangian field theory requires a choice of boundary condition, and that
there is a natural class of boundary conditions corresponding to the different 
spectral coordinate systems $\widetilde{Z}_\gamma$; thus in a general theory the equality
\eqref{eq:nrs-proposal} could indeed hold, with both sides depending 
on this choice of boundary condition.
We hope to develop this story more fully in the future.

\appendix

\section{Computations of spectral coordinates} \label{app:computations}

In this appendix we give some computations omitted from the main text.

\subsection{Computations for the cubic potential}

{\bf Computation of \eqref{eq:spectral-coords-cubic}.}
We will only describe the computation for $\cX_A$; that for $\cX_B$ is similar.

We need to compute the parallel transport of $\nabla^\ab$ along a path in 
the homology class $\gamma_A$.
To compute concretely
it is convenient to work relative to bases 
of $\nabla^\ab$-flat sections in each domain.
Each local $\nabla^\ab$-flat section
corresponds to a local $\nabla$-flat section, and by 
continuation we can think of all these
local flat sections as lying in a single $2$-dimensional vector space $V$,
the space of global $\nabla$-flat sections over the plane.
See \autoref{fig:cubic-computations}.
\insfigscaled{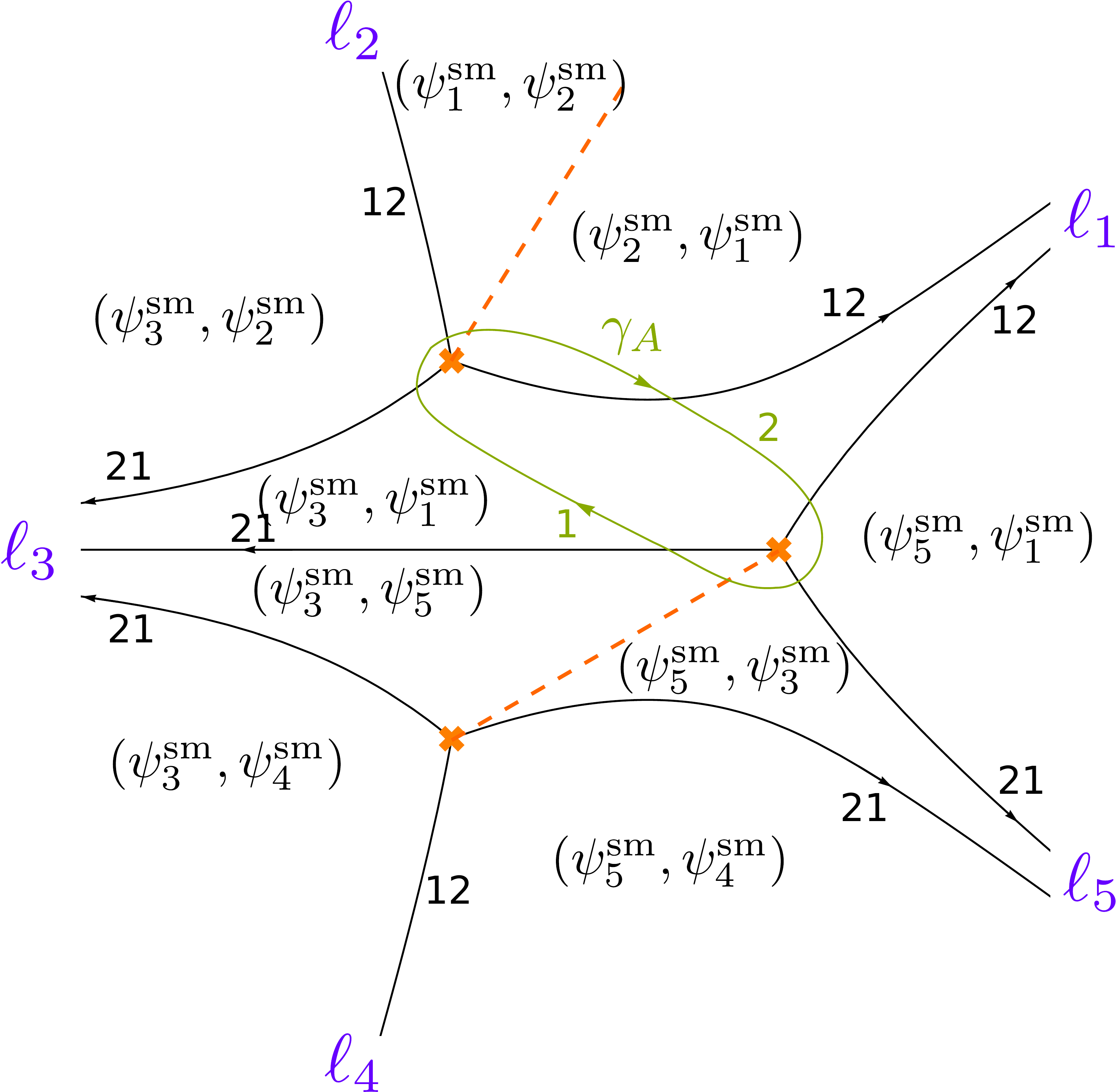}{0.26}{The Stokes graph from \autoref{fig:cubic-network},
with the local WKB bases shown in each domain. To write the basis concretely as an ordered pair of
solutions we have used the trivialization of the double cover $\Sigma$ away from branch cuts; 
thus, in a domain
containing a branch cut, we write two versions of the basis, one on each side of the cut.}

Relative to these local bases, 
the parallel transport within each domain is just represented
by $1$, and the only nontrivial part is the gluing factor
from \eqref{eq:gluing-1}:
\begin{itemize}
\item When we cross a single wall of type $ij$ 
on sheet $i$, from side $L$ to side $R$, we get a factor
\begin{equation}
\frac{[\psi_i^L, \psi_j^L]}{[\psi_i^R, \psi_j^L]}.
\end{equation}

\item When we cross a single wall of type $ij$ on sheet $j$,
we also get a gluing factor, but this
factor is just $1$ if $\psi_j^L = \psi_j^R$, 
which it always is in this example.
\end{itemize}

The representative of $\gamma_A$ shown in \autoref{fig:cubic-computations}
crosses six walls;
multiplying the factors for these six crossings, starting from the eastmost region, gives
\begin{equation}
	\cX_A = \frac{[\psi_5^\smsol, \psi_1^\smsol]}{[\psi_5^\smsol, \psi_3^\smsol]} \times 1 \times 1 \times \frac{[\psi_3^\smsol, \psi_2^\smsol]}{[\psi_1^\smsol, \psi_2^\smsol]} \times 1 \times 1
\end{equation}
matching \eqref{eq:spectral-coords-cubic} as desired.

\subsection{Computations for the Mathieu equation}

{\bf Computation of \eqref{eq:XB-mathieu}.} We need to compute the parallel
transport of $\nabla^\ab$ along a path in the homology class $\gamma_B$.
We use the path given in \autoref{fig:mathieu-sn-1}. 

As above,
it is convenient to work relative to bases 
of $\nabla^\ab$-flat sections in each domain.
See \autoref{fig:computation-1}.
Again by
continuation we think of all these
local flat sections as lying in a single $2$-dimensional vector space $V$.
In this case there is an added technical difficulty: the monodromy around $z = 0$
means there are no global $\nabla$-flat sections. Instead we identify $V$ as
the space of $\nabla$-flat sections on the complement of
the blue dashed line (``monodromy cut'').

\insfigscaled{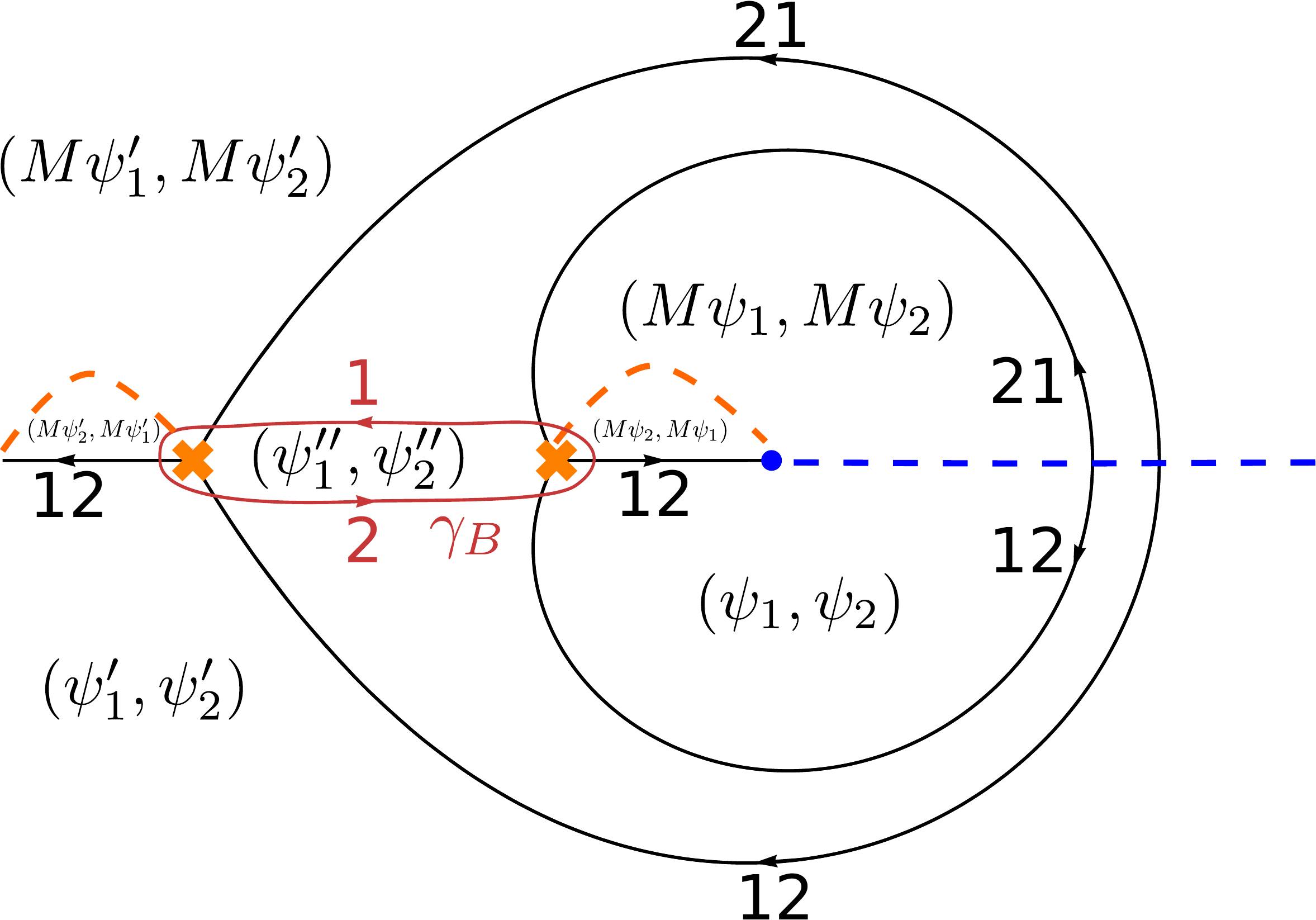}{0.3}{The Stokes graph from \autoref{fig:mathieu-sn-1}, with the
local WKB bases shown in each domain. As before, to 
write the basis concretely as an ordered pair of
solutions we have used the trivialization of the double cover $\Sigma$ away from branch cuts; 
thus, in a domain
containing a branch cut, we write two versions of the basis, one on each side of the cut.
When we cross the monodromy cut, the local WKB basis of $\nabla^\ab$-flat sections 
does not change, but the way we identify them with elements of $V$ does jump, by
the action of the monodromy $M$.
}
Again the only nontrivial part of the parallel transport is the gluing factors
appearing in
\eqref{eq:gluing-1}, \eqref{eq:gluing-2},
When we cross a double wall on sheet $i$, from side $L$ to
side $R$, we get a factor
\begin{equation}
	\sqrt{\frac{[\psi_i^L , \psi_j^L]}{[\psi_i^R , \psi_j^R]} \frac{[\psi_i^L , \psi_j^R]}{[\psi_i^R , \psi_j^L]}},
\end{equation}
and when we cross a single wall of type $ij$ on sheet $i$, from side $L$ to side $R$, 
we get a factor
\begin{equation}
\frac{[\psi_i^L , \psi_j^L]}{[\psi_i^R , \psi_j^L]}.
\end{equation}

We can further simplify these factors by 
choosing bases with $[\psi_1, \psi_2] = 1$, $[\psi'_1 , \psi'_2] = 1$, $[\psi''_1, \psi''_2] = 1$. Then
starting from the southwest corner, 
the gluing factors we encounter are
\begin{equation}
\cX_B = 
\sqrt{\frac{[\psi'_2 , \psi''_1]}{[\psi''_2 , \psi'_1]}} \times
\sqrt{\frac{[\psi''_2 , \psi_1]}{[\psi_2 , \psi''_1]}} \times
   1 \times  
\sqrt{\frac{[M\psi_1 , \psi''_2]}{[\psi''_1 , M \psi_2]}} \times
\sqrt{\frac{[\psi''_1 , M \psi'_2]}{[M \psi'_1 , \psi''_2]}} \times
   1.
\end{equation}
Using $M \psi''_1 = \mu \psi''_1$, $M \psi''_2 = \mu^{-1} \psi_2$, and $M \psi_1 = \psi_2$, $M \psi'_1 = \psi'_2$, 
this reduces to
\begin{equation}
\cX_B = \frac{[\psi_1 , \psi''_2]}{[\psi_1 , \psi''_1]} \frac{[\psi'_1 , \psi''_1]}{[\psi'_1 , \psi''_2]}
\end{equation}
which matches the desired \eqref{eq:XB-mathieu}.

\medskip

{\bf Computation of \eqref{eq:mathieu-coordinates-eg1}.}
Just as above, all we need to compute are the gluing factors
along the paths $\gamma_A$ and $\gamma_B$, with respect
to the bases shown in \autoref{fig:computation-2}.
\insfig{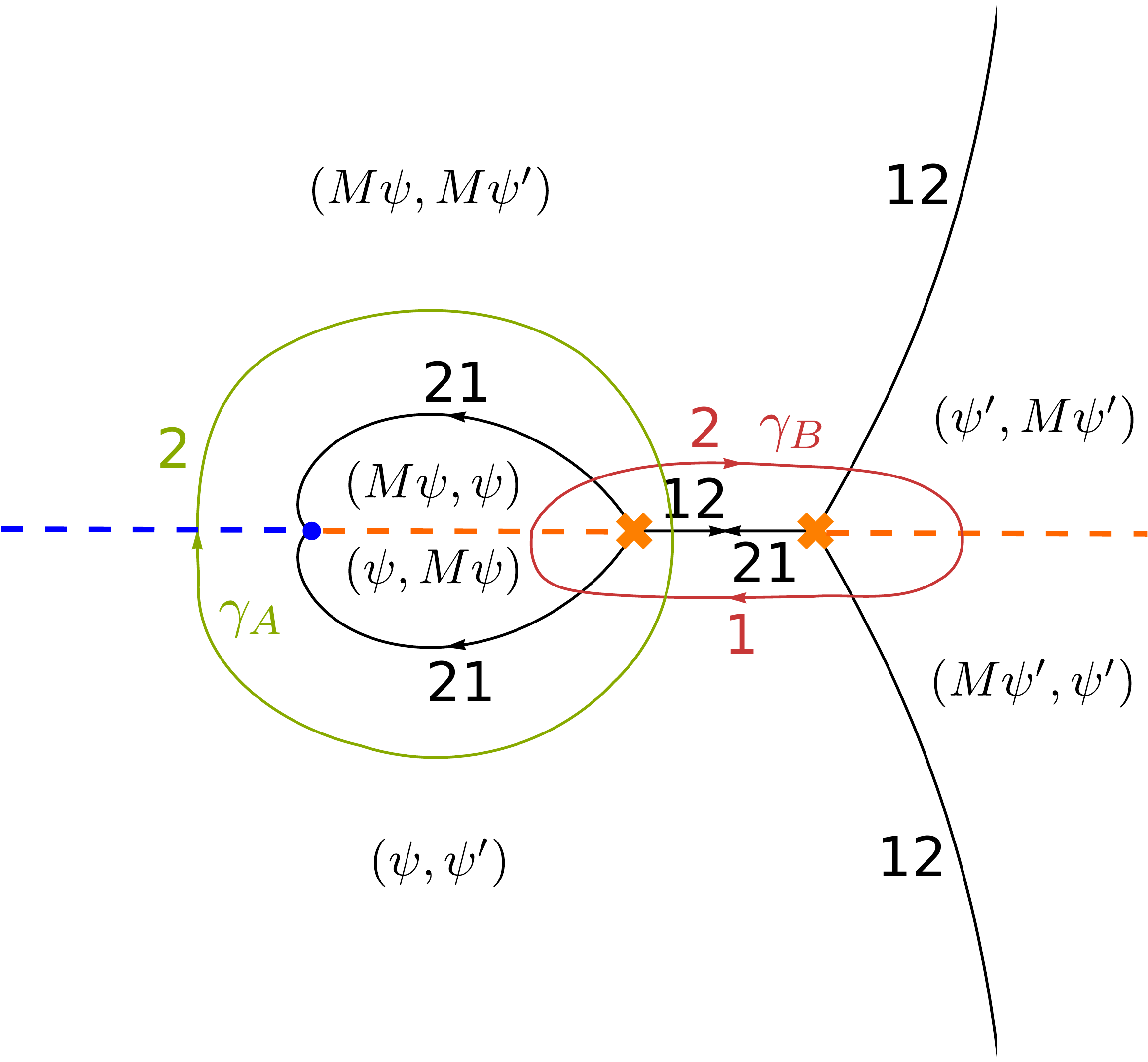}{The Stokes graph from \autoref{fig:mathieu-sn-3}, with local 
WKB bases shown in each domain. All notation is as in \autoref{fig:computation-1} above.}
We can choose $[\psi , \psi'] = 1$ to simplify.
In going around $\gamma_A$ we only meet one wall,
with the gluing factor
\begin{equation} \label{eq:mathieu-xa-app}
	\cX_A = \pm \sqrt\frac{[M \psi' , \psi]}{[\psi' , M \psi]}.
\end{equation}
To fix the branch we would need to carefully implement 
the WKB prescription from \autoref{sec:stokes-gluing}, which we
do not do here.

For $\gamma_B$ the product of gluing factors, starting from
the southeast, is
\begin{equation} \label{eq:mathieu-xb-app}
	\cX_B = \frac{[M \psi' , \psi']}{[\psi , \psi']} \times 1 \times \frac{[M \psi , \psi]}{[M \psi , M \psi']} \times 1 = \frac{[M \psi' , \psi'][M \psi , \psi]}{[\psi , \psi']^2}.
\end{equation}

The results \eqref{eq:mathieu-xa-app}, \eqref{eq:mathieu-xb-app} 
match the desired \eqref{eq:mathieu-coordinates-eg1}.

\subsection{Computations for the \texorpdfstring{$T_3$}{T3} equation}

{\bf Abelianizations and adapted bases.}
Suppose we have a $\cW$-abelianization of the $T_3$
equation. Then we can choose bases compatible with the
$\cW$-abelianization in the various 
domains of \autoref{fig:computations-3}, as shown.

\insfig{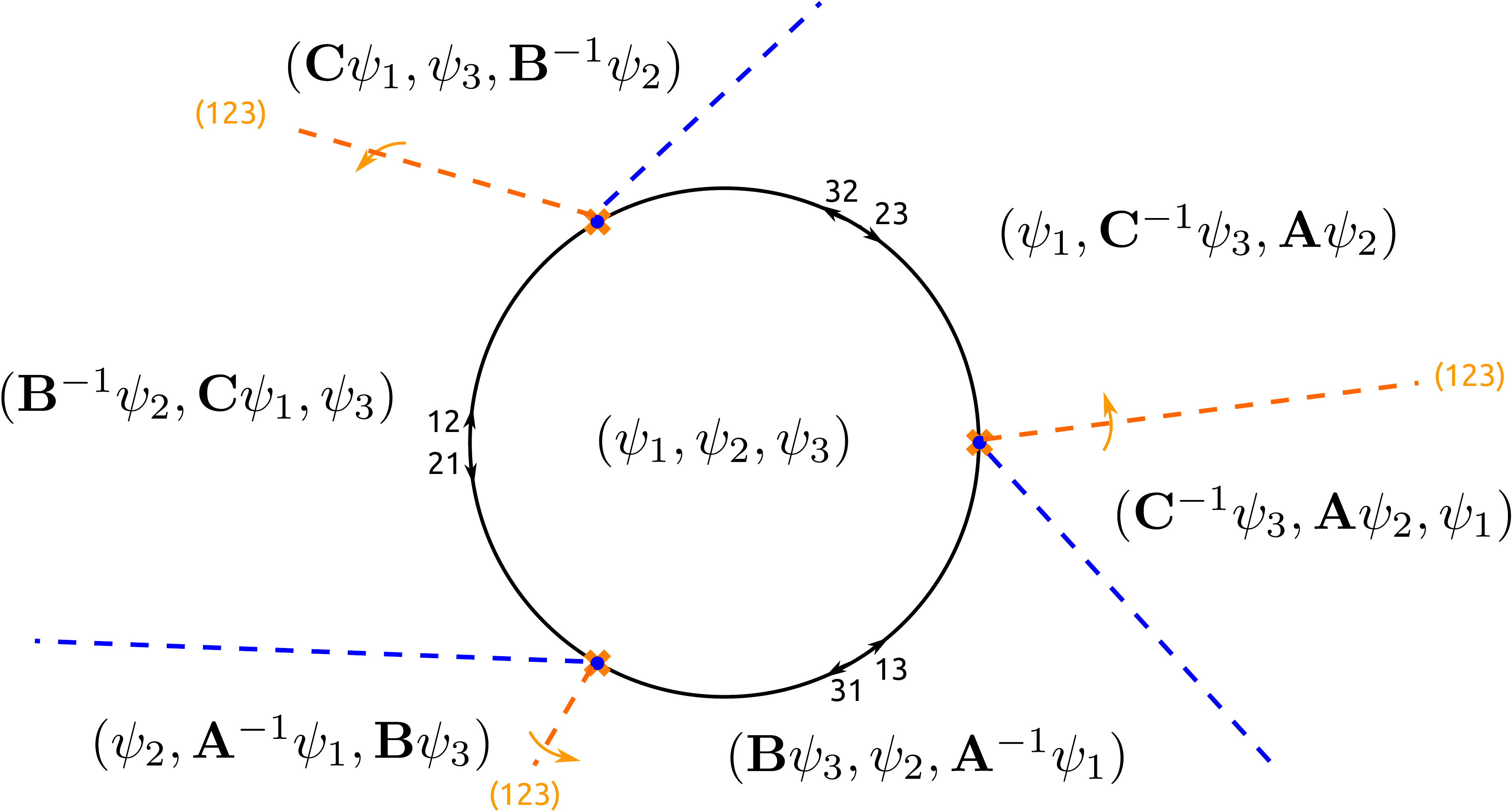}{The Stokes graph from \autoref{fig:circle-network}, with local WKB
bases shown in each domain. The notation is as in the figures above.}

In writing the form of these bases we began by
labeling the basis in the middle as $(\psi_1,\psi_2,\psi_3)$
and then used the facts that:
\begin{itemize}
\item According to \eqref{eq:gluing-2-rank3} 
the $k$-th projective basis element does 
not change when we cross
a wall of type $ij$ and $ji$ (this implies e.g.
that the first basis element in the northeast region
must be $\psi_1$),
\item Crossing a branch cut of the covering $\Sigma \to C$ 
(orange in \autoref{fig:computations-3})
permutes the projective basis elements,
\item The projective bases on the two sides of a monodromy cut (blue in \autoref{fig:computations-3}) differ by the monodromy ($\bA$, $\bB$ or $\bC$) attached to the cut.
	\end{itemize}
One key fact remains to be used: 
again by \eqref{eq:gluing-2-rank3}, for a wall of 
type $ij$ and $ji$, the plane spanned by
the $i$-th and $j$-th basis elements is 
the same on both sides of the wall.
Applying this to the northeast
wall, which is of type $23$ and $32$,
leads to the condition that
\begin{equation}
\IP{\psi_2,\psi_3} = \IP{\bC^{-1} \psi_3, \bA \psi_2}, 
\end{equation}
 which is \eqref{eq:planarity-1}; doing similarly for the other two walls
gives the other two parts of 
\eqref{eq:planarity-relations}.
Thus, the basis $(\psi_1, \psi_2, \psi_3)$ is indeed
a basis in special position. 
Conversely, given a basis $(\psi_1, \psi_2, \psi_3)$ in special position,
the local bases shown in \autoref{fig:computations-3}
give a $\cW$-abelianization. This shows the claimed
identification between $\cW$-abelianizations
and bases in special position.

\medskip

{\bf Computation of \eqref{eq:spectral-coords-t3}.}
As above, all we need to compute are the gluing factors
along the paths representing $\gamma_A$ and $\gamma_B$ shown
in \autoref{fig:circle-network-cycles}.
These factors are given by \eqref{eq:gluing-2-rank3}:
for a wall of type $ij$ and $ji$, and a path on
sheet $i$, the factor is
\begin{equation}
	\sqrt{\frac{[\psi_i^L , \psi_j^L , \psi_k^L]}{[\psi_i^R , \psi_j^R , \psi_k^L]} \frac{[\psi_i^L , \psi_j^R , \psi_k^L]}{[\psi_i^R , \psi_j^L , \psi_k^L]}}.
\end{equation}
Since all the walls are double, we will not need to use
\eqref{eq:gluing-1-rank3} anywhere.

For $\gamma_A$ the computation is particularly simple: only
two of the four crossings give a nontrivial factor,
namely the places where the path crosses the $23$-$32$
wall. This gives
directly 
\begin{align}
\cX_A &= \sqrt{\frac{[\psi_2 , \bA \psi_2 , \psi_1]}{[\bA \psi_2 , \bC^{-1} \psi_3 , \psi_1]} \frac{[\psi_2 , \psi_3 , \psi_1]}{[\bA \psi_2 , \psi_2 , \psi_1]}} \times \sqrt{\frac{[\psi_3 , \bC^{-1} \psi_3 , \psi_1]}{[\bC^{-1} \psi_3 , \bA \psi_2 , \psi_1]} \frac{[\psi_3 , \psi_2 , \psi_1]}{[\bC^{-1} \psi_3 , \psi_3 , \psi_1]}} \\
&= \frac{[\psi_2 , \psi_3 , \psi_1]}{[\bC^{-1} \psi_3 , \bA \psi_2 , \psi_1]}
\end{align}
matching \eqref{eq:XA-t3} as desired. 
The computation giving $\cX_B$ is similar but
a little longer since three of the four crossings give nontrivial
factors: thus we have altogether $6$ factors in
numerator and denominator; one common factor cancels, leaving
the desired \eqref{eq:XB-t3}.

\bibliographystyle{utphys}
\bibliography{sl3-oper-wkb}

\end{document}